\documentclass{svjour3} %
\smartqed %
\usepackage[utf8]{inputenc}
\usepackage[letterpaper,margin=3cm]{geometry}
\usepackage{amsmath}
\usepackage{amsfonts}
\usepackage{amssymb}
\usepackage{listings}
\usepackage{mathrsfs}
\usepackage{times}
\usepackage{float}
\usepackage{color}
\usepackage{setspace}
\usepackage{color,soul}

\usepackage{amsmath,amsfonts,amsthm,eucal}
\usepackage{enumerate}
\usepackage{float}
\usepackage[title]{appendix}
\usepackage{color}
\usepackage{microtype}
\usepackage{siunitx}
\usepackage[
pdfstartview=xyz,
bookmarks=true,
colorlinks=true,
linkcolor=blue,
urlcolor=blue,
citecolor=blue,
pdftex,
bookmarks=true,
linktocpage=true,   %
hyperindex=true
]{hyperref}

\usepackage{lipsum}
\usepackage{lineno}
\usepackage[figbotcap,tabtopcap,bf,tight]{subfigure}

\usepackage{natbib}
\usepackage[pdftex]{graphicx}
\usepackage{epstopdf}
\usepackage{booktabs} 
\usepackage{tikz,mathpazo}
\usetikzlibrary{shapes.geometric, arrows}

\usepackage{caption}

\newcommand{\tensor}[1]{\ensuremath{\boldsymbol{#1}}}

\DeclareMathOperator{\grad}{\nabla^{\tensor x}}
\DeclareMathOperator{\diver}{\nabla^{\tensor x}\cdot}
\DeclareMathOperator{\curl}{\nabla^{\tensor x} \times}

\DeclareMathOperator{\lap}{\nabla^{2}}
\DeclareMathOperator{\sym}{sym}

\theoremstyle{remark}

\renewcommand{\vec}[1]{\ensuremath{\boldsymbol{#1}}}
\DeclareRobustCommand{\myotimes}{%
  \mathbin{\text{%
    \scalebox{0.9}{%
      $
       \mspace{1.5mu}
       \overline{\mspace{-1.5mu}\otimes\mspace{-1.5mu}}
       \mspace{1.5mu}
      $%
    }%
  }}%
}

\newcommand{\reals} {{\mathbb{R}}}
\usepackage{mathtools}
\usepackage{ulem}
\usepackage{array}
\newcolumntype{H}{>{\setbox0=\hbox\bgroup}c<{\egroup}@{}}
\newcommand{\wrt}{with respect to }
\newcommand{\enn}{equivariant neural network }
\newcommand{\ennend}{equivariant neural network.}
\usepackage{paralist}
\usepackage{color}
\definecolor{darkblue}{rgb}{0.0, 0.0, 0.8}
\definecolor{darkred}{rgb}{0.8, 0.0, 0.0}
\definecolor{darkgreen}{rgb}{0.0, 0.8, 0.0}

\usepackage[ruled,linesnumbered]{algorithm2e}
\usepackage[noend]{algpseudocode}

\title{Equivariant geometric learning for digital rock physics: estimating formation factor and effective permeability tensors from Morse graph}
\author{Chen Cai \and Nikolaos Vlassis  \and Lucas Magee \and Ran Ma \and Zeyu Xiong  \and Bahador Bahmani \and Teng-Fong Wong \and Yusu Wang \and WaiChing Sun}
\date{\today}

\titlerunning{Equivariant geometric learning for digital rock physics}

\institute{ \at
Chen Cai, Lucas Magee, Yusu Wang
Halicioglu Data Science Institute, 
University of California, San Diego, California
\and 
Nikolaos N. Vlassis, Ran Ma, Zeyu Xiong, Bahador Bahmani,  WaiChing Sun \at
Department of Civil Engineering and Engineering Mechanics, 
Columbia University, New York, New York  
\and
Teng-fong Wong
Department of Geosciences,
Stony Brook University, Stony Brook, New York
}

\begin{document}

\maketitle

\begin{abstract}
We present a SE(3)-equivariant graph neural network (GNN) approach that directly predicting the formation factor and effective permeability from micro-CT images. 
FFT solvers are established to compute both the formation factor and effective permeability, while the topology and geometry
of the pore space are represented by a persistence-based Morse graph. Together, they constitute the database for training, validating, and testing the neural networks. 
While the graph and Euclidean convolutional approaches both employ neural networks to generate low-dimensional latent space to represent the features of the micro-structures for forward predictions, the SE(3) equivariant neural network is found to generate more accurate predictions, especially when
the training data is limited. Numerical experiments have also shown that the new SE(3) approach leads to predictions that fulfill the material frame indifference whereas the predictions from classical convolutional neural networks (CNN) 
may suffer from spurious dependence on the coordinate system of the training data. Comparisons among predictions inferred from training the CNN and those from graph convolutional neural networks (GNN) with and without the equivariant constraint indicate that the equivariant graph neural network seems to perform better than the CNN and GNN without enforcing equivariant constraints. 
\end{abstract}


\section{Introduction}
Formation factor and effective permeability are both important hydraulic and engineering properties of porous media. While formation factor measures the relative electric resistance of the void space of the porous media to the reference fluid, the 
effective permeability measures the ability of the fully fluid-saturated porous media to transmit fluid under a given pressure gradient. 
Predicting effective permeability and conductivity from micro-structures has been a crucial task for numerous science and engineering disciplines. Ranging from biomedical engineering (cf. \citet{cowin2007tissue}), geotechnical engineering (e.g. \citet{terzaghi1996soil}), nuclear waste disposal to petroleum engineering, the knowledge on permeability may affect the economical values of reservoirs, the success of tissue regeneration, and the likelihood of landslides and earthquakes \citep{white2006calculating, sun2011multiscale, sun2014modeling, kuhn2015stress, paterson2005experimental, suh2021immersed}. 

Effective permeability can be estimated via the establishment of porosity-formation-factor-permeability relationship (cf. \citet{jaeger2009fundamentals, bosl1998study}). This relationship can be completely empirical and established via regression analysis or they can be based on theoretical justification such as fractal geometry assumption \cite{costa2006permeability} and percolation threshold \citep{mavko1997effect}. 
The recent advancement of tomographic imaging techniques has made it possible to obtain 3D images of microstructures of various porous media \citep{arns2004virtual, fredrich2006predicting, sun2018prediction, sun2011connecting}. These newly available images provide a new avenue to establish a fuller and precise picture of the relationship between the pore geometry and macroscopic transport properties. If image segmentation is conducted properly, the pore space inferred from micro-CT images can be directly used to obtain formation factor and effective permeability through inverse problems. Finite volume, finite element as well as recently lattice Boltzmann and Fast Fourier transform solvers can all be employed to solve Poisson's and Stokes' equation to obtain formation factor and effective permeability. These image-based simulations and inverse problems have led to an emerging new research area often referred to as Digital Rock Physics (DRP) \citep{andra2013digitala, andra2013digitalb}. By using images to predict physical properties, the digital rock physics framework could be viewed as a digital twin that enables a non-destructive way to infer material properties in a more repeatable and cost-efficient manner compared to physical tests of which the results are regarded as the ground truth. 

Nevertheless, conducting 3D simulations on images can be quite costly, as the computational time and memory requirement both scale up with the domain size and become the technical barriers for the digital rock physics approach. 
The recent advancements on data science and convolutional neural networks have led to new research direction that focuses on determining the relationships between spatial data set (e.g. microCT images, molecular structures) and mechanical properties (e.g. effective permeability \citep{srisutthiyakorn2016deep, chen2020machine}, and elastic modulus \citep{kim2019first}, glass transition temperature \citep{tao2021machine}) and the development of digital twin solution that 
incorporates neural network and big data predictions \citep{alber2019integrating, kumar2020inverse}. 

To formulate the machine learning problem properly, an effective feature engineering is important
to help deducing low-dimensional features that can be used by the regressor to make predictions \citep{banerjee2019graph, xu2015machine}. 
A conventional digital rock approach that may not involve neural network is to directly represent 
the pore space as a pore network (a weighted undirected graph). 
For instance, \citet{van2016machine, van2018computational} and \citet{van2019preferential} extend the graph theoretic analysis for granular materials \citep{walker2017spatial} to deduce topological features of pore-network that improves the permeability prediction. 

Another approach to facilitate the learning problem is to use convolutional neural networks to capture the spatial and sometimes temporal dependencies among images through the application of relevant filters \citep{vlassis2020geometric,meng2020composite, raissi2019physics, raissi2020hidden, li2018transfer}.
Among the earlier works dedicated for porous media, \citet{srisutthiyakorn2016deep} shows the application of CNN and multi-layer perceptron (MLP) frameworks for predicting scalar-valued permeability of rock samples directly from the 2D and 3D images of rock samples. 
 \citet{wu2018seeing} proposed a novel CNN architecture that utilizes other material descriptors such as porosity in the CNN's fully connected layer. They show that extra physical or geometrical features may enhance the prediction capacity. \citet{sudakov2019driving} study the effect of feature engineering among various geometrical descriptors on the accuracy of the permeability regression task. They empirically show that the conventional 3D CNN outperforms 2D CNN and MLP for permeability prediction. The diffusivity of synthetic 2D porous rock samples is successfully predicted for a wide range of porosity values via a CNN architecture enhanced by incorporating field knowledge (the same idea as \citet{wu2018seeing}), pre-processing of the image input, and customizing the loss function \citep{wu2019predicting}. These earlier studies focus on developing surrogate models for the scalar-valued quantities, while the permeability is usually an anisotropic tensor-valued quantity for heterogeneous rock samples.

 \citet{santos2020poreflow} introduce a convolutional neural network framework called PoreFlow-Net that directly predicts flow velocity field at each voxel of the input image then infer the effective permeability via inverse problems. 
They incorporate additional geometrical and physical features, e.g., Euclidean distance map, in their proposed CNN pipeline to inform the neural network about local and global boundary conditions. They use the L1-norm loss function that makes prediction more robust to outliers compared to the mean squared root loss function. They show that the proposed model generalizes well in dealing with complex geometries not seen in the training stage. 

While the proposed work shows good matches between predicted scalar permeability and the benchmark values, the framework has yet to extend to predict ansiotropic permeability which may require more simulation data and predictions of more complex flow patterns. Furthermore, 
as the classical CNN architecture cannot admit rotational symmetry groups to conform with the frame-indifference property of the flow field, an intractable amount of rotated data set may be required to augmented in 3D applications. 
Finallhy, recent work by \citet{batzner2021se} shows the application of an equivariant neural network that strongly satisfies underlying symmetry properties for learning interatomic potentials with high accuracy even in the limited data regime.

Despite the aforementioned progress, training classical convolutional neural networks that leads to robust property predictions is not trivial. First, a fundamental issue that may lead to erroneous prediction is that the classical CNN is not designed to give predictions independent of coordinate frames. For instance, if the micro-CT image is rotated by a rotation tensor $\tensor{R}$, then the same permeability tensor  represented by the old and new coordination systems should be related by 
\begin{equation}
\tensor{k}' = \tensor{R} k \tensor{R}^{T} 
\label{eq:krotated}    
\end{equation}
whereas the eigenvalues and the three invariants of the permeability tensors %
should be independent of the frame. Furthermore, the effective permeability is expected to remain constant when the images are in rigid body translation. However, while the earlier designs of convolutional neural network do exhibit translational equivariant behavior, the predictions may suffer from spurious dependence due to the rotation of the observer \citep{cohen2016group}. 
As such, the resultant  
predictions on effective permeability and formation factor tensors may therefore exhibit sensitivity on the frame of reference and hence reduce the quality of the predictions. 
Another important issue that deserves attention is the representation of the pore structures. While it is common to use the binary indicator function the Euclidean voxel space of the binary image to represent the pore structures, such an approach may not be most efficient and convenient. For example, the Euclidean representation may incorrectly incorporate ring artifacts into the encoded feature vector.
Furthermore, the micro-CT images are often sufficiently large that the 3D convolutional layer may 
demand significant memory and GPU time and hence make the training expensive \citep{vlassis2020geometric, frankel2019predicting}. 

The goal of this research is to overcome these two major technical barriers to enable faster and more accurate and rigorous predictions on the hydraulic properties (effective permeability and formation factor). 
In particular, we introduce three new techniques that have yet to be considered for geomechanics predictions, (1) the representation of pore geometry via Morse graph, (2) the equivariant neural network that generates predictions equivariant with respect to 3D rotation, and (3) the graph convolutional neural network that generates low-dimensional features to aid predictions for the formation factor and effective permeability. Our results show that the graph neural network (GNN) consistently performs better than the classical convolutional neural network counterpart and the SE(3) \enn performs the best among other options. 

This paper is organized as follows. We first provide a brief account of how the database is generated. Our goal is to introduce supervised learning where the inputs are the pore space represented by either a binary indicator function in the Euclidean physical domain or a Morse graph that represents the topology of the pore space, and the outputs are the formation factor and permeability. 

\section{Database generation via direct numerical simulations with FFT solvers} \label{sec:database}
\begin{figure}[htbp]
\begin{center}
\includegraphics[scale=.5]{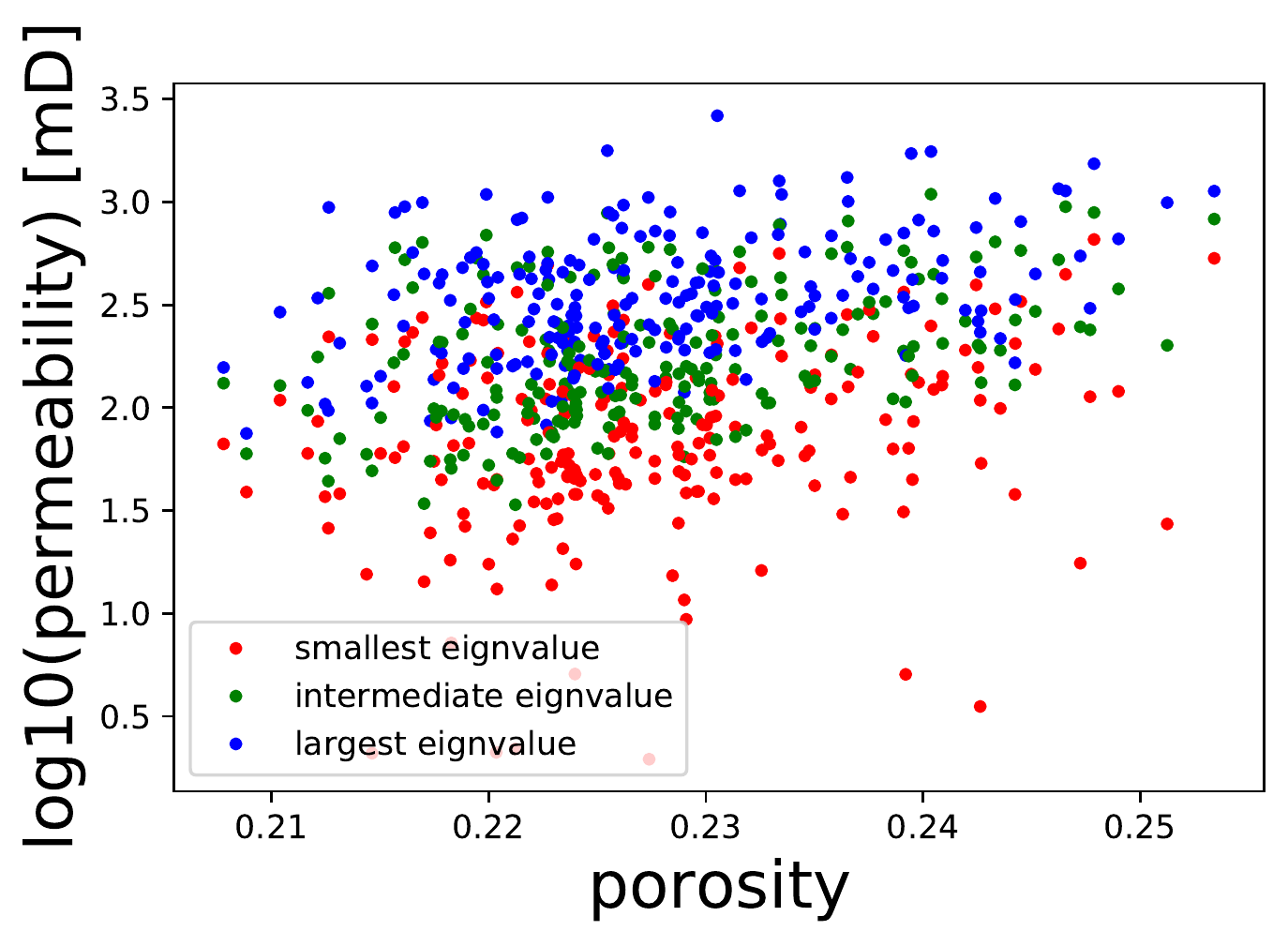}
\includegraphics[scale=.5]{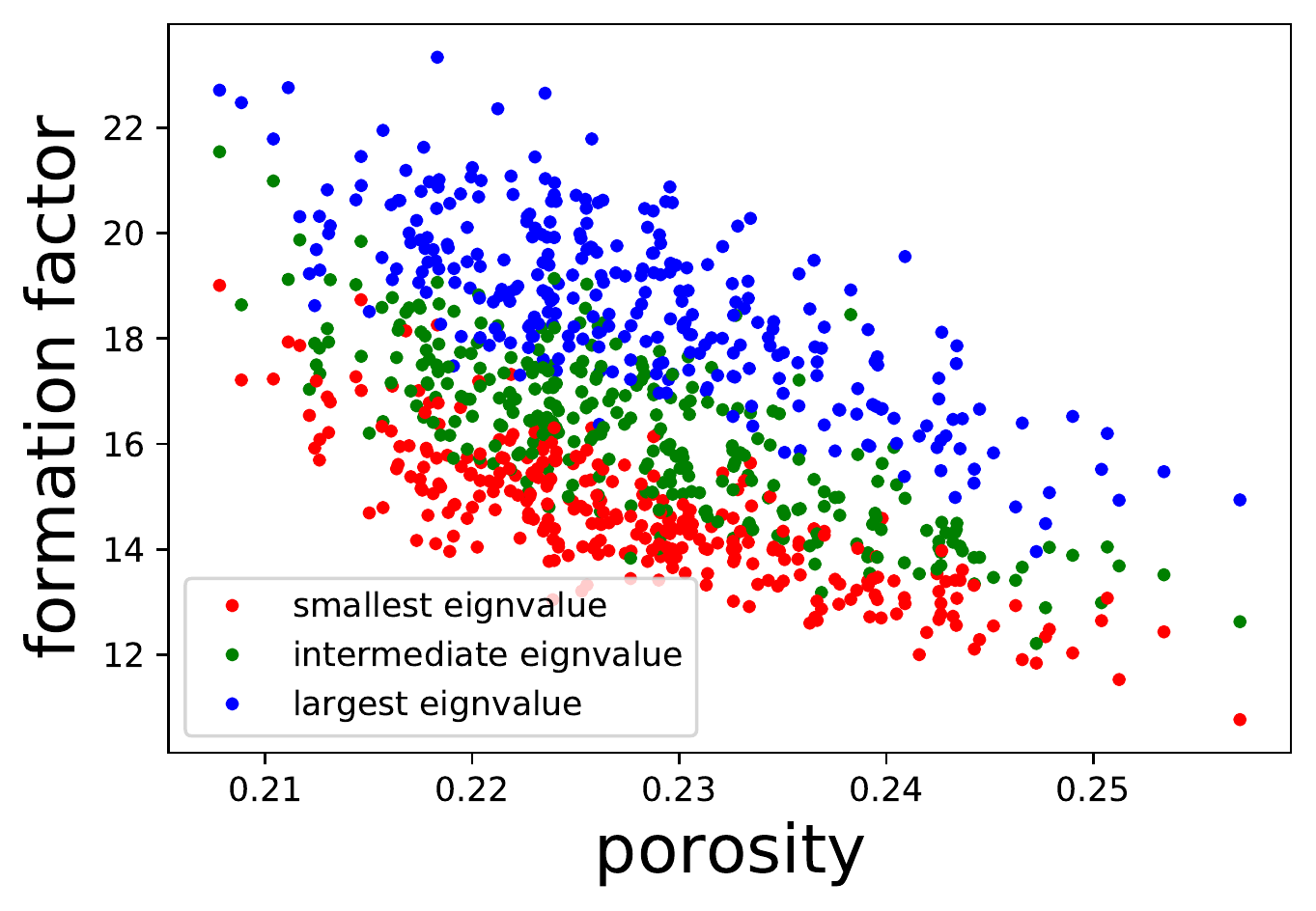}
\caption{Eigenvalues distribution for permeability and formation factor inferred from FFT simulations performed on sub-volumes of a sandstone microCT image. The image voxel resolution is 8-micron and the sub-volume is of 150X150X150 voxels.}
\label{fig:summary}
\end{center}
\end{figure}

To generate the database, we use Fast Fourier Transform Solver to infer both effective permeability and formation factor from the micro-CT images. 
For completeness, the procedures to obtain the formation factor and effective permeability are outlined in the two sub-sections. The results are summarized in Figure \ref{fig:summary}.

\subsection{Effective permeability inferred from micro-CT images}
Consider a unit cell $\Omega = \prod_{\alpha} [-Y_{\alpha}, Y_{\alpha}]$ defined
in the Euclidean space $\mathbb{R}^d$ with $\vec{e}_{\alpha}$ being
the $\alpha$-th basis vector.
A scalar-valued function $f: \mathbb{R}^d \rightarrow \mathbb{R}$ is called a periodic function if
\begin{equation}
f \left( \vec{x} + \sum_{\alpha=1}^{d} Y_{\alpha} k_{\alpha} \vec{e}_{\alpha} \right) = f(\vec{x}), \quad
\forall \vec{x} \in \mathbb{R}^d \textrm{ and } \vec{k} \in \mathbb{Z}^d.
\end{equation}
Furthermore, the Sobolev space of periodic functions $H_{\#}^{s}(\mathbb{R}^d)$ is defined as
\begin{equation}
H_{\#}^{s} (\mathbb{R}^d) = \lbrace f \vert f \in H^{s}(\mathbb{R}^d), f \textrm{ is periodic} \rbrace.
\end{equation}

The unit cell is further divided into the solid skeleton $\Omega_s$ and the fluid-filled void $\Omega_l$,
with $\Omega_f \cup \Omega_s = \Omega$ and $\Omega_f \cap \Omega_s = \varnothing$.
The solid skeleton is a closed set with a periodic pattern, i.e.\ $\bar{\Omega}_s = \Omega_s$.
The local pressure $p:\Omega \rightarrow \mathbb{R}$ within the unit cell is split into the average part and a fluctuation term
\begin{equation}
\grad p(\vec{x}) = \tensor{G} + \grad p^{*}(\vec{x}), \quad
\langle \grad p^{*}(\vec{x}) \rangle_{\Omega}
= \frac{1}{V} \int_{\Omega} \grad p^{*}(\vec{x}) \textrm{ d}V  = \vec{0}
\end{equation}
where $\tensor{G} \in \mathbb{R}^d$ is the macroscopic pressure gradient
and $p^{*}(x):\Omega \rightarrow \mathbb{R}$ is the microscopic pressure perturbation.
Suppose that the incompressible viscous fluid flows within the void region with the flow velocity $\vec{v}: \Omega \rightarrow \mathbb{R}^d$,
such that the velocity field and the pressure field fulfill the incompressible Stokes equation
\begin{equation}
\label{eq:Stokes}
\left\{\begin{matrix}
\mu_f \lap \vec{v} (\vec{x}) + \grad p (\vec{x}) = 0, \quad
\diver \vec{v} = 0 &,& \quad \forall \vec{x} \in \Omega_f
\\ 
\vec{v} (\vec{x}) = 0 &,& \quad \forall \vec{x} \in \Omega_s,
\end{matrix}\right.
\end{equation}
where $\mu_f$ is the viscosity.
Then, the permeability tensor $\tensor{\kappa}$ of the porous media is computed as
\begin{equation}
\langle \vec{v} \rangle_{\Omega} = - \frac{\tensor{\kappa}}{\mu_f} \tensor{G}.
\end{equation}

An FFT-based method is utilized to solve the Stokes equation \eqref{eq:Stokes} within
the porous media \citep{monchiet2009fft}.
A fourth order compliance tensor $\tensor{M}(x)$ is first introduced
\begin{equation}
\sym \left( \grad \vec{v} \right) = \tensor{M}(x) : \tensor{\sigma}, \quad
\tensor{M}(x) =
\left\{\begin{matrix}
\frac{1}{2 \mu_f} \left( \mathbb{I} - \frac{1}{3}\tensor{I} \otimes \tensor{I} \right)
&,& \quad \forall \vec{x} \in \Omega_f
\\ 
\tensor{0} &,& \quad \forall \vec{x} \in \Omega_s,
\end{matrix}\right.
\end{equation}
where $\mathbb{I}$ is the fourth order identity tensor,
$\tensor{I}$ is the second order identity tensor,
and a small value $\zeta$ is introduced to maintain the stress continuity
within the solid skeleton.
The incompressible Stokes equation \eqref{eq:Stokes} is equivalently written as
\begin{equation}
\label{eq:LS_stokes}
\tensor{\Gamma} (\vec{x}) \ast \left[ \tensor{M} (\vec{x}) : \left( \tensor{\sigma}_0 (\vec{x})
+ \delta \tensor{\sigma} (\vec{x}) \right) \right] = \tensor{0}
\Rightarrow
\tensor{\Gamma} (\vec{x}) \ast \left( \tensor{M} (\vec{x}) : \delta \tensor{\sigma} (\vec{x}) \right)
= - \tensor{\Gamma} (\vec{x}) \ast \left( \tensor{M} (\vec{x}) : \tensor{\sigma}_0 (\vec{x}) \right)
\end{equation}
where $\ast$ denotes the convolution operator.
The fourth order Green's tensor $\tensor{\Gamma}$ is defined in the frequency space as \citep{nguyen2013fourier}:
\begin{equation}
\hat{\tensor{\Gamma}} = 2 \mu_0 \left( \tensor{\beta} \otimes \tensor{\beta}
+ \tensor{\beta} \myotimes \tensor{\beta} \right), \quad
\tensor{\beta} = \tensor{I} - \frac{1}{\vert \vec{\xi}^2 \vert} \vec{\xi} \otimes \vec{\xi}, \quad
(\tensor{a} \myotimes \tensor{b})_{ijkl} = \frac{1}{2} \left( a_{ik} b_{jl} + a_{il} b_{jk} \right).
\end{equation}
where $\xi_{\alpha} = k_{\alpha} \slash 2 Y_{\alpha}, \vec{k} \in \mathbb{Z}^d$,
and $\mu_0$ is a reference viscosity.
For $\vec{\xi} = \vec{0}$, $\hat{\tensor{\Gamma}} = \tensor{0}$ because of the zero-mean property.
For an arbitrary second-order tensor field $\tensor{T}(\vec{x})$,
$\tensor{\Gamma} \ast \tensor{T}$ is divergence free and its spatial average vanishes.
Also, $\tensor{\Gamma} \ast \tensor{\varepsilon} = 0$ for any
compatible strain field $\tensor{\varepsilon}(\vec{x}) = \sym(\grad \vec{u})$.
The macroscopic pressure gradient $\vec{G}$ is introduced through the initial guess of
the stress distribution $\tensor{\sigma}_0$ \citep{monchiet2009fft,nguyen2013fourier}:
\begin{equation}
\tensor{\sigma}_0 = - \tensor{\Lambda} (\vec{x}) \ast \vec{f} (\vec{x}),
\end{equation}
where $\vec{f} (\vec{x})$ extends the macroscopic pressure gradient $\vec{G}$ from
the void $\Omega_s$ to the unit cell $\Omega$,
and $\tensor{\Lambda} (\vec{x})$ projects $\vec{f}$ to a divergence-free stress field.
The third order tensor field $\tensor{\Lambda} (\vec{x})$ and
the vector field $\vec{f} (\vec{x})$ is derived as
\begin{equation}
\tensor{\Lambda}_{ijk} (\vec{\xi}) = 
\left\{\begin{matrix}
\frac{i}{\vert \vec{\xi} \vert^4} \left[ \left( \delta_{ij} \xi_{k}
+ \delta_{ik} \xi_{j} + \delta_{jk} \xi_{i} \right) \vert \vec{\xi} \vert^2
- 2 \xi_i \xi_j \xi_k \right] &,& \quad \vec{\xi} \neq 0
\\ 
0 &,& \quad \vec{\xi} = 0,
\end{matrix}\right.
\end{equation}
\begin{equation}
\vec{f} (\vec{x}) =
\left\{\begin{matrix}
\tensor{G} &,& \quad \forall \vec{x} \in \Omega_f
\\ 
- \frac{1 - c_s}{c_s} \tensor{G} &,& \quad \forall \vec{x} \in \Omega_s.
\end{matrix}\right.
\end{equation}
where $c_s$ defines the volume fraction of the solid phase.

\begin{figure}[htbp]
\begin{center}
\subfigure[porosity=0.257]{
\includegraphics[width=0.45\textwidth]{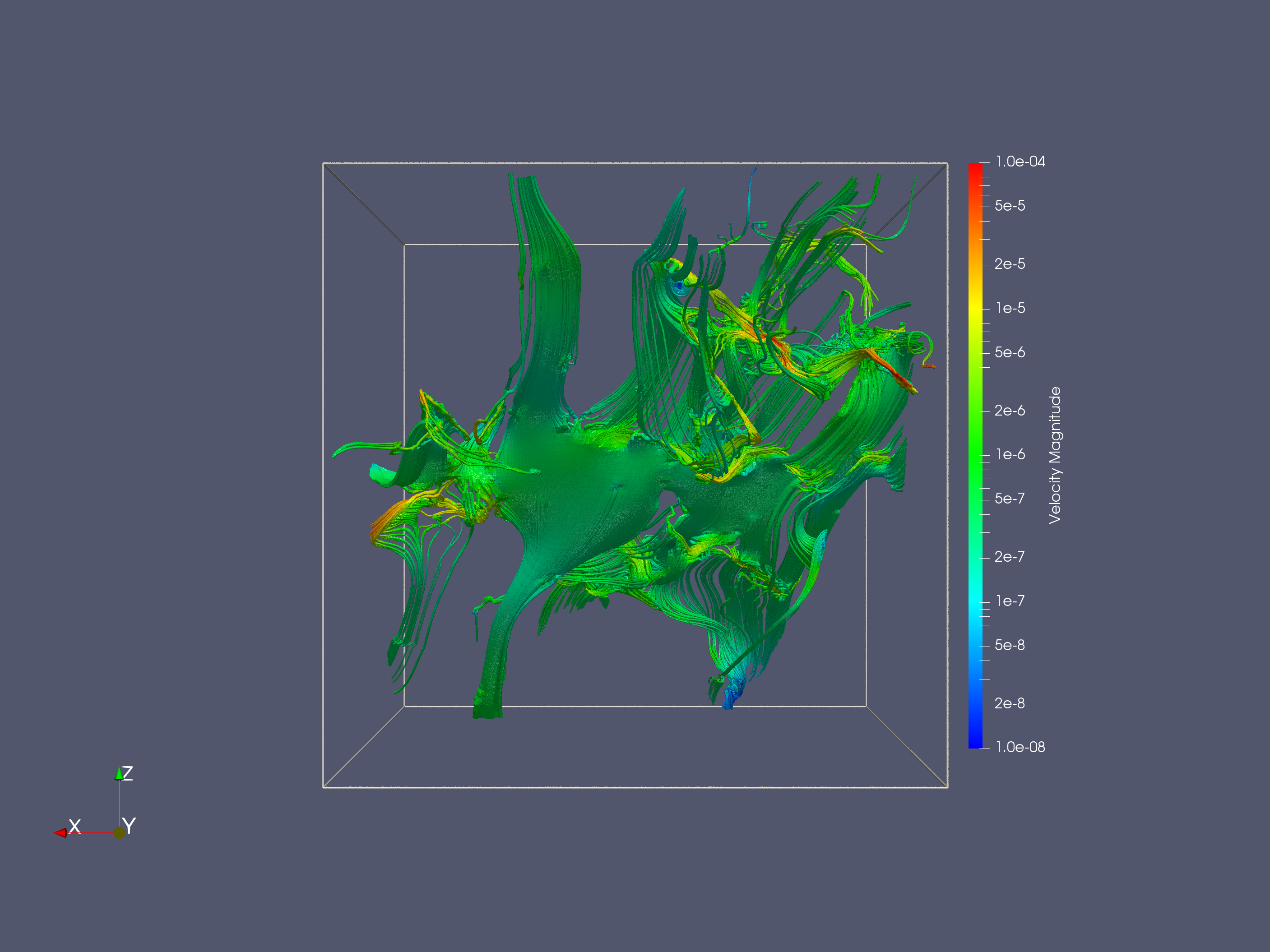}
}
\subfigure[porosity=0.228]{
\includegraphics[width=0.45\textwidth]{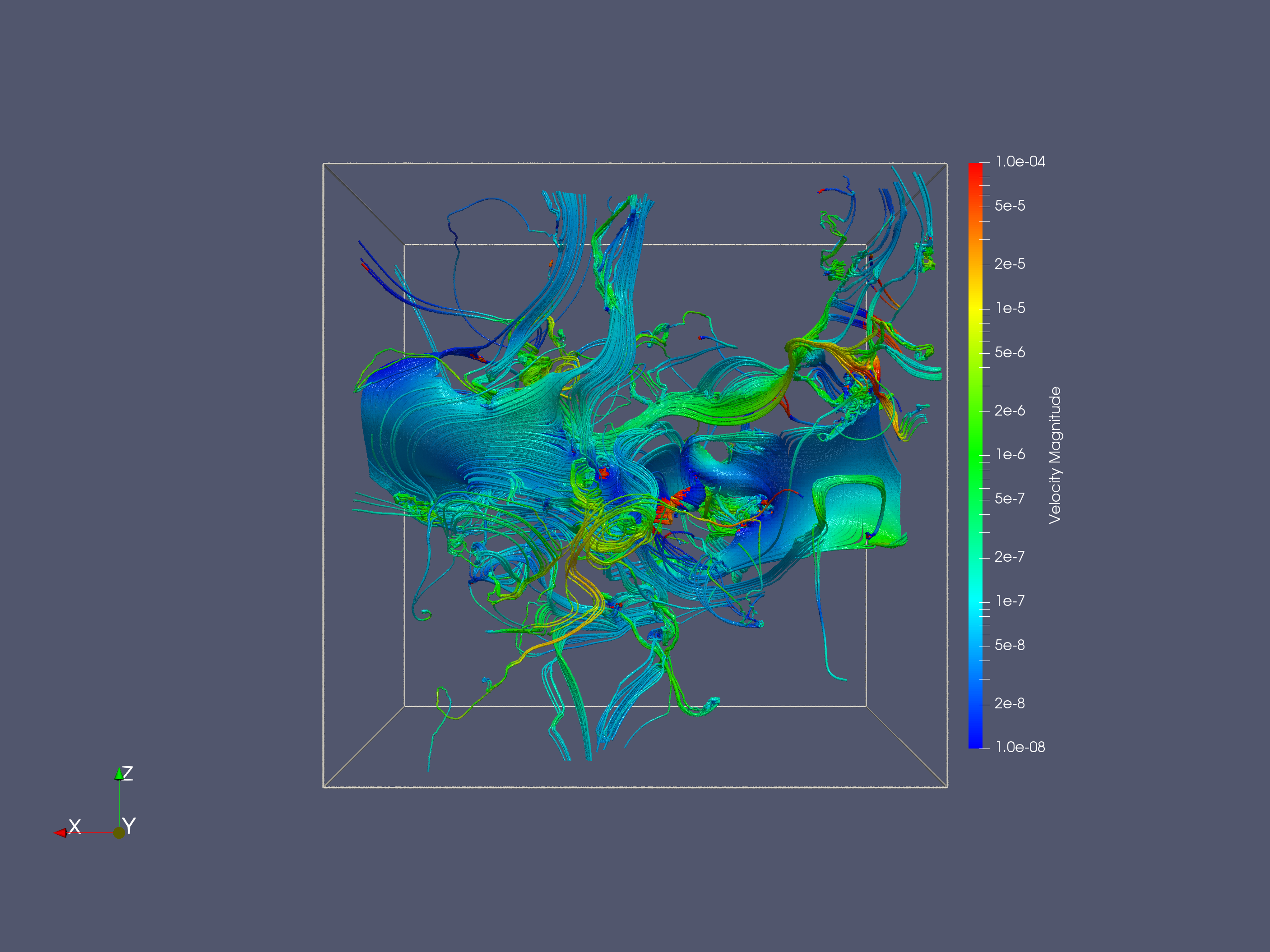}
}
\subfigure[porosity=0.223]{
\includegraphics[width=0.45\textwidth]{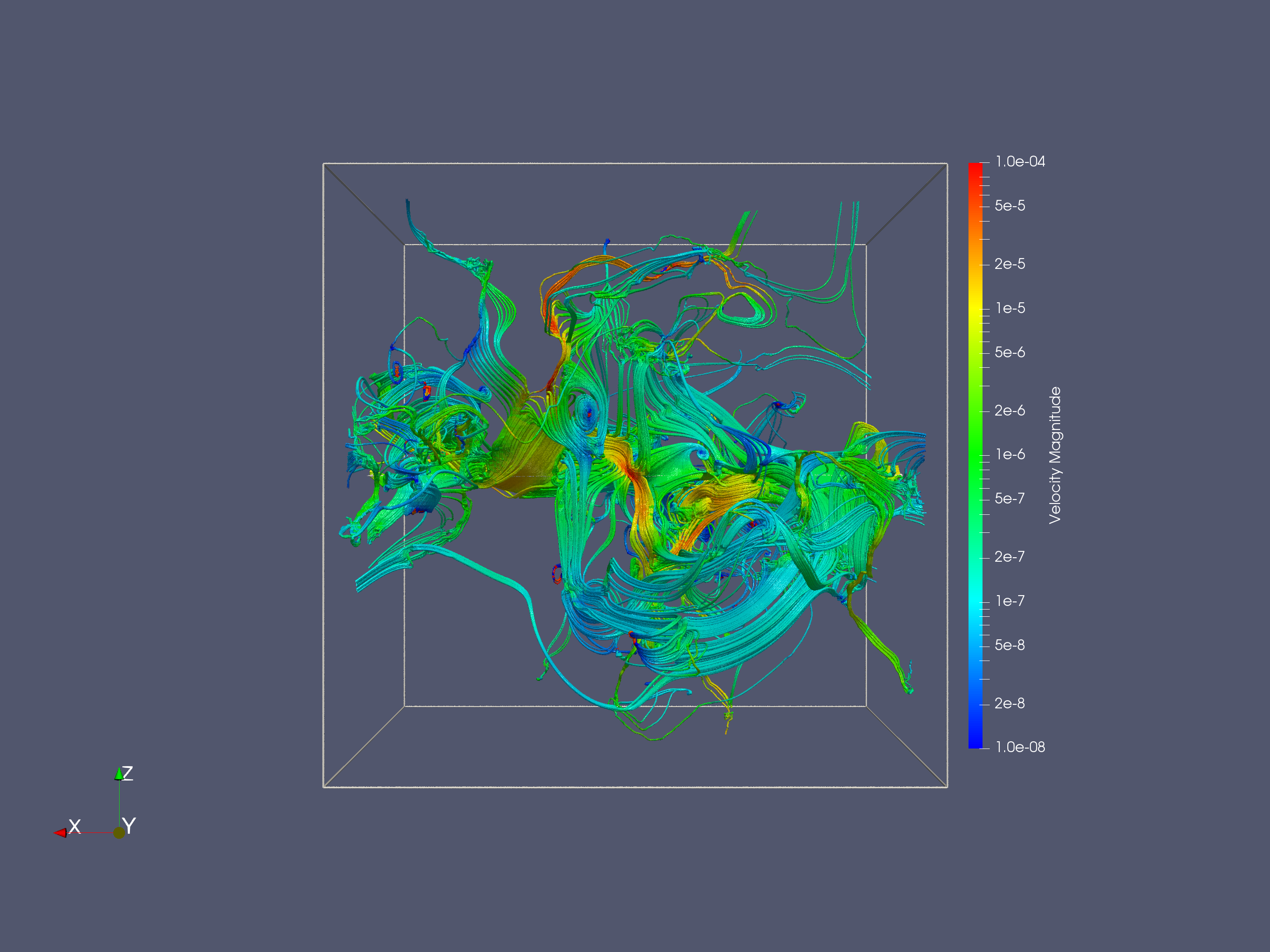}
}
\subfigure[porosity=0.208]{
\includegraphics[width=0.45\textwidth]{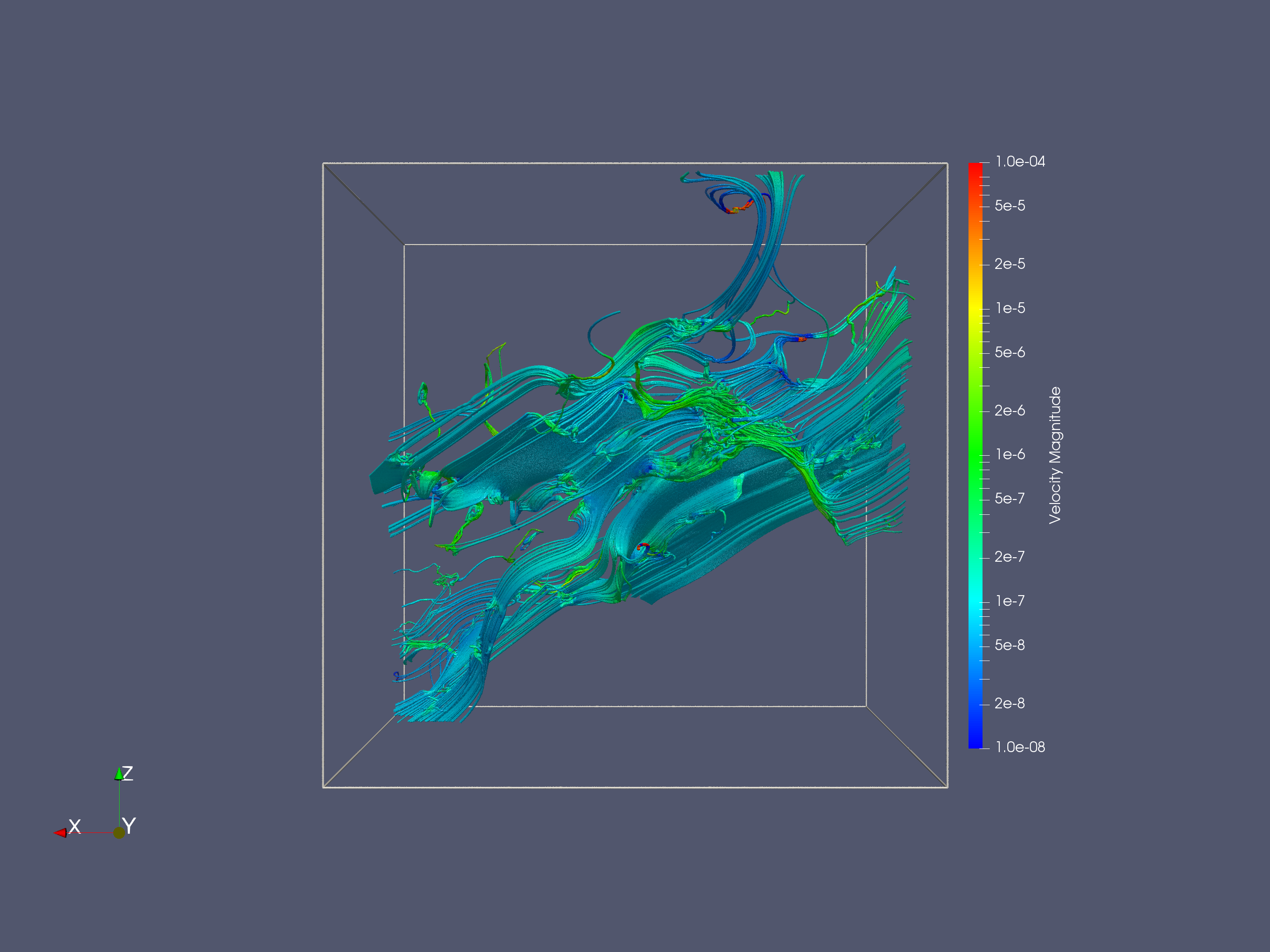}
}
\caption{Streamline of the flow velocity for four RVE with porosity = 0.257 (UPPER LEFT), 0.228 (UPPER RIGHT), 0.223 (LOWER LEFT) and 0.208 (LOWER RIGHT) obtained from FFT direct numerical simulations in which the effective permeabilities are extracted.}
\label{fig:dnspermeability}
\end{center}
\end{figure}

Equation \eqref{eq:LS_stokes} is linear and is solved by the conjugate gradient method.
Figure \ref{fig:dnspermeability} shows the streamlines of flow velocity obtained four FFT simulations performed on RVE with porosity ranging from 0.208 to 0.257. To aid the visualization, 
the color map of the streamline is plotted on the logarithmic scale. 
Note that the 
true flow velocity is equal to
Darcy's velocity divided by the porosity (assuming that the area and volume fraction of void are equal) \citep{bear2013dynamics}. Hence, if two microstructures are of the same effective permeability but with different porosities, the one with the lower porosity is expected to have higher flow velocity. 
What we observed in the streamlines plotted in Figure \ref{fig:dnspermeability} nevertheless indicates that the flow speed is lower in the RVE of lower porosity. This observation is consistent with the porosity-permeability relationship established in previous work such as \citet{andra2013digitalb}.

\subsection{Formation factor inferred from micro-CT images}
The formation factor $F$, which is defined as the ratio between the fluid conductivity
and the effective conductivity of the saturated porous rock,
is used to quantify the electric resistivity of a porous media relative to that of water. 
The formation factor is considered as a physical property that solely depends on the geometry and 
topology of the pore space in the porous media \citep{archie1942electrical}. 
Previous studies have shown that the formation factor is significantly influenced
by the electric double layer between the solid phase and the liquid phase,
and the surface conductivity should be taken into consideration
to perform a micro-CT image-based formation factor prediction at the pore scale \citep{zhan2010pore}.

Suppose that all the related physical quantities, including the local conductivity $\tensor{A}$,
the current density $\vec{J}$, and the electrostatic potential gradient $\vec{e}$,
are defined within the periodic unit cell $\Omega$, the charge conservation equation reads:
\begin{equation}
\label{eq:conductivity}
\curl \vec{e}(\vec{x}) = \vec{0}, \quad
\vec{J}(\vec{x}) = \tensor{A}(\vec{x}) \cdot \vec{e}(\vec{x}), \quad
\diver \vec{J}(\vec{x}) = 0.
\end{equation}
The local electrostatic potential gradient field $\vec{e}(\vec{x})$ is further split into
its average $\vec{E}$ and a fluctuation term $\vec{e}^{*}(\vec{x}) \in H_{\#}^{1} (\mathbb{R}^d)$:
\begin{equation}
\vec{e}(\vec{x}) = \vec{E} + \vec{e}^{*}(\vec{x}), \quad
\langle \vec{e}^{*}(\vec{x}) \rangle_{\Omega} = \vec{0}.
\end{equation}

The charge conservation equation \eqref{eq:conductivity} defined within
the periodic domain $\Omega$ is equivalently written as the periodic Lippmann-Schwinger equation:
\begin{equation}
\label{eq:LS_FF}
\mathbb{G}(\vec{x}) \ast \left[ \tensor{A} (\vec{x}) \cdot \left( \vec{E}
+ \delta \vec{e} (\vec{x}) \right) \right] = \vec{0}
\Rightarrow
\mathbb{G} (\vec{x}) \ast \left[ \tensor{A} (\vec{x}) \cdot \delta \vec{e} (\vec{x}) \right]
= - \mathbb{G} (\vec{x}) \ast \left[ \tensor{A} (\vec{x}) \cdot \vec{E} \right].
\end{equation}
The fourth order Green's operator $\mathbb{G}$ is provided in the frequency space
as \citep{vondvrejc2014fft}:
\begin{equation}
\hat{\mathbb{G}}(\vec{\xi}) =
\frac{\vec{\xi} \otimes \vec{\xi}}{\vec{\xi} \cdot \tensor{A}_0 \cdot \vec{\xi}}.
\end{equation}
where $\tensor{A}_0$ is a reference conductivity.
For $\vec{\xi} = \vec{0}$, $\hat{\mathbb{G}} = \tensor{0}$ because of the zero-mean property.
By applying the averaging electric potential gradient $\vec{E}$ upon the unit cell,
the local current density $\vec{J}$ is obtained by solving Equation \eqref{eq:LS_FF}
with conjugate gradient method.

\begin{figure}[htbp]
\begin{center}
\includegraphics[width=0.7\textwidth]{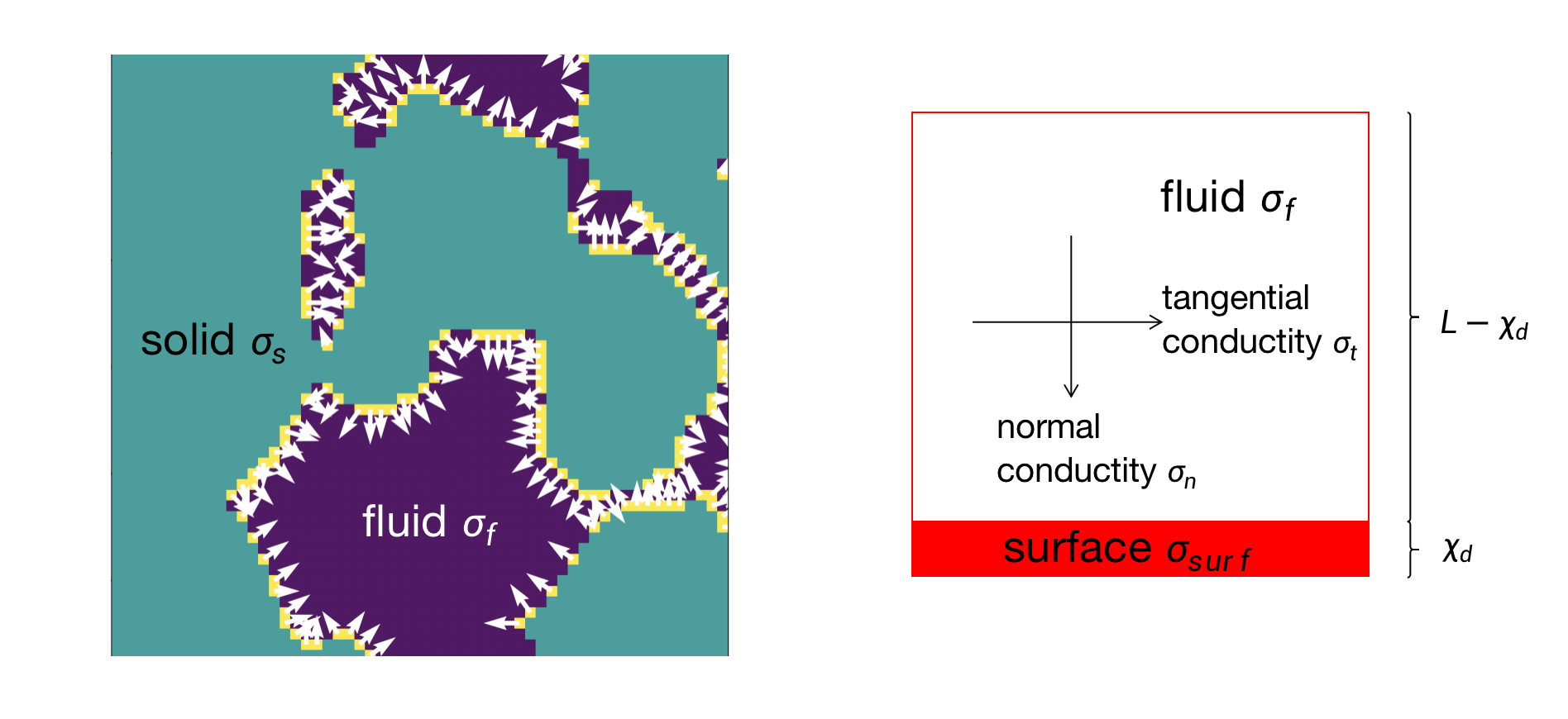}
\caption{Surface and unit normal vectors between solid and fluid phases (LEFT) and structure of electric double layer (RIGHT).}\label{fig:surface_conductivity}
\end{center}
\end{figure}

It has been demonstrated that the anisotropic conductivity within the interfacial region
has a strong influence on the formation factor due to the presence of electric double layers between the solid and liquid constituents \citep{weller2013relationship, bussian1983electrical}. However, the thickness of the double layer is usually smaller than the voxel length. As such, running the FFT electric conductance simulations on the three-phase porous media consisting of solid, double layer and water is not feasible. In this work, we extend the upscaling approach in \citep{zhan2010pore} to first compute the tensorial effective conductivity of the surface voxel such that the voxels at the solid-fluid interfacial regions are upscaled to effective values analytically. 
Then, the FFT simulation is run to determine the effective conductivity of the entire RVE (see Figure \ref{fig:surface_conductivity}). 

The interface was first identified from the three-dimensional binary image, which was labeled as yellow voxels with the unit normal vectors in Figure \ref{fig:surface_conductivity} (LEFT). To extract the interface, each voxel was analyzed along with the 26 closest neighboring voxels in the 3x3x3 window centered around it. The isolated voxels, whose neighbors are all in the opposite phase, were eliminated. Then the solid voxels neighboring the fluid were selected to represent the interface if they have at least 9 fluid neighboring voxels. Furthermore, as shown in Figure \ref{fig:surface_conductivity} (RIGHT), the interface voxels were modeled as electric double layers, consisting of a layer of fluid and a layer with conductivity $\sigma_{surf}$. The normal component $\sigma_n$ of the interface conductivity was calculated as two electrical resistors in series while the tangential component $\sigma_t$ was computed by resistors in parallel as shown in \eqref{eq:surface_conductivity_1}.
\begin{equation}
\label{eq:surface_conductivity_1}
\sigma_t=\frac{L-\chi_d}{L}\sigma_{f}+\frac{\chi_d}{L}\sigma_{surf} \; , \; 
\sigma_n=[(\frac{L-\chi_d}{L}\sigma_{f})^{-1}+(\frac{\chi_d}{L}\sigma_{surf})^{-1}]^{-1}
\end{equation}
where $L$ is the voxel length, $\chi_d$ is the thickness of the surface layer, $\sigma_{f}$ is the fluid conductivity.
To determine the normal vectors of the interface, the signed distance function was generated by an open-source software ImageJ (cf. \citet{abramoff2004image}), in which the value of each voxel represents its shortest distance to the interface. The normal vectors $\boldsymbol{n}_1$ were then determined by the gradient of the signed distance function as shown in Figure \ref{fig:surface_conductivity} (LEFT). The upscaled electric conductivity tensors $\boldsymbol{\sigma}_3$ of the interface voxel are calculated via 
\begin{equation}
\label{eq:surface_conductivity_2}
\boldsymbol{\sigma}_3=\sigma_n\boldsymbol{n}_1\otimes\boldsymbol{n}_1+\sigma_t\boldsymbol{n}_2\otimes\boldsymbol{n}_2+\sigma_t\boldsymbol{n}_3\otimes\boldsymbol{n}_3. 
\end{equation}
where $\vec{n}_{1}$ is the normal vector of the solid surface and $\vec{n}_2$ and $\vec{n}_3$ 
are two orthogonal unit vectors that spans the tangential plane of the surface.
This approach is different from \citet{zhan2010pore} in the sense that the analytical electric conductivity at the solid-fluid boundary is tensorial and anisotropic instead of volume-averaged and isotropic. 
Such a treatment enables more precise predictions of the anisotropy of the surface effect on the electric conductivity that is important for estimating the conductivity tensor.

Once all voxels in the spatial domain are assigned with the corresponding electrical conductivity, we obtain a heterogeneous domain where electric conductivity may vary spatially. The effective conductivity tensor 
can be obtained through computational homogenization via the FFT solver described above.
The effective conductivity $\tensor{\sigma}_{\textrm{eff}}$ is defined as a second-order tensor
which projects the applied electric potential gradient $\vec{E}$ to the homogenized
current density as:
\begin{equation}
\label{eq:effective_conductivity}
\langle \vec{J}(\vec{x}) \rangle_{\Omega} = \tensor{\sigma}_{\textrm{eff}} \cdot \vec{E}.
\end{equation}
Since the charge conservation equation \eqref{eq:conductivity} is linear,
the effective conductivity $\tensor{\sigma}_{\textrm{eff}}$ is constant for each micro-CT image.
Then, the formation factor $\tensor{F}$ is defined as the ratio between the fluid conductivity $\sigma_{f}$ and the effective conductivity $\tensor{\sigma}_{\textrm{eff}}$ as:
\begin{equation}
\label{eq:formation_factor}
\tensor{F} = \sigma_{f}  \tensor{\sigma}_{\textrm{eff}}^{-1}.
\end{equation}

\begin{figure}[htbp]
\begin{center}
\subfigure[porosity=0.257]{
\includegraphics[width=0.45\textwidth]{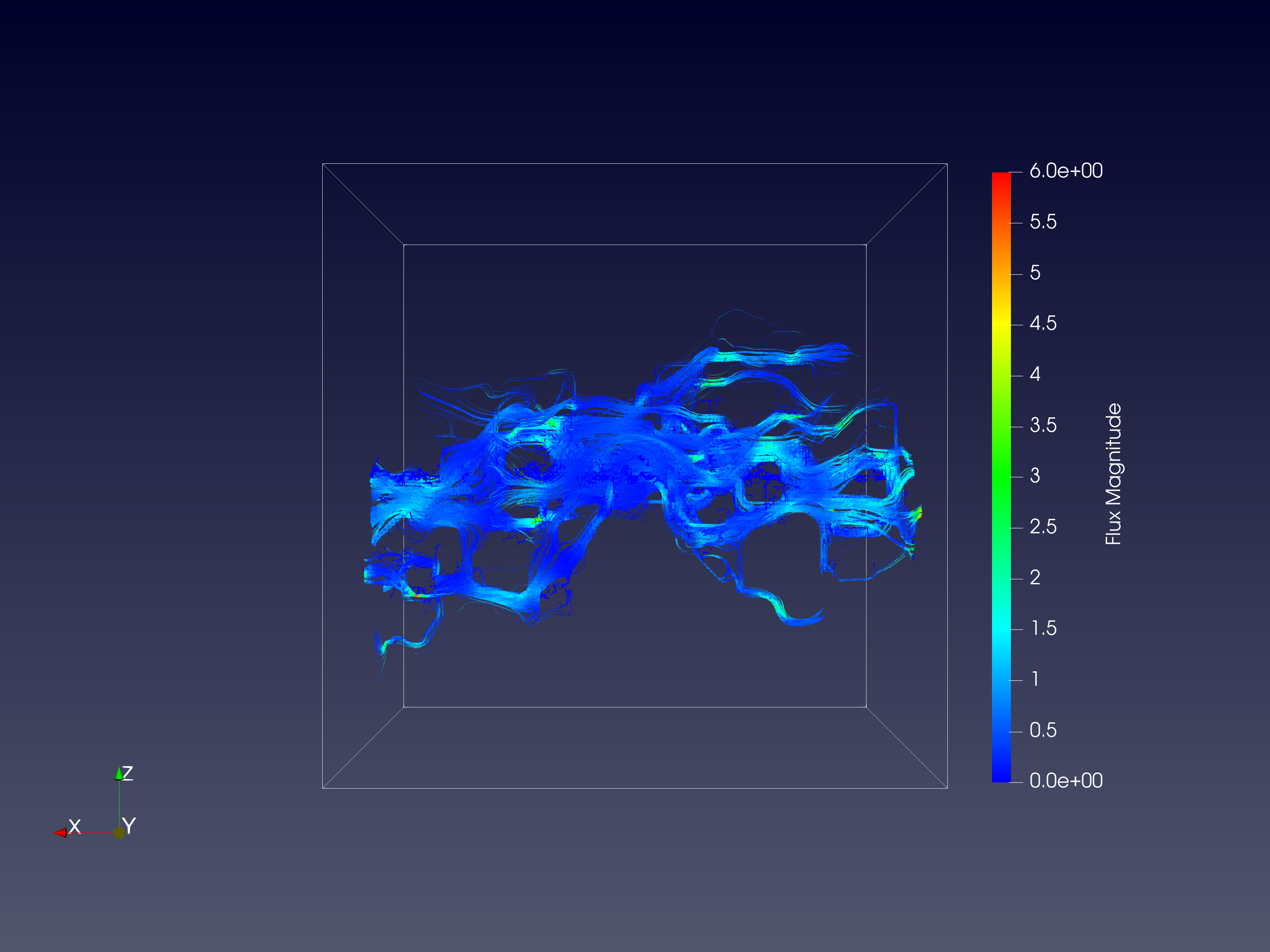}
}
\subfigure[porosity=0.228]{
\includegraphics[width=0.45\textwidth]{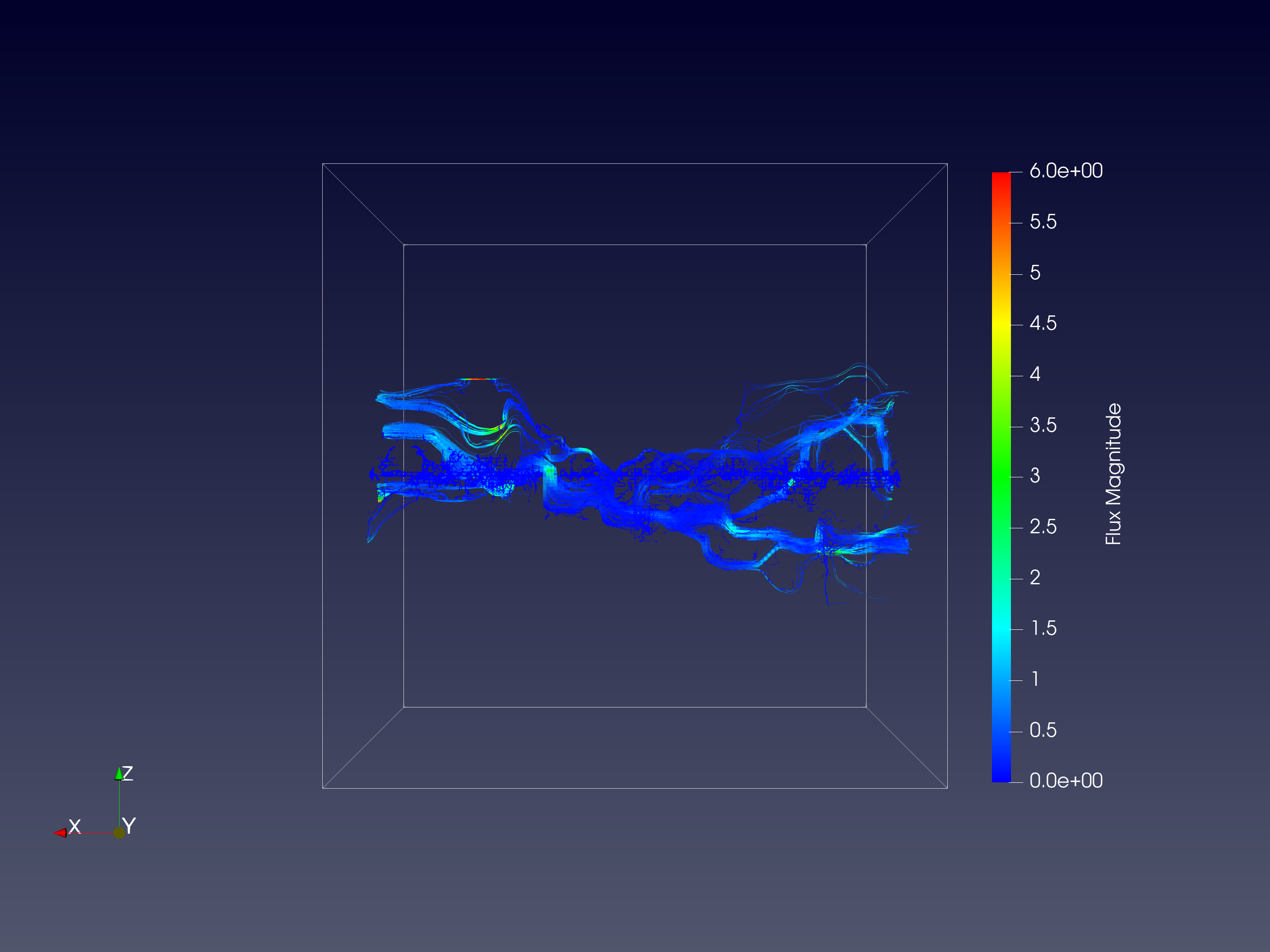}
}
\subfigure[porosity=0.223]{
\includegraphics[width=0.45\textwidth]{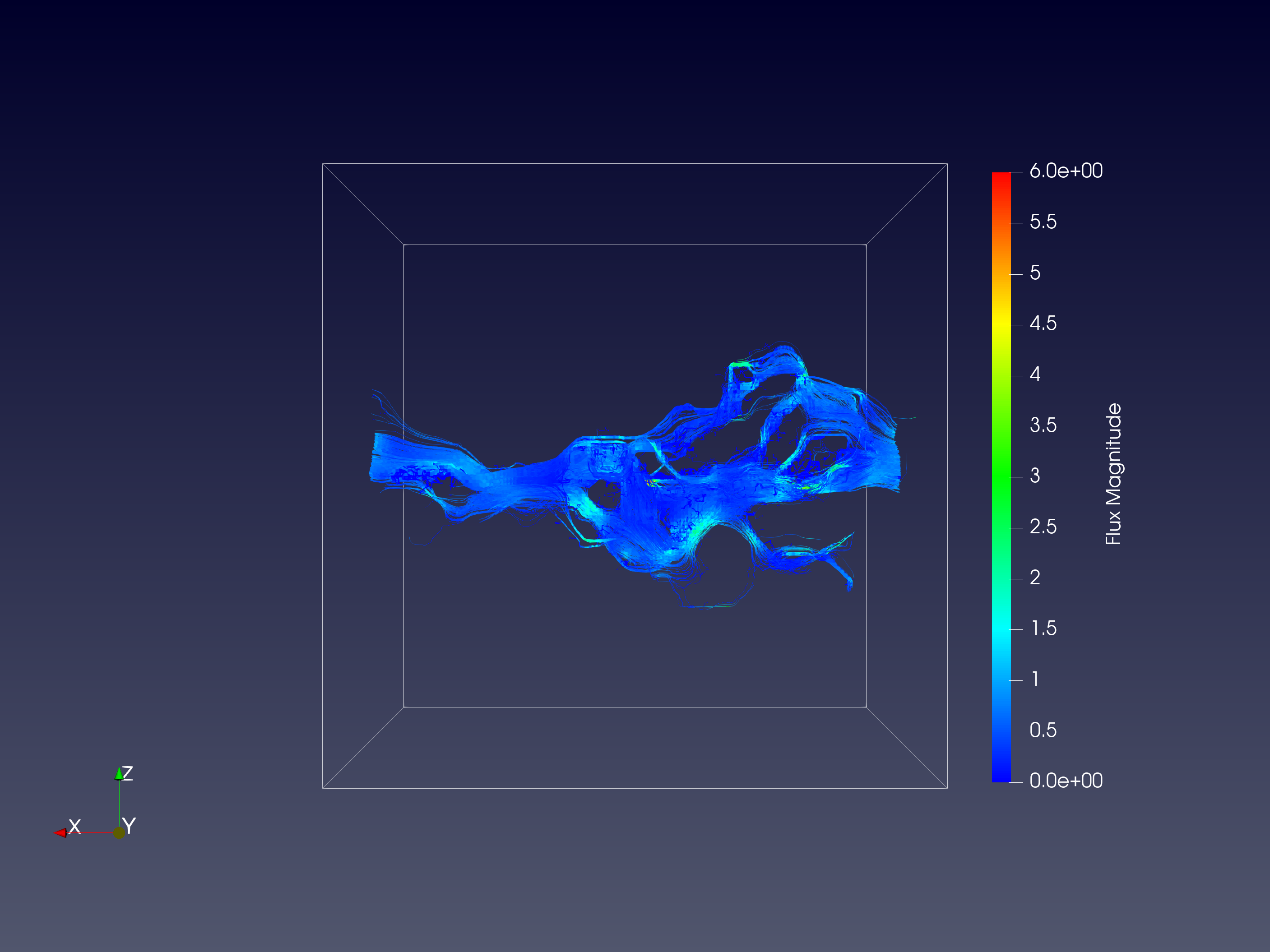}
}
\subfigure[porosity=0.208]{
\includegraphics[width=0.45\textwidth]{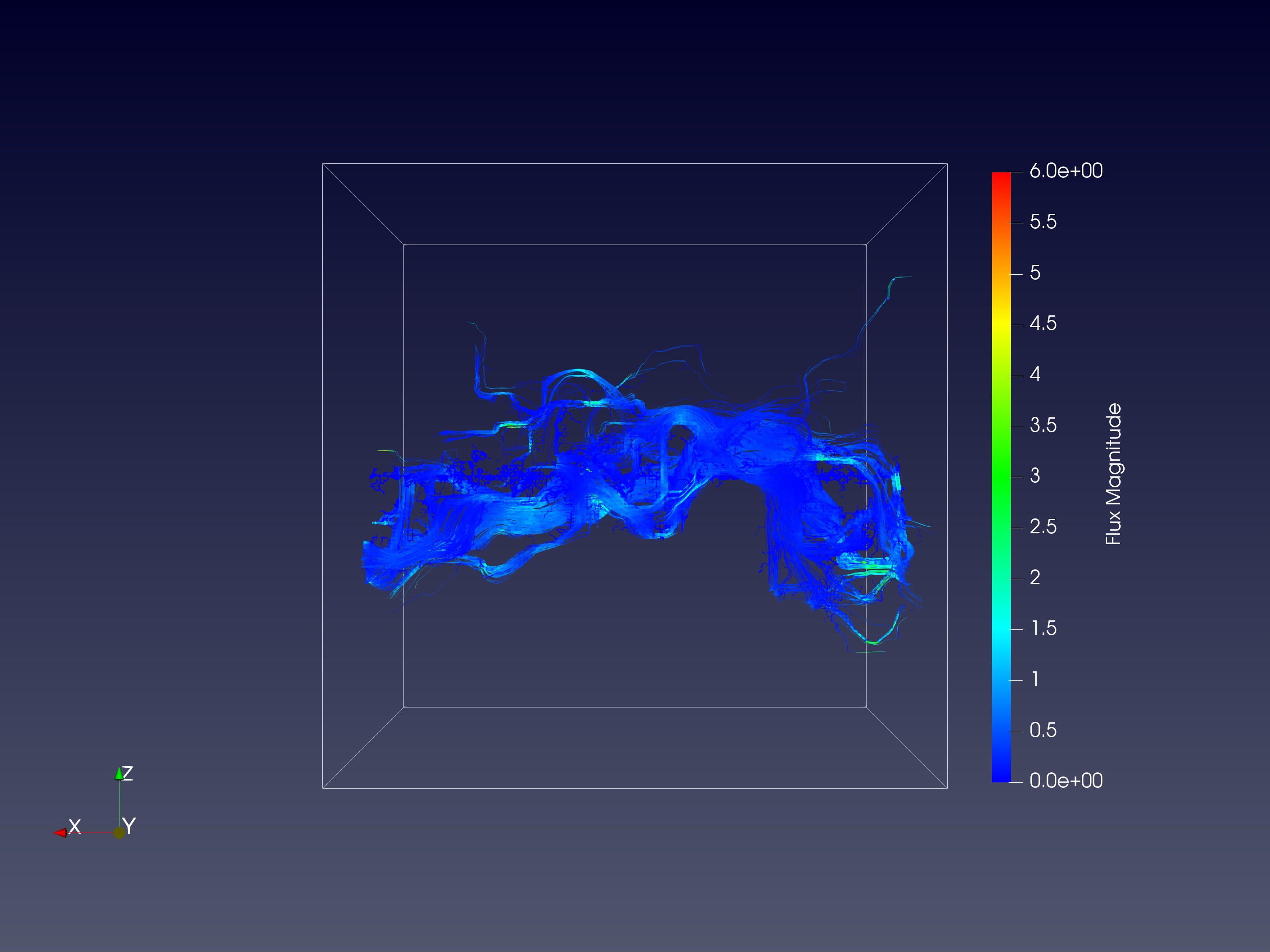}
}
\caption{Streamline of the gradient of electric current for four RVE with porosity = 0.257 (UPPER LEFT), 0.228 (UPPER RIGHT), 0.223 (LOWER LEFT), and 0.208 (LOWER RIGHT) obtained from FFT direct numerical simulations in which the effective conductivities are extracted.}
\label{fig:dnsconduct}
\end{center}
\end{figure}

Figure \ref{fig:dnsconduct} shows the simulated streamlines of the electric current along the x-axis for the same RVEs plotted in Figure \ref{fig:dnspermeability}. As expected, the RVEs with larger porosity tend to be more conductive. However, it is also obvious that the streamlines in Figures \ref{fig:dnspermeability} and \ref{fig:dnsconduct} also exhibit significantly different patterns. This difference in the streamline 
pattern could be attributed to the surface effect of the electric conductivity and could be 
affected by the periodic boundary condition enforced by the FFT solver.

Regardless of the reasons, the streamline patterns of the inverse problems may dictate the difficulty of the neural network predictions of the formation factor and permeability and will be explored further in Section \ref{sec:result}. 

\remark{The homogenized permeability $\tensor{\kappa}$ is non-symmetric in general
since the macroscopic pressure gradient $\tensor{G}$ and
the homogenized velocity $\langle \vec{v} \rangle$ are not power conjugate to each other.
In fact, the homogenized stress $\langle \tensor{\sigma} \rangle$ and the homogenized
velocity gradient $\langle \grad \vec{v} \rangle$ are power conjugate
and their inner product equals to the input power due to the Hill-Mandel condition.
Similar results are also reported previously \citep{white2006calculating,sun2018prediction}.
On the other hand, the formation factor $\tensor{F}$ is symmetric,
since the homogenized current density $\langle \vec{J} \rangle$ and
the applied electric potential gradient $\vec{E}$ are power conjugate to each other.}

\section{Discrete Morse Graph Generation}
To predict permeability, one could directly apply a CNN-type architecture on the input 3D images. However, recognizing that the effective permeability is in fact determined by the pores among grains, it is thus intuitive that a representation capturing such a pore network is more informative and provides better inductive bias for the learning algorithm. 

Our framework will first construct a graph-skeleton representation of the ``pore" networks from the 3D images, using the discrete Morse-based algorithm of \citep{DWW18}, which we refer to as DM-algorithm herein. 
Below we first briefly provide the intuition behind this algorithm. We then describe the enrichment of such a DM-graph representation, which we will be used as input for a GNN architecture. Note that our experimental results (presented in Tables \ref{tab:comppermeability} and \ref{tab:compformationfactor}) show that such a graph skeleton + GNN framework indeed leads to much better prediction accuracy compared to an image + CNN-based framework. 

\subsection{Discrete Morse based graph skeletonization}

\begin{figure}[thbp]
\begin{center}
\includegraphics[width = 1\linewidth]{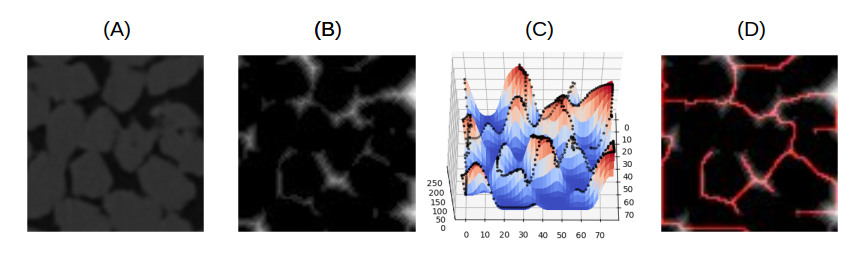}
\end{center}
\vspace*{-0.1in}
\caption{An example of discrete Morse graph reconstruction on 2D data. (A) A single image from our 3D stack. (B) The signed distance from the boundary of (A).  (C)  The image (B) is converted into a triangulation and density function. The discrete Morse graph reconstruction then captures the mountain ridges of the density function. (D) These ridges capture maximal distance from the boundary of (A).}
\label{fig:Morse}
\end{figure}
To provide the intuition of the DM-algorithm of \citet{WWL15, DWW18}, first consider the continuous case where we have a density function $\rho: \reals^d\to \reals$. In our setting, an image can be thought of as a discretization of a function on $\reals^3$. 
If we view the graph of this density function as a terrain in $\reals^{d+1}$ (see Figure \ref{fig:Morse} A), then its graph skeleton can be captured by the ``mountain ridges" of this density terrain, as intuitively, the density on such ridges is higher than the density off the ridges. 
Mathematically, such mountain ridges can be described by the so-called 1-stable manifolds from Morse theory, which intuitively correspond to curves connecting maxima (mountain peaks) to saddles (of the index ($d-1$)) and to other maxima, forming boundaries separating different basins/valleys in the terrain. 
To simplify the terrain and only capture ``important" mountain ridges, we use the persistent homology theory \citep{zomorodian2005computing}, one of the most important developments in the field of topological data analysis in the past two decades, and simplify those less important mountain peaks/valleys (deemed as noise).
As the graph is computed via the global mountain ridge structures, this approach is very robust to gaps in signals, as well as non-uniform distribution of signals.

In the discrete setting, our input is a cell complex as a discretization of the continuous domain. For example, our input is a 3D image, which is a cubic complex decomposition of a subset of $\reals^3$, consisting of vertices, edges, square faces, and cube cells. The 1-stable manifolds in Morse theory are differential objects, and sensitive to discretization. Instead, one can use the \emph{discrete Morse theory} developed by Robin Forman \citep{For95}, which is not a discretization of the classical Morse theory, but rather a combinatorial analog of it. Due to the combinatorial nature, the resulting algorithm is very stable and simplification via persistent homology can easily be incorporated. The final algorithm is conceptually clean and easy to implement. 
Some theoretical guarantees of this algorithm are presented in \citep{DWW18} and it has already been applied to applications, e.g, in geographic information system and neuron-image analysis \citep{WWL15, DWW17, DWW19, BMWL20}.

\subsection{Morse Graph representation of pore space}\label{sec:morse}

For each simulation, we have a corresponding 150 x 150 x 150 binary image stack representing the rock boundary.   We first compute the signed distance to the boundary on this domain, shift all non-zero values by 255 minus the maxima signed distance in the domain (this assures the new maxima signed distance is 255), and apply a Gaussian filter with $\sigma=1$.  We then run the discrete Morse algorithm with a persistence threshold $\delta = 48$.

\begin{figure}[thbp]
\begin{center}
\includegraphics[width = 1\linewidth]{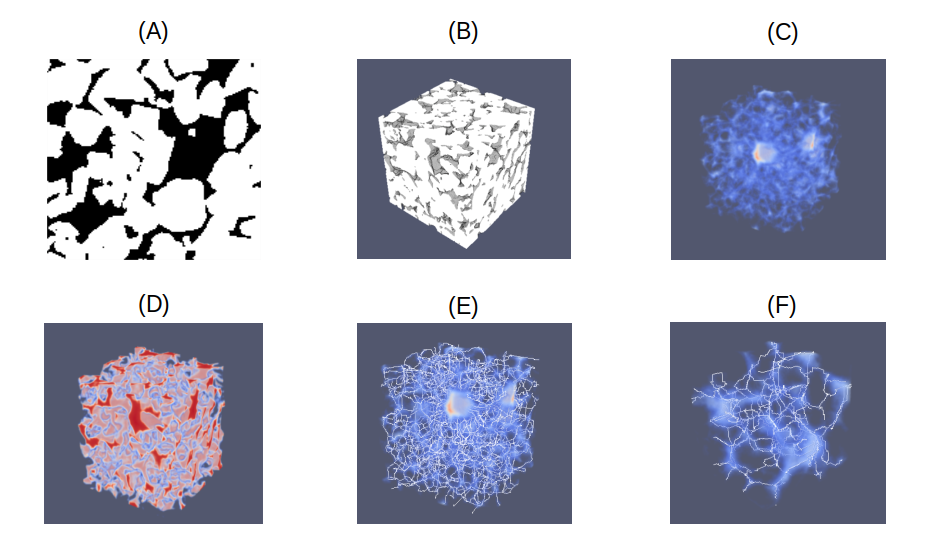}
\end{center}
\vspace*{-0.1in}
\caption{The Morse graph pipeline we use on our data. (A) A single image from our 3D stack. (B) The 3D volume of the entire image stack. (C)  The signed distance function from the boundary in (B). (D) The signed distance function is shifted by 255 - the max value of the signed distance function at all locations is nonzero.  Then a Gaussian filter with sigma equal to 1 is applied.  This is the input function for discrete Morse graph reconstruction. (E) The discrete Morse graph reconstruction output (persistence threshold = 48) overlayed on the original signed distance function. (F) The discrete Morse graph reconstruction output on the first octant of the original signed distance function. By zooming in we see that the graph is accurately capturing the signed distance function. }
\label{fig:Morse_Pipeline}
\end{figure}

On the output Morse graph for each simulation, we assign features used later by the GNN.  The features assigned to each node are its coordinates $(x, y, z)$, its (gaussian smoothed) signed distance value, its original signed distance value, whether or not the vert is a maximum, and three node flow values.  Node flow value is calculated from the underlying vector field computed during the Morse graph calculation.  Every node in the domain will be connected to a single node in the Morse output in the vector field. The first node flow feature simply counts the number of nodes that flow into that node 
the second counts the number of non-boundary nodes that flow into the node, and the thirds sums the function values of nodes that flow into the node. The features assigned to each edge are whether or not it is a saddle, and its persistence value if it is a saddle (all non-saddle edges have a persistence value $\leq \delta$, and we simply assign -1), length, and minima, average, and total function value of vertices along the edge for both the input Morse function and the original signed distance function.  Prior to removing degree two nodes discussed in the next paragraph, each edge has length 1, and zero is assigned to the minima, average, and total function features.  These values are properly updated at every removal of a degree two node. 

Finally, for the sake of computational efficiency in training the GNN, we reduce the number of nodes and edges in the graph by removing nodes of degree two that are not maxima or adjacent to a saddle edge.  The only node features from the original graph that need to be updated are the node flow feature.  For each node removed in the simplified graph, it lies along a unique edge connecting to vertices in the simplified graph.  We add the removed node's flow features to the node flow features of the node in the direction of the maxima on this edge.  The edge features mentioned previously are also updated to capture the information lost by removing nodes.

\section{Equivariant GNN}

We now introduce the notation of equivariance and strategies to build equivariant neural networks. The equivariant neural networks we designed here take the Morse graph described in Section \ref{sec:morse} as input and outputs 
the formation factor and effective permeability tensors.
One of the key challenges of predicting property tensor lies in the frame indifference. Given a frame $B_1$, we can represent the position of each graph node \wrt $B_1$ and obtain the prediction from the model \wrt $B_1$. If we change the frame from $B_1$ to a different frame $B_2$, the model prediction \wrt $B_2$ will certainly be different in terms of numbers but in principle, two predictions under different frames represent the same geometric object. The requirement that model prediction is independent of arbitrary-chosen frames is natural and can be mathematically formulated as \textbf{equivariance}. 

However, the traditional neural network has no such inductive bias built-in, and therefore not ideal for permeability prediction. To achieve this goal of frame indifference, we will adapt the recent progress on the equivariant neural network, a set of neural networks that is equivariant to symmetry transformations. We first outline the notion of the group, its representation, and feature types, and then introduce equivariance and strategies of building an \ennend

\textbf{Group, its representation, and feature types:}
Let $G$ be a set. We say that $G$ is a group with law of composition
$\star$ if the following axioms hold:
$1)$ closure of $\star$: the assignment $(g, h) \rightarrow g\star h \in G$, defines a function $G \times G \rightarrow G$. We call $g\star h$ the product of $g$ and $h$. 
$2)$ existence of identity: there exists a $e_G \in G$ such that for every $g \in G$, we have $g\star e_G = g = e_G\star g$. We call $e_G$ an identity of $G$.
$3)$ existence of inverse: for every $g \in G$, there exists an $h \in G$ such that $h\star g = e_G = g\star h$. We call such $h$ an inverse of $g$.
$4)$ associativity: for any $g, h, k \in G$ we have $(g\star h)\star k = g\star (h\star k)$.

A group representation $\rho: G \rightarrow GL(N)$ is a map from a group $G$ to the set of $N \times N$ invertible matrices $GL(N)$. Critically $\rho$ is a group homomorphism; that is, it satisfies the following property $\rho(g_1\star g_2) = \rho(g_1)\rho(g_2)$ for any $g_1, g_2 \in G$ where the multiplication on the right side of equality denotes the matrix multiplication. Specifically for 3D rotations group $SO(3)$, we have a few interesting properties: 1) its representations are orthogonal matrices, 2) all representations can be decomposed as 

\begin{equation}
\rho(g)=\mathbf{Q}^{\top}\left[\bigoplus_{\ell} \mathbf{D}_{\ell}(g)\right] \mathbf{Q},
\end{equation}

where $\mathbf{Q}$ is an orthogonal, $N\times N$ change-of-basis matrix called Clebsch-Gordan coefficients. $D_l$ for $l = 0, 1, 2, ...$ is a $(2l+ 1) \times (2l+ 1)$ matrix known as a Wigner-D matrix. $\oplus$ is the direct sum of matrices along the diagonal. Features transforming according to $D_l$ are called type-$l$ features. Type-0 features (scalars) are invariant under rotations and type-1 features (vectors) rotate according to 3D rotation matrices. A rank-2 tensor decompose into representations of dimension 1 (trace), 3 (anti-symmetric part), and 5 (traceless symmetric part). In the case of symmetric tensors such as permeability and formation factor, they are $0\oplus2$ features.

\textbf{Equivariance: }
Given a set of transformations $T_{g}: \mathcal{V} \rightarrow \mathcal{V}$ for $g\in G$ where $G$ is a group. a function $\phi: \mathcal{V} \rightarrow \mathcal{Y}$ is equivariant if and only if for every $g\in G$ there exists a transformation$S_{g}: \mathcal{Y} \rightarrow \mathcal{Y}$ such that 
$$
S_{g}[\phi(v)]=\phi\left(T_{g}[v]\right) \quad \textit{for all } g \in G, v \in \mathcal{V}
$$
In this paper, we are interested in building neural networks that are equivariant \wrt the 3D rotation group $SO(3)$. 

Efforts have been made to build equivariant neural networks for different types of data and symmetry groups such as 3D rotation group SO(3) \citep{thomas2018tensor, weiler20183d, fuchs2020se, cohen2019gauge, finzi2020generalizing, weiler2019general, cohen2019gauge, worrall2019deep} and permutation group \citep{zaheer2017deep, qi2017pointnet, keriven2019universal, maron2018invariant, maron2019provably, bevilacqua2021equivariant}. In the case of 3D roto-translations tensor field network (TFN) \citep{thomas2018tensor} is a good choice of parametrizing map $\phi$. It by construction satisfies the equivariance constraint and recently has been shown universal \citep{dym2020universality}, i.e., any continuous equivariant function on point clouds can be approximated uniformly on compact sets by a composition of TFN layers.

\textbf{Equivariant Neural Network}: Just as convolution neural network (CNN) is composed of linear, pooling and nonlinear layers, one also has to follow the same principle to build an equivariant neural network. The major challenge is to characterize the equivariant linear map and design the architecture to parametrize such maps. We sketch how to build linear equivariant layers below. Note that the input of \enn is a Morse graph constructed from 3D voxel images. For simplicity of presentation, we follow the original paper Tensor Field Network (see below) and use the point cloud as our input. Adding connection between point clouds can be easily done by modifying Equations \ref{tfn-conv}.

\textbf{Tensor Field Network}:
Tensor Field Network (TFN) is one of the equivariant neural networks targeted for point clouds. 
Tensor Field Network maps feature fields on point clouds to feature fields on point clouds under the constraint of SE(3) equivariance, the group of 3D rotations and translations. For the point clouds $\{\mathbf{x}_{i}\}_{i=1}^{N}$ of size $N$, the input is a field $f: \reals^3 \rightarrow \reals^d$ of the form
\begin{equation}
\label{eq:tfn}
f(\mathbf{x})=\sum_{j=1}^{N} f_{j} \delta\left(\mathbf{x}-\mathbf{x}_{j}\right)
\end{equation}
where $\{f_j\}$ are node features such as atom types or degrees of Morse graphs, $\delta$ is the Dirac delta function, and $\{\mathbf{x}_{j}\}$ are the 3D point coordinates. In order to satisfy the equivarance constraint, each $f_j \in \mathbb{R}^{d}$ has to be a concatenation of vectors of different types, where a subvector of type-$l$ is denoted as $f^l_j$. A TFN layer takes type-$k$ feature to type-$l$ feature via the learnable kernel $\mathbf{W}^{\ell k}: \reals^3 \rightarrow \reals^{(2l+1)(2k+1)}$. The type $l$ output at position $\mathbf{x}_i$ is 

\begin{figure}[htbp]
\begin{center}
\includegraphics[scale=1]{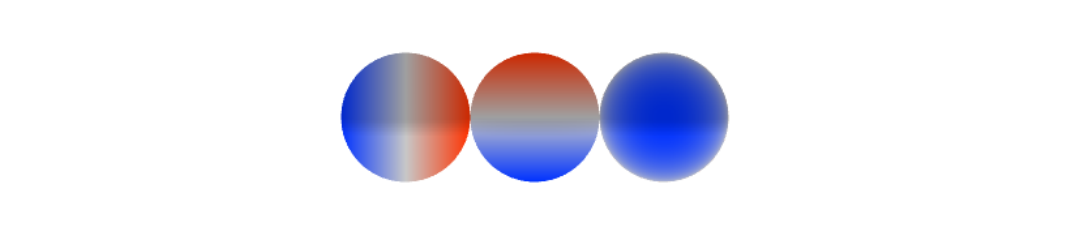}
\includegraphics[scale=1]{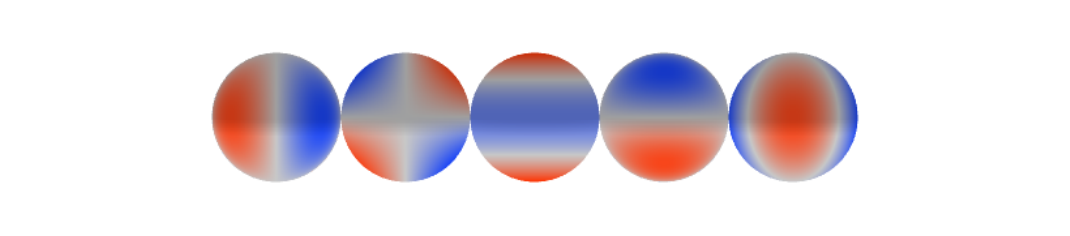}
\includegraphics[scale=1]{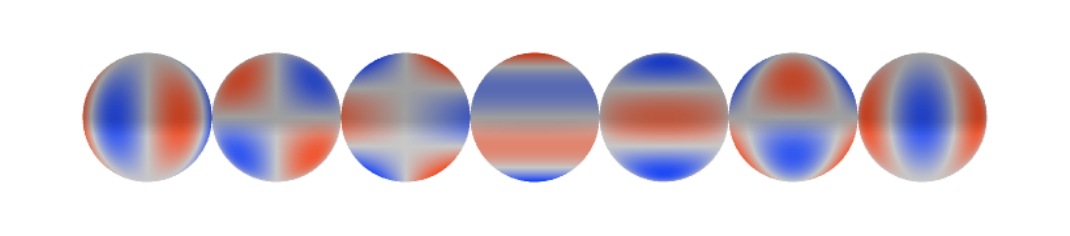}
\caption{We visualize the real-valued spherical harmonics $Y_{1}, Y_{2}, Y_{3}$ in each row, where the color and density indicates the sign and absolute value of the functions. Spherical harmonics are solutions of Laplace equations for functions on the sphere. They form a complete set of orthogonal functions and are used to parametrize the weights of an \ennend}
\end{center}
\end{figure}

\begin{equation}
\label{tfn-conv}
\mathbf{f}_{\mathrm{out}, i}^{\ell}=\sum_{k \geq 0} \underbrace{\int \mathbf{W}^{\ell k}\left(\mathbf{x}^{\prime}-\mathbf{x}_{i}\right) \mathbf{f}_{\mathrm{in}}^{k}\left(\mathbf{x}^{\prime}\right) \mathrm{d} \mathbf{x}^{\prime}}_{k \rightarrow \ell \text { convolution }}=\sum_{k \geq 0} \sum_{j=1}^{n} \underbrace{\mathbf{W}^{\ell k}\left(\mathbf{x}_{j}-\mathbf{x}_{i}\right) \mathbf{f}_{\mathrm{in}, j}^{k}}_{\text {node } j \rightarrow \text { node } i \text { message }}
\end{equation}

It has been shown that kernel $\mathbf{W}^{\ell k}$ \citep{weiler20183d, kondor2018n, thomas2018tensor} has to lie in the span of an equivariant basis $\left\{\mathbf{W}_{J}^{\ell k}\right\}_{J=|k-\ell|}^{k+\ell}$. Mathematically, 
\begin{equation}
\mathbf{W}^{\ell k}(\mathbf{x})=\sum_{J=|k-\ell|}^{k+\ell} \varphi_{J}^{\ell k}(\|\mathbf{x}\|) \mathbf{W}_{J}^{\ell k}(\mathbf{x}), \quad \text { where } \mathbf{W}_{J}^{\ell k}(\mathbf{x})=\sum_{m=-J}^{J} Y_{J m}(\mathbf{x} /\|\mathbf{x}\|) \mathbf{Q}_{J m}^{\ell k}.
\end{equation}

Each basis kernel $\mathbf{W}_{J}^{\ell k}: \mathbb{R}^{3} \rightarrow \mathbb{R}^{(2 \ell+1) \times(2 k+1)}$ is formed by taking the linear combination of Clebsch-Gordan matrices $\mathbf{Q}_{J m}^{\ell k}$ of shape $(2 \ell+1) \times(2 k+1)$, where the $J, m^{\mathrm{th}}$ linear combination coefficient
is the $m^{\mathrm{th}}$ dimension of the $J^{\mathrm{th}}$ spherical harmonic. Note that the only learnable part in $\mathbf{W}^{\ell k}$ is the radial function $\varphi_{J}^{\ell k}(\|\mathbf{x}\|)$. $\mathbf{Q}_{J m}^{\ell k}$ and $Y_{J m}$ (and therefore $\mathbf{W}_{J}^{\ell k}(\mathbf{x})$) are precomputed and fixed. See \citep{thomas2018tensor, fuchs2020se} for more details.

\textbf{CNN}: The non-Euclidean graph neural network architectures' performance is tested against a classically used Euclidean 3D convolutional neural network. The 3D convolutional network has been previously employed in the identification of material parameters \citep{santos2020poreflow}. Convolutional layers have successfully been implemented for feature extraction from both 2D and 3D images. However, they can be prone to noise, grid resolution issues and do not guarantee frame indifference in general. %

The architecture employed in this control experiment predicts the formation factor and permeability tensors directly from the 3D microstructure image. The input of this architecture is a 3D binary voxel image ($150 \times 150 \times 150$ pixels) and the output is either the formation factor tensor or the permeability tensor. The architecture consists of five 3D convolution layers with ReLU activation functions. Each convolutional layer is followed by a 3D max pooling layer. The output of the last pooling layer is flattened and then fed into two consecutive dense layers (50 neurons each) with ReLU activations for final prediction. %

\textbf{GNN}: We also build a baseline GNN to compare against equivariant GNN. There are various choices of Graph Convolution layers, and we experiment popular choices such at GCN \citep{kipf2016semi}, GraphSage \citep{hamilton2017inductive}, GAT \citep{velivckovic2017graph}, CGCNN \citep{xie2018crystal}, and GIN \citep{xu2018powerful}. Empirically we find GIN works the best. The building block of our graph neural networks is based on the modification of Graph Isomorphism Network (GIN) that can handle both node and edge features. In particular, we first linear transform both node feature and edge feature to be vectors of the same dimension. At the $k$-th layer, GNNs update node representations by

\begin{equation}
h_{v}^{(k)}=\operatorname{ReLU}\left(\operatorname{MLP}^{(k)}\left(\sum_{u \in \mathcal{N}(v) \cup\{v\}} h_{u}^{(k-1)}+\sum_{e=(v, u): u \in \mathcal{N}(v) \cup\{v\}} h_{e}^{(k-1)}\right)\right)
\end{equation}
where $\mathcal{N}(v)$ is a set of nodes adjacent to $v$, and $e = (v; v)$ represents the self-loop edge. MLP stands for multilayer perceptron. Edge features $h_e^{(k-1)}$ is the same across the layers. 

We use average graph pooling to obtained the graph representation from node embeddings, i.e., $h_{G}=\operatorname{MEAN}\left(\left\{h_{v}^{(K)} | v \in G\right\}\right)$. We set the number of layers to be 20 and the embedding dimension to be 50. %

\textbf{Other experiment details}: The metric used for training and test to measure the agreement of prediction and true permeability is loss($\hat{y}$, $y$) = $\frac{||y - \hat{y}||_F}{||y||_F}$, where $y, \hat{y}$ stands for the true permeability and predicted permeability tensor and $||y||_F$ stands for Frobenius norm, i.e., $||y||_F = \sqrt{\sum_{i=1}^{3} \sum_{j=1}^{3}\left|y_{i j}\right|^{2}} = \sqrt{trace(y^*y)}$. Note that $||y||_F = ||Uy||_F = ||yU||_F$ for any unitary matrix $U$. We use 67\% data for training and the rest for test. We try 5 different random split to obtain more accurate result. To tune the hyperparameters (learning rate, number of layers...) of different models, We take 1/3 of training data for validation and pick the best hyperparameter combinations on validation data. We use software e3nn \citep{e3nn_2020_3723557} for Tensor Field Network and pytorch geometric \citep{fey2019fast} for baseline graph neural network. 

\textbf{Equivaraince Error}:
Following \cite{fuchs2020se}, we use $\Delta_{E Q}=\left\|L_{s} \Phi(f)-\Phi L_{s}(f)\right\|_{2} /\left\|L_{s} \Phi(f)\right\|_{2}$ to measure equivariance error where $L_s, \Phi$ and $f$ denotes the group action on output feature field, neural network, and input feature field respectively. we measure the exactness of equivariance by applying uniformly sampled SO(3)-transformations to input and output. The distance between the two, averaged over samples, yields the equivariance error. 
The $\Delta_{E Q}$ for TFN is $1.1*10^{-7}$, indicating TFN is equivariant \wrt SO(3) up to the small error.

\begin{figure}[htbp]
\begin{center}
\includegraphics[scale=.2]{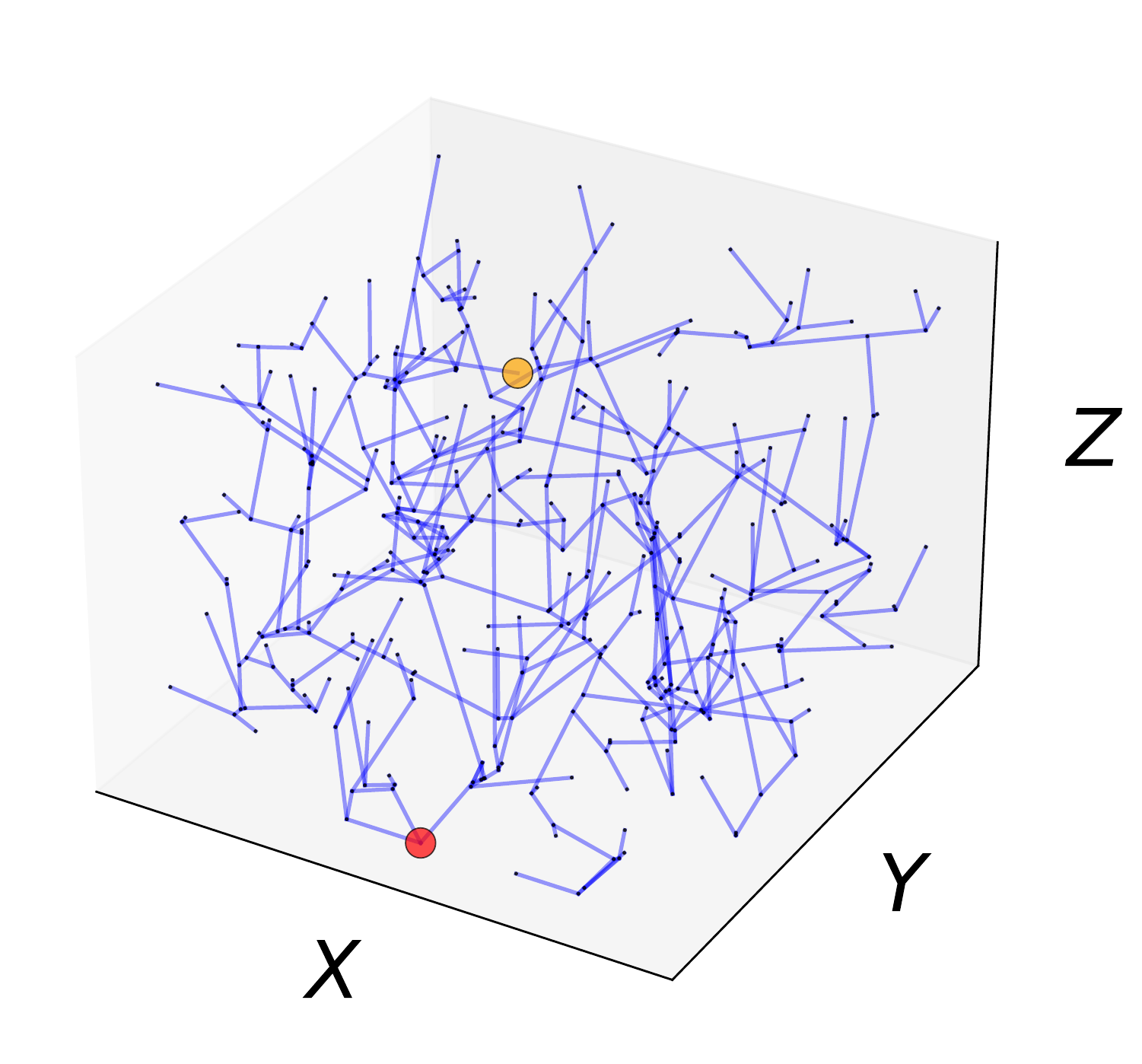}
\includegraphics[scale=.2]{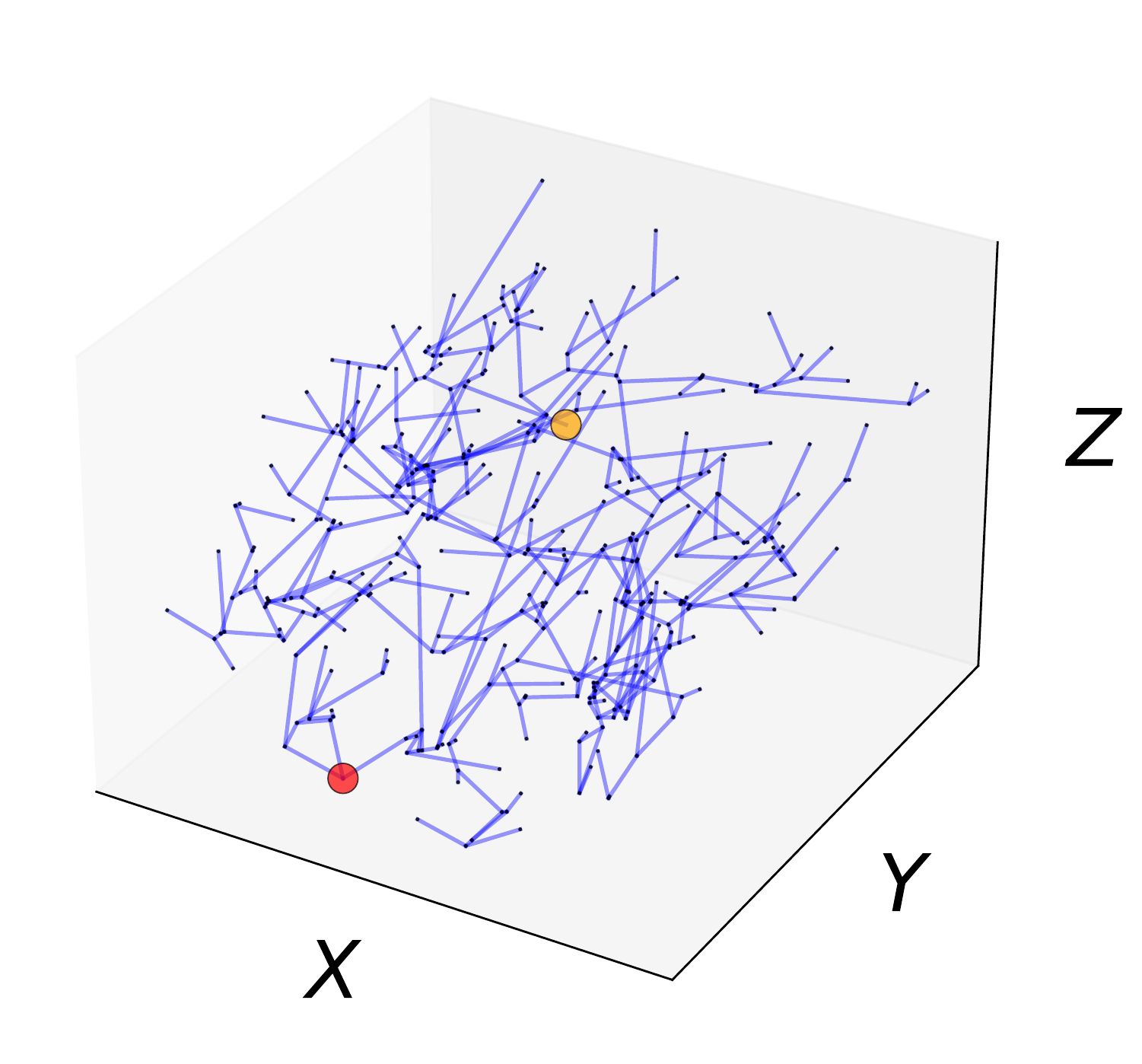}
\includegraphics[scale=.2]{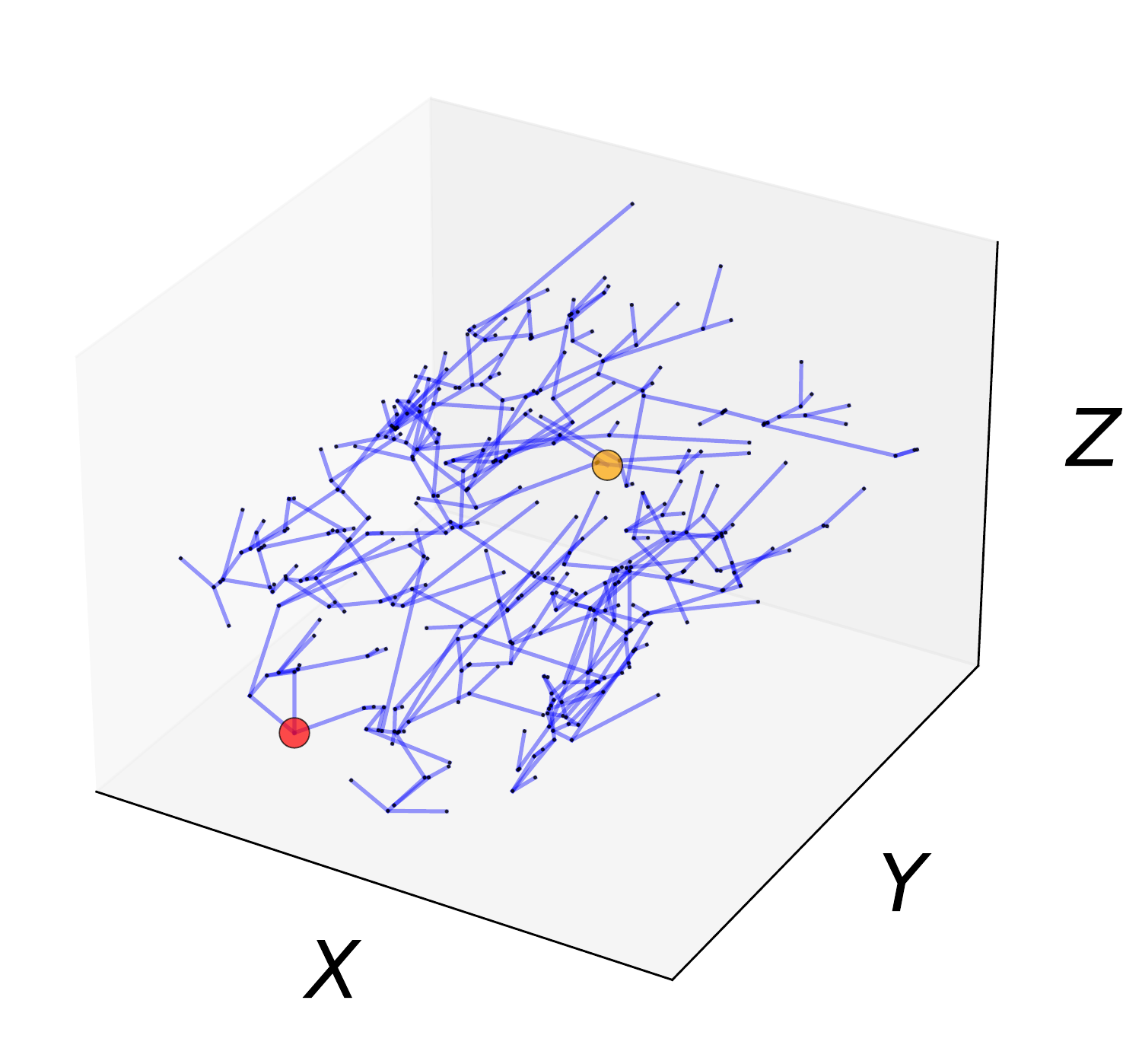}
\includegraphics[scale=.2]{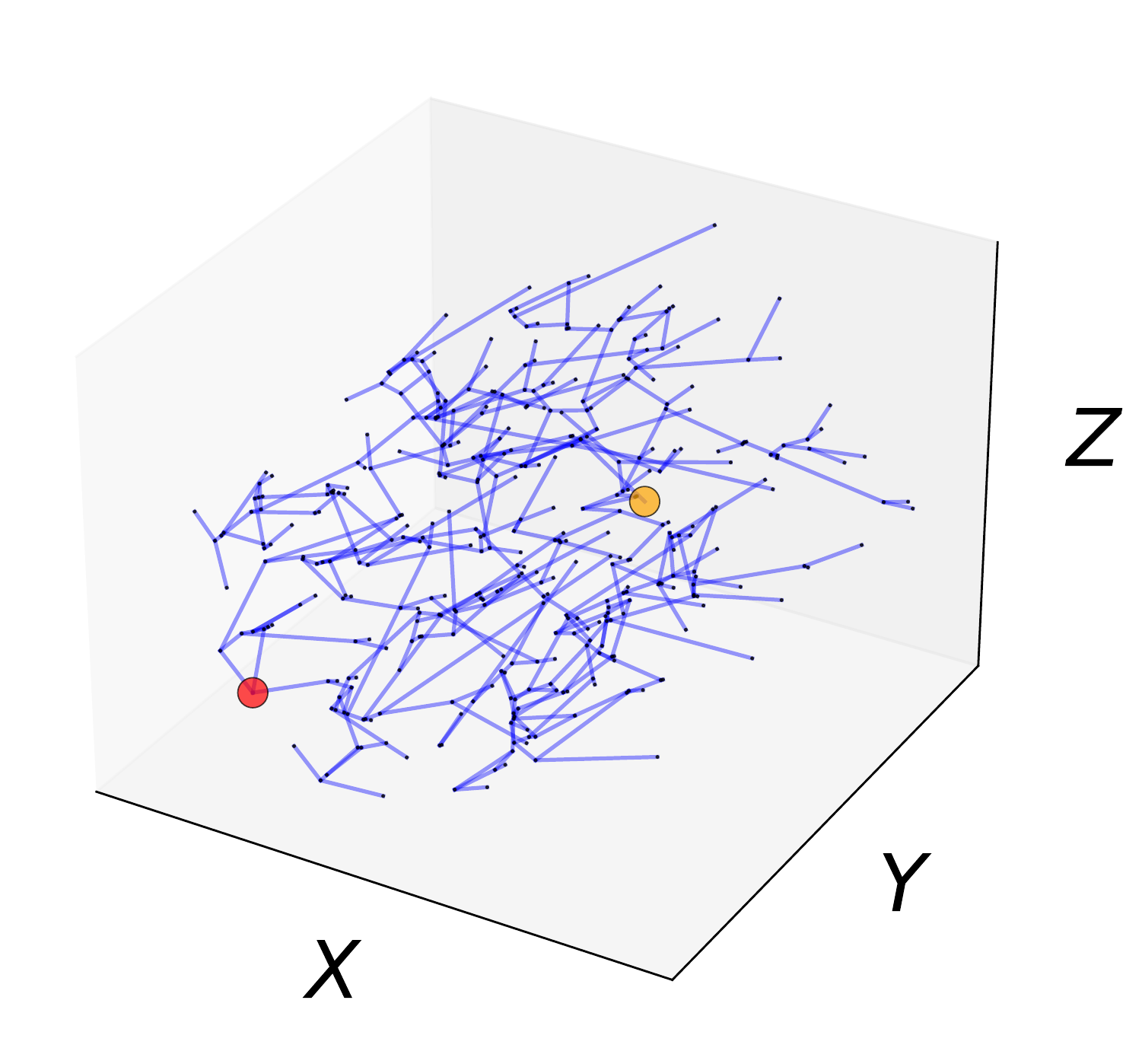}
\includegraphics[scale=.2]{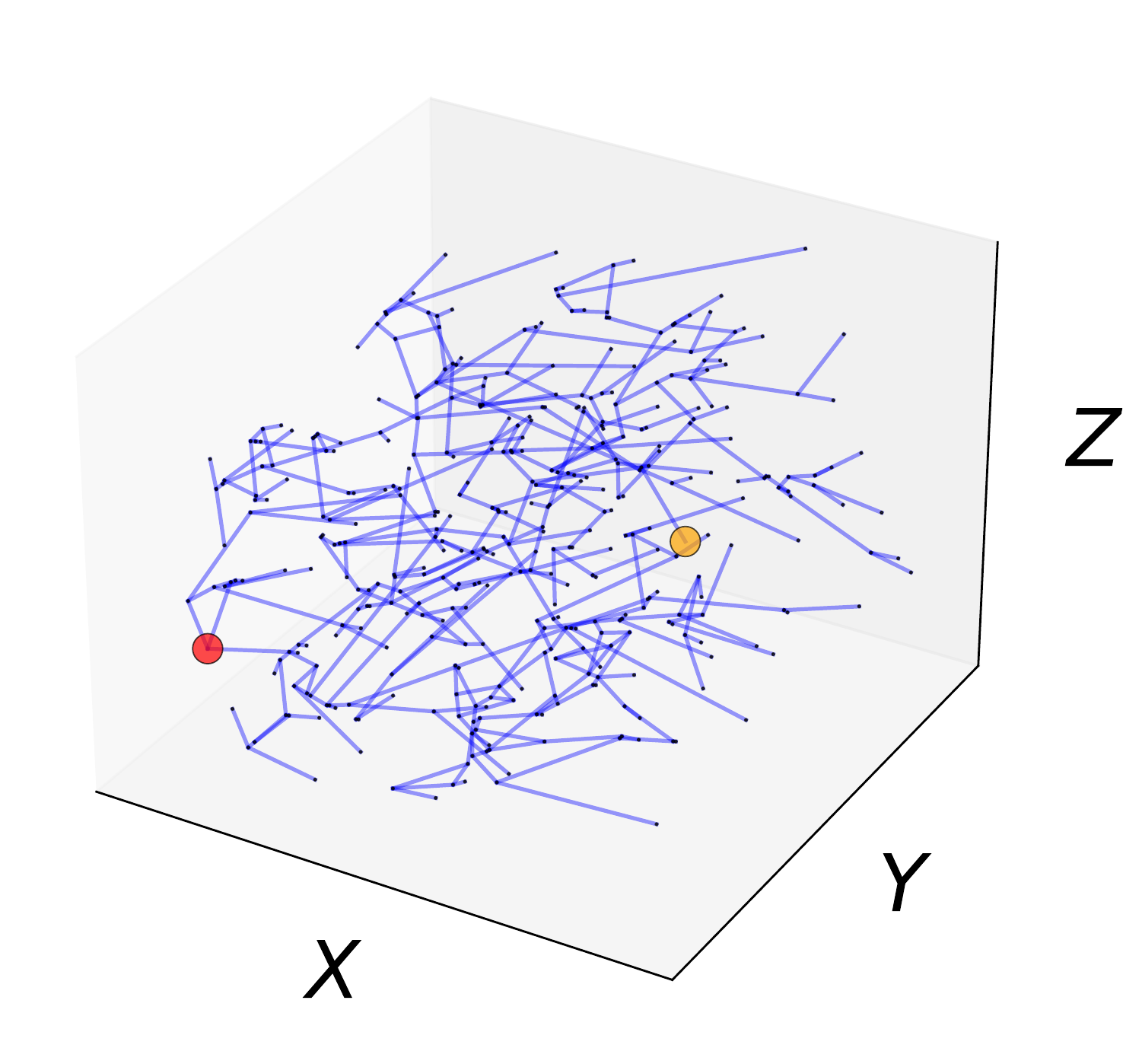}

\includegraphics[scale=.2]{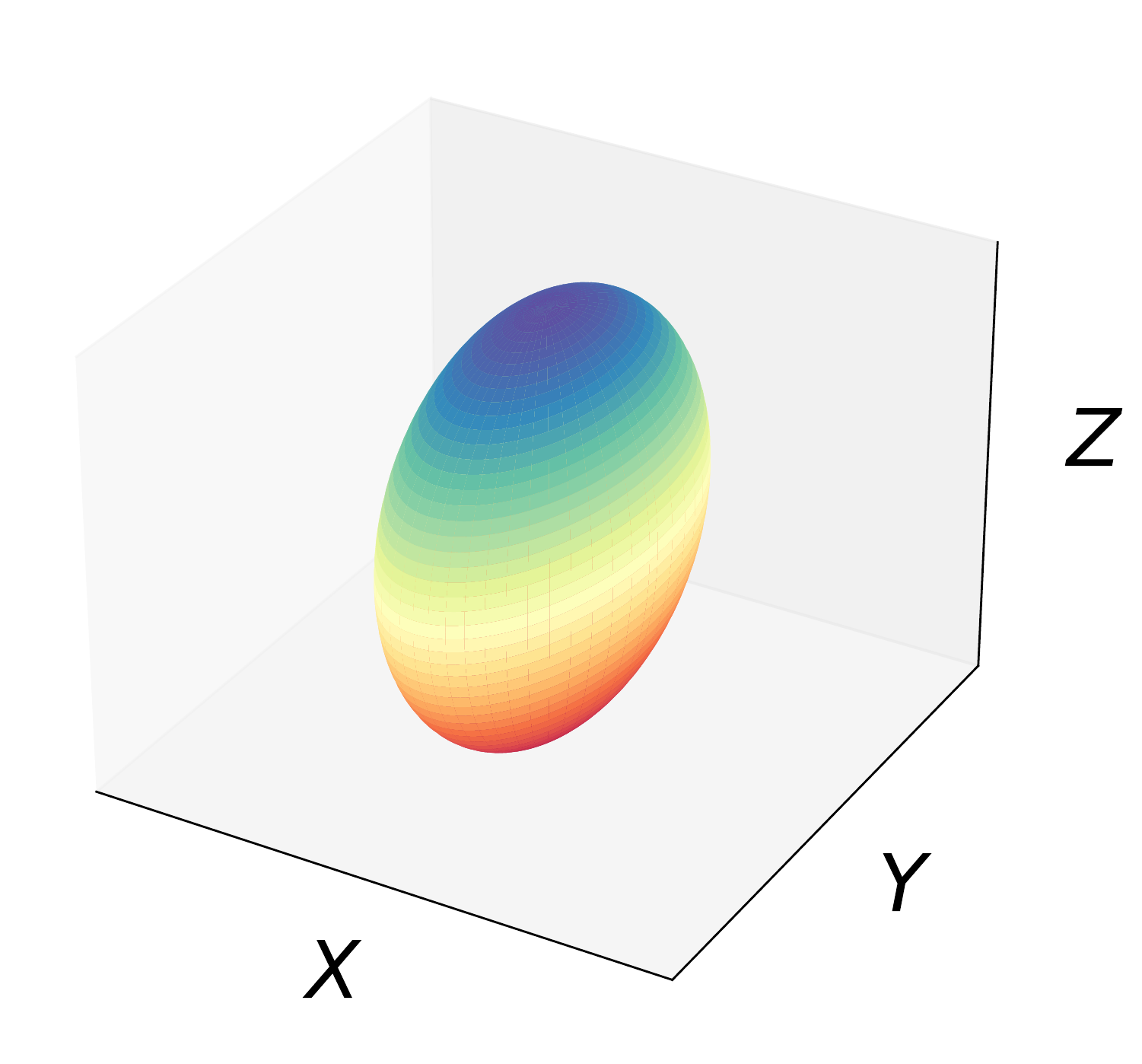}
\includegraphics[scale=.2]{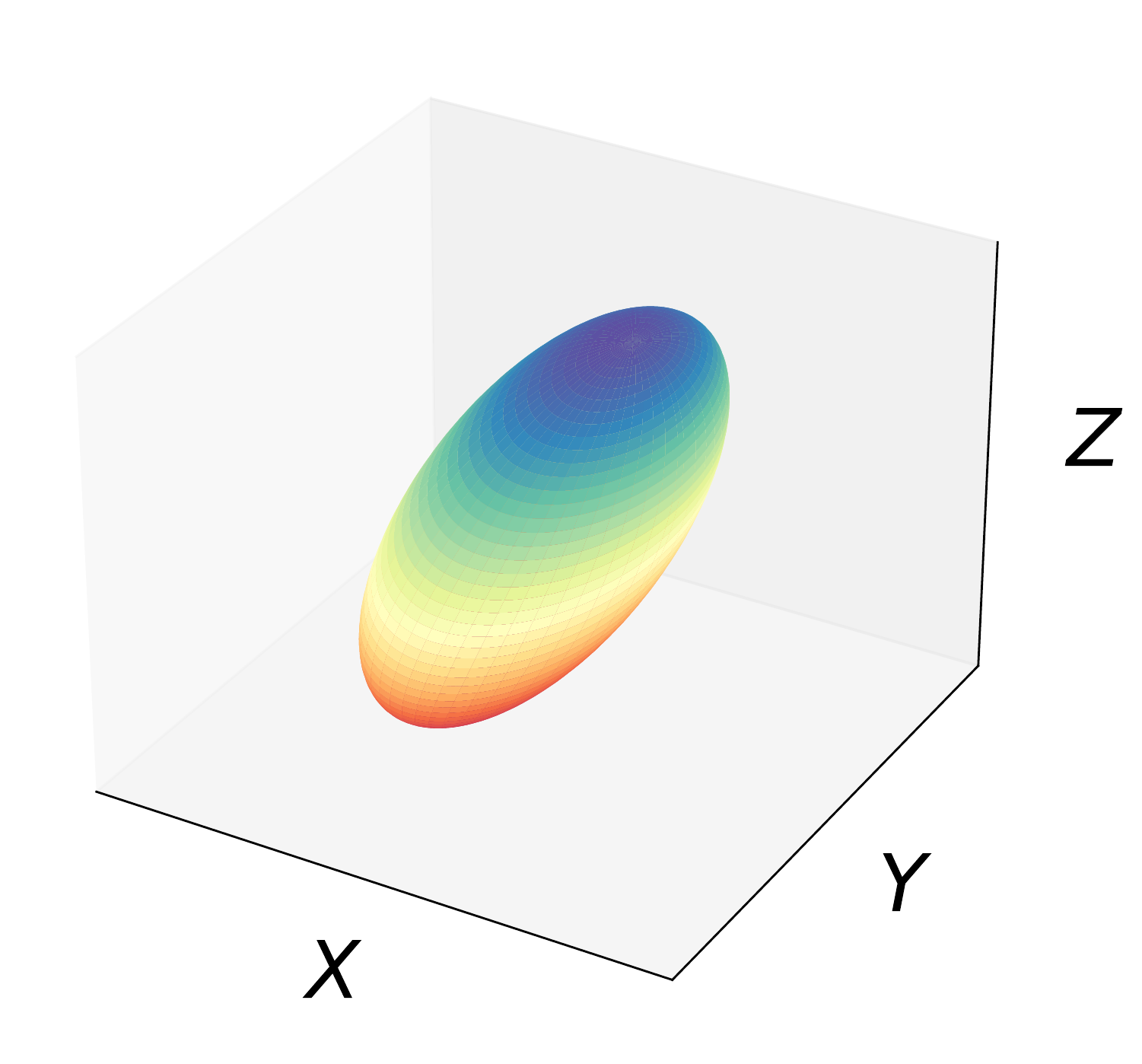}
\includegraphics[scale=.2]{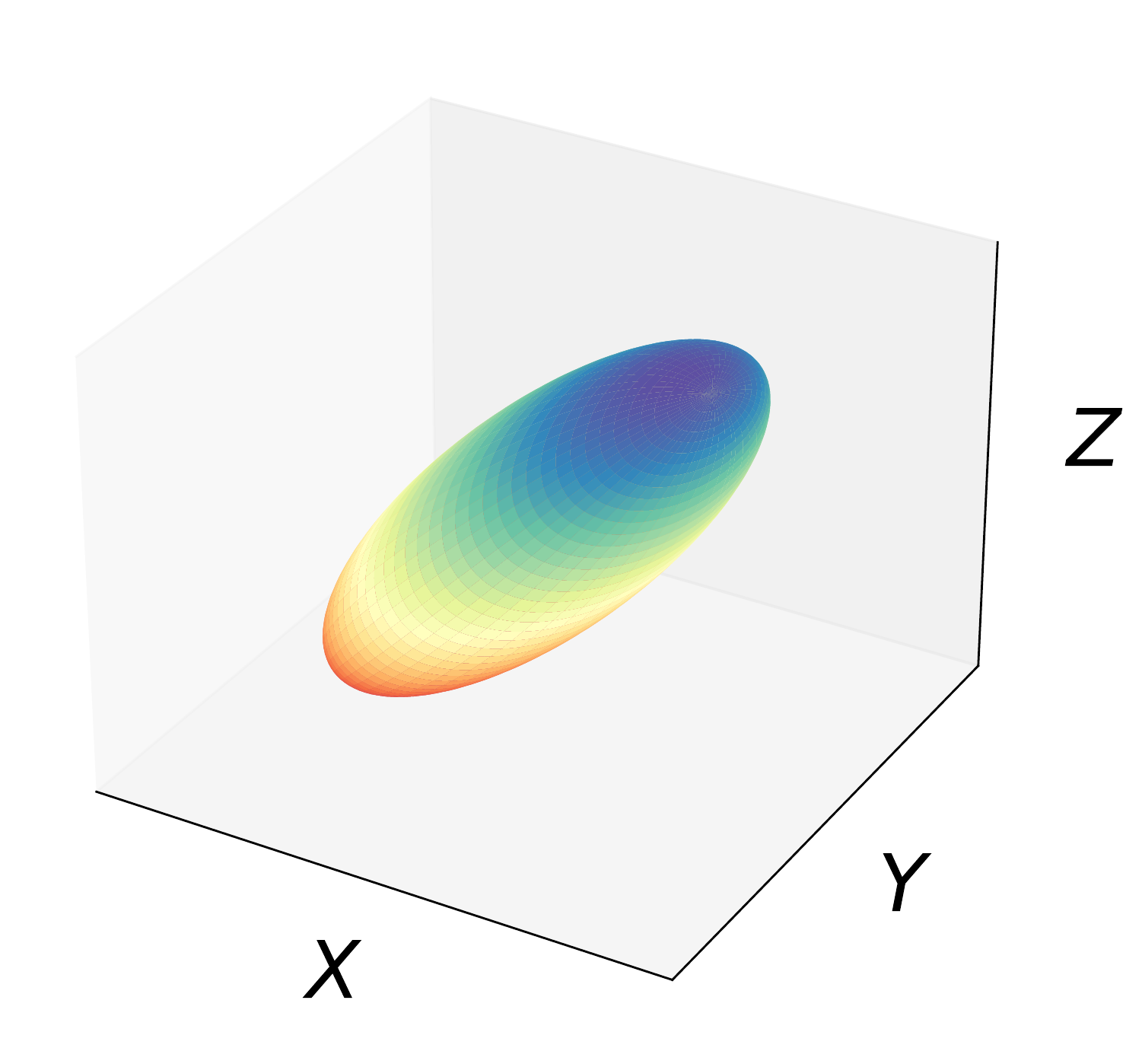}
\includegraphics[scale=.2]{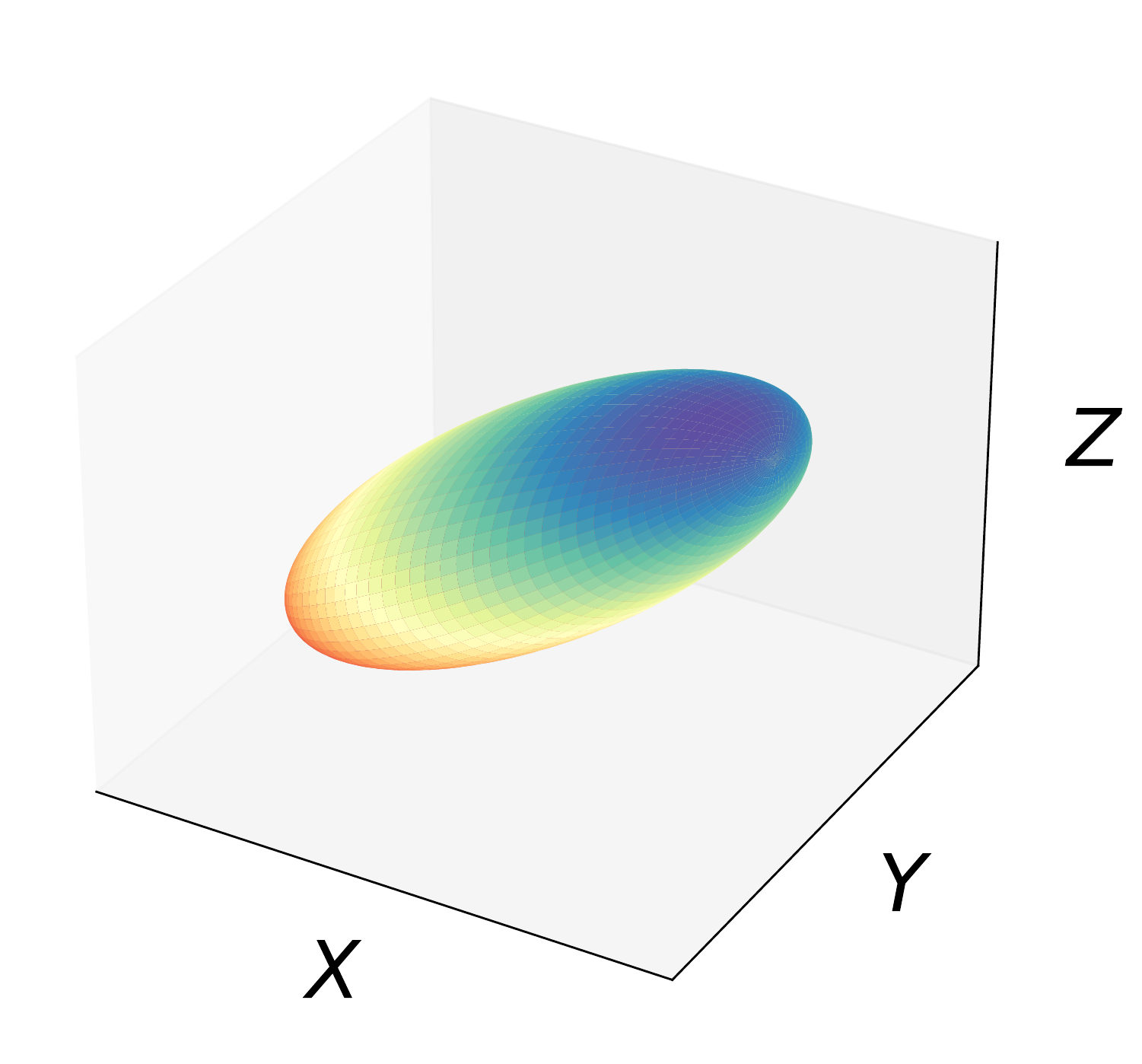}
\includegraphics[scale=.2]{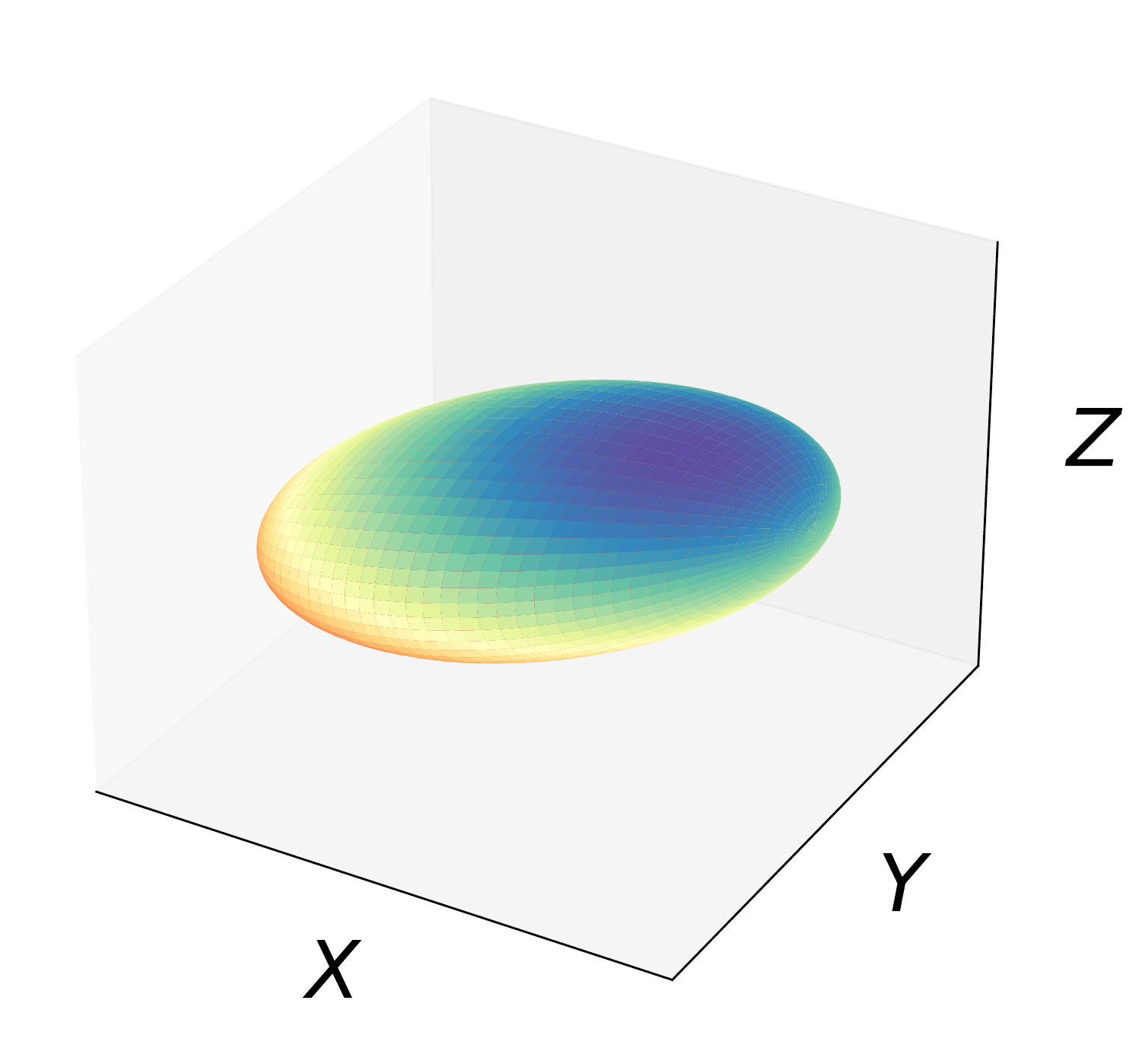}
\caption{Rotating Morse graph along y-axis (see top figures) will make the prediction of 
equivariant neural networks rotate accordingly (see bottom figures).
The second-order tensor predicted by the \enn is visualized as an ellipsoid. }
\label{fig:rot-y}
\end{center}
\end{figure}

\begin{figure}[htbp]
\begin{center}
\includegraphics[scale=.2]{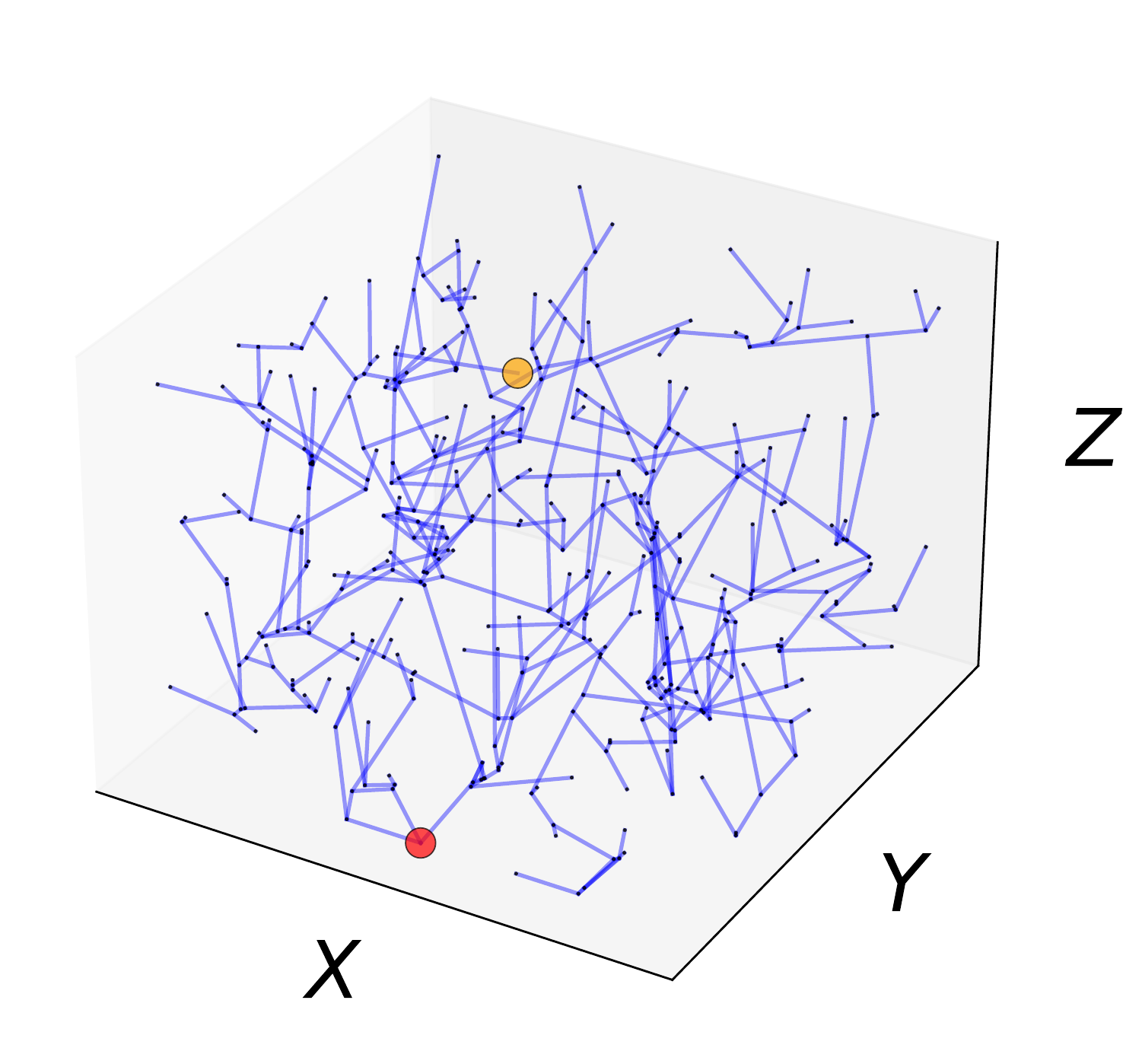}
\includegraphics[scale=.2]{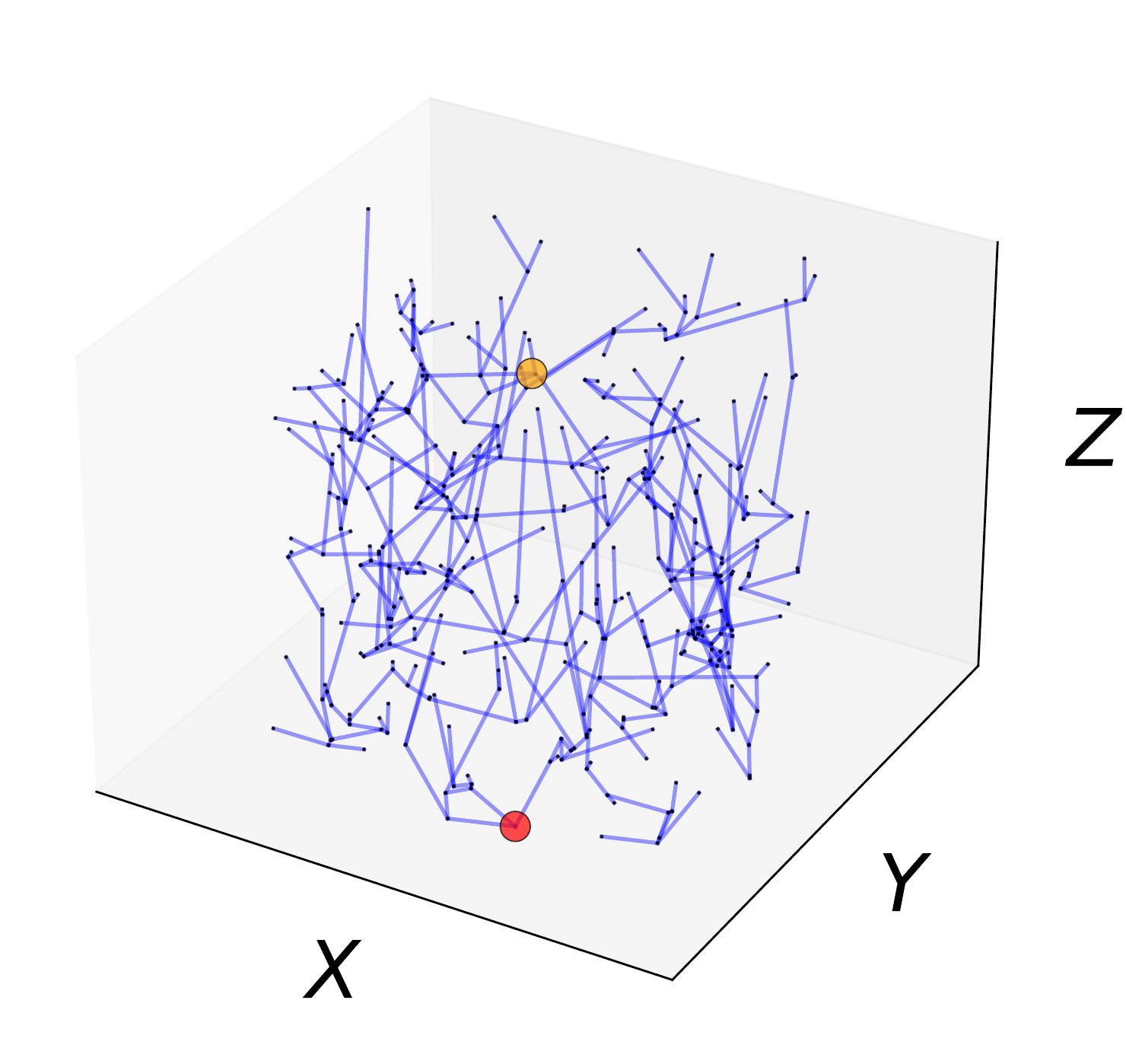}
\includegraphics[scale=.2]{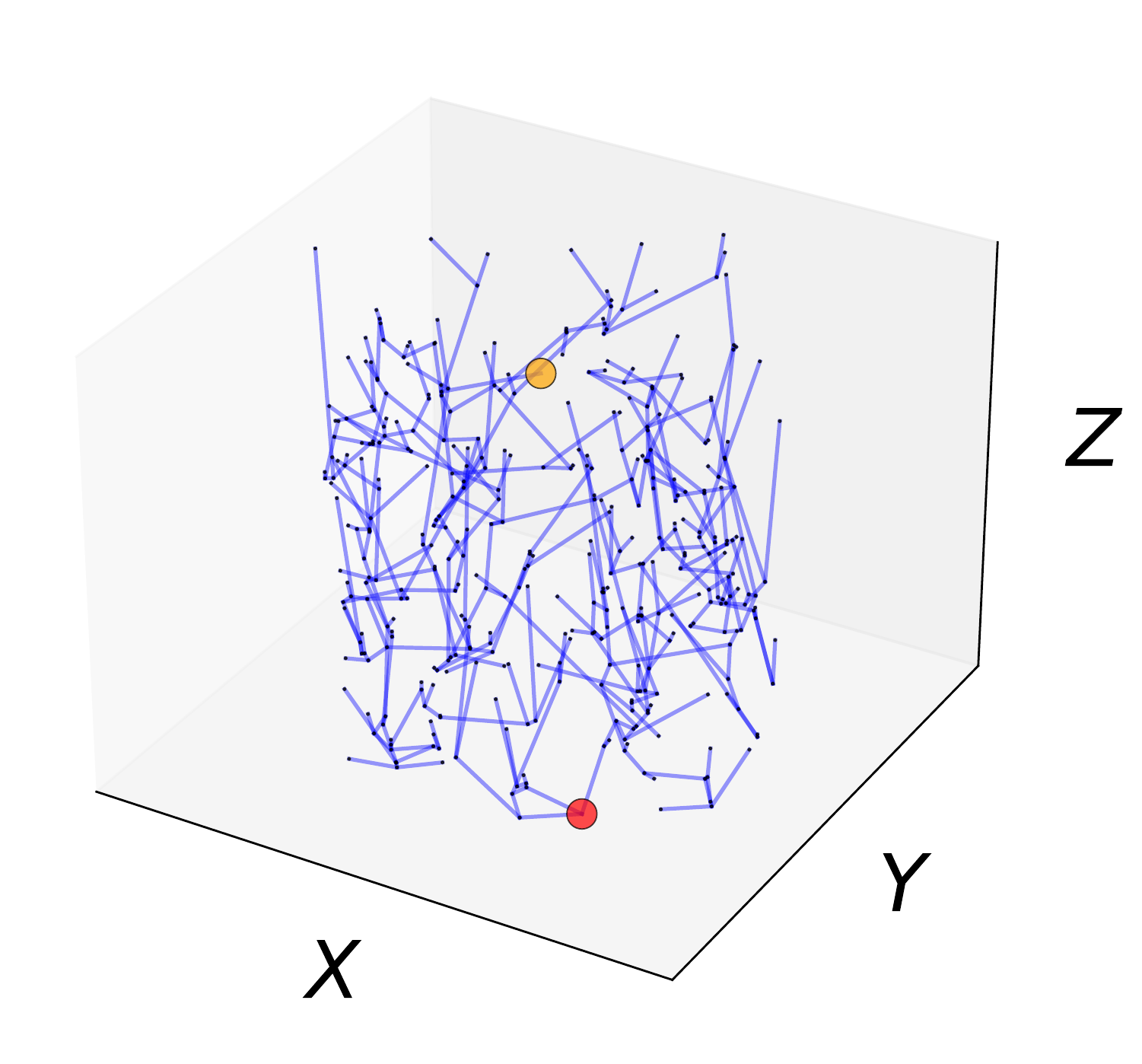}
\includegraphics[scale=.2]{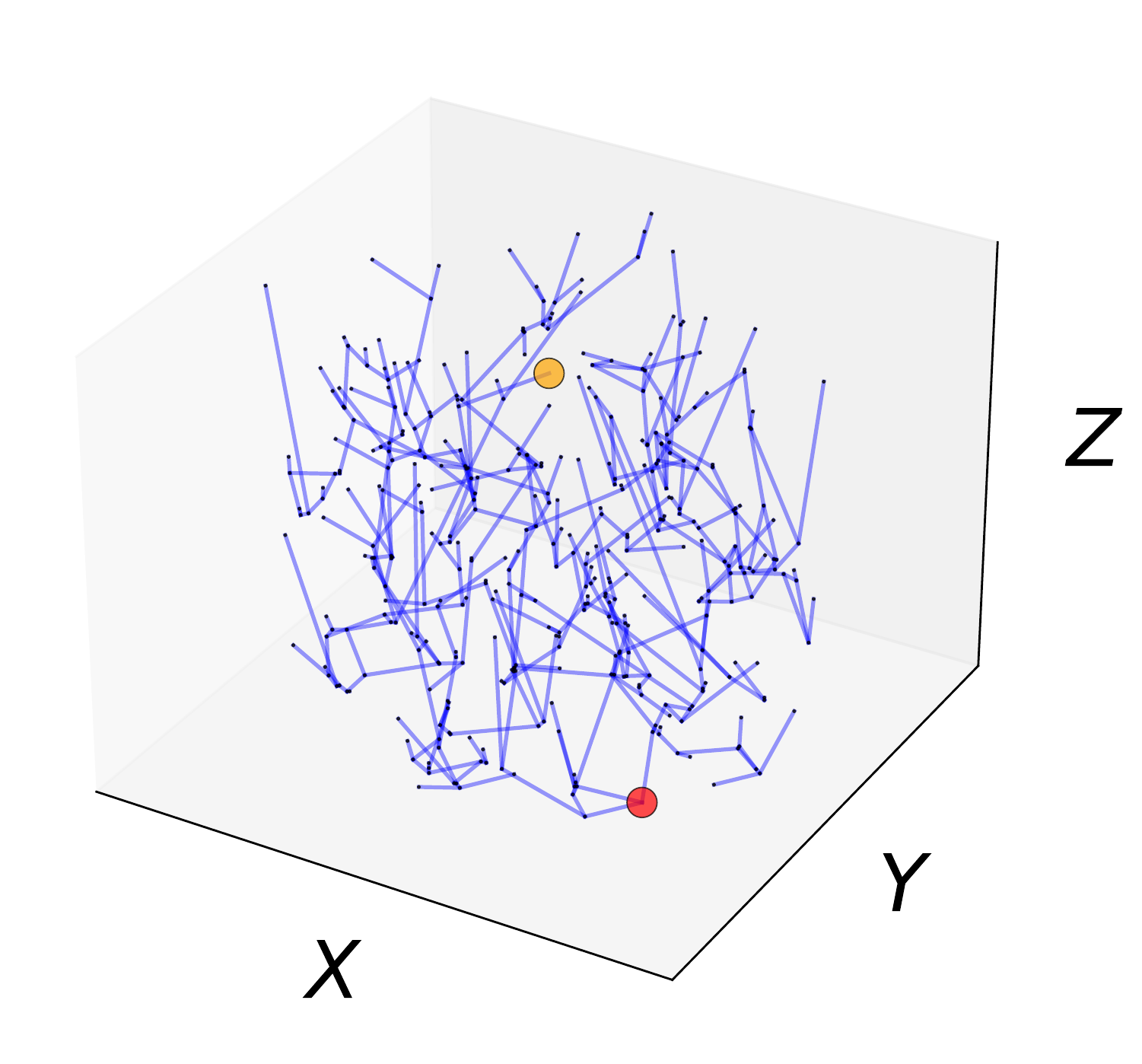}
\includegraphics[scale=.2]{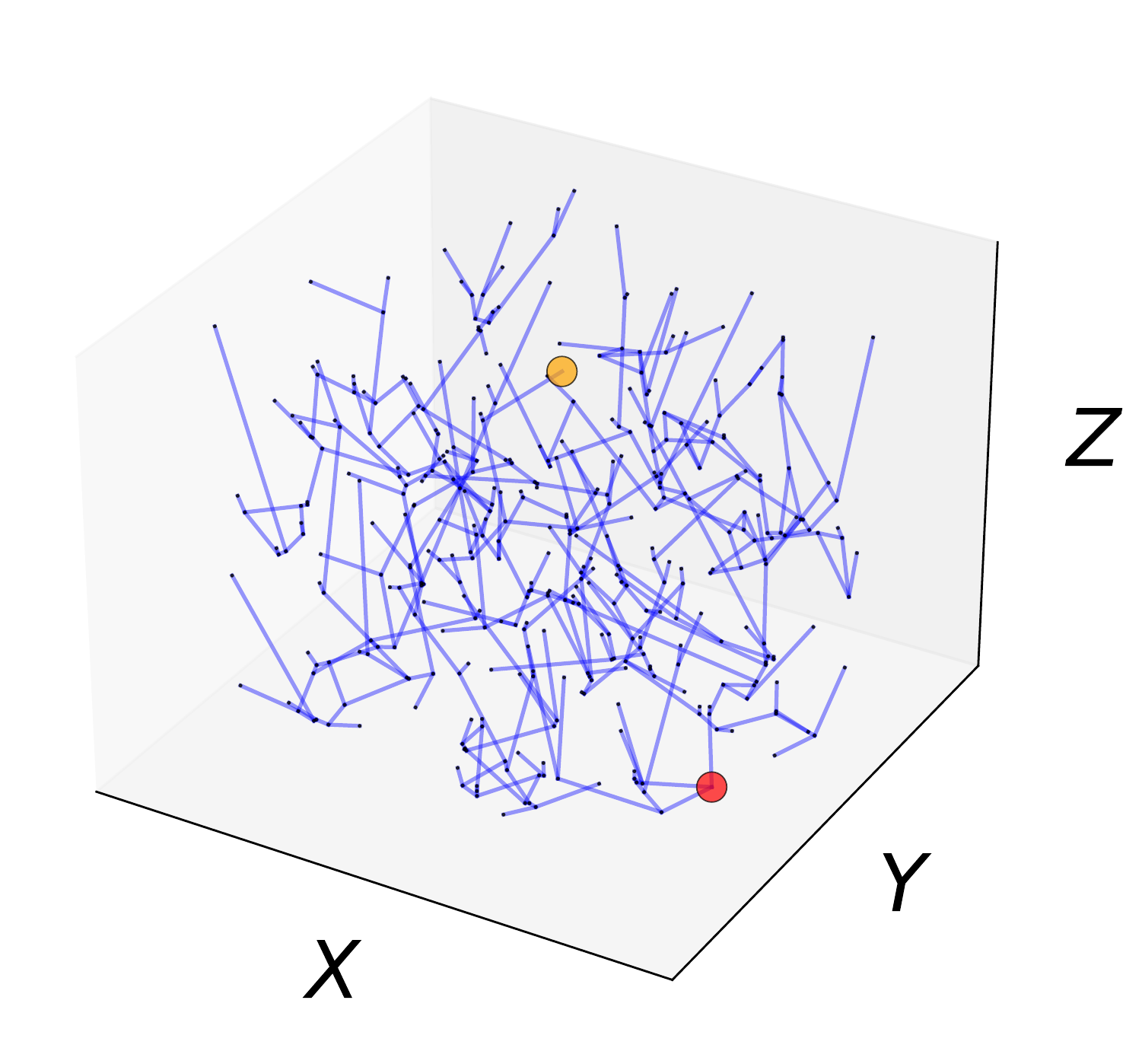}

\includegraphics[scale=.2]{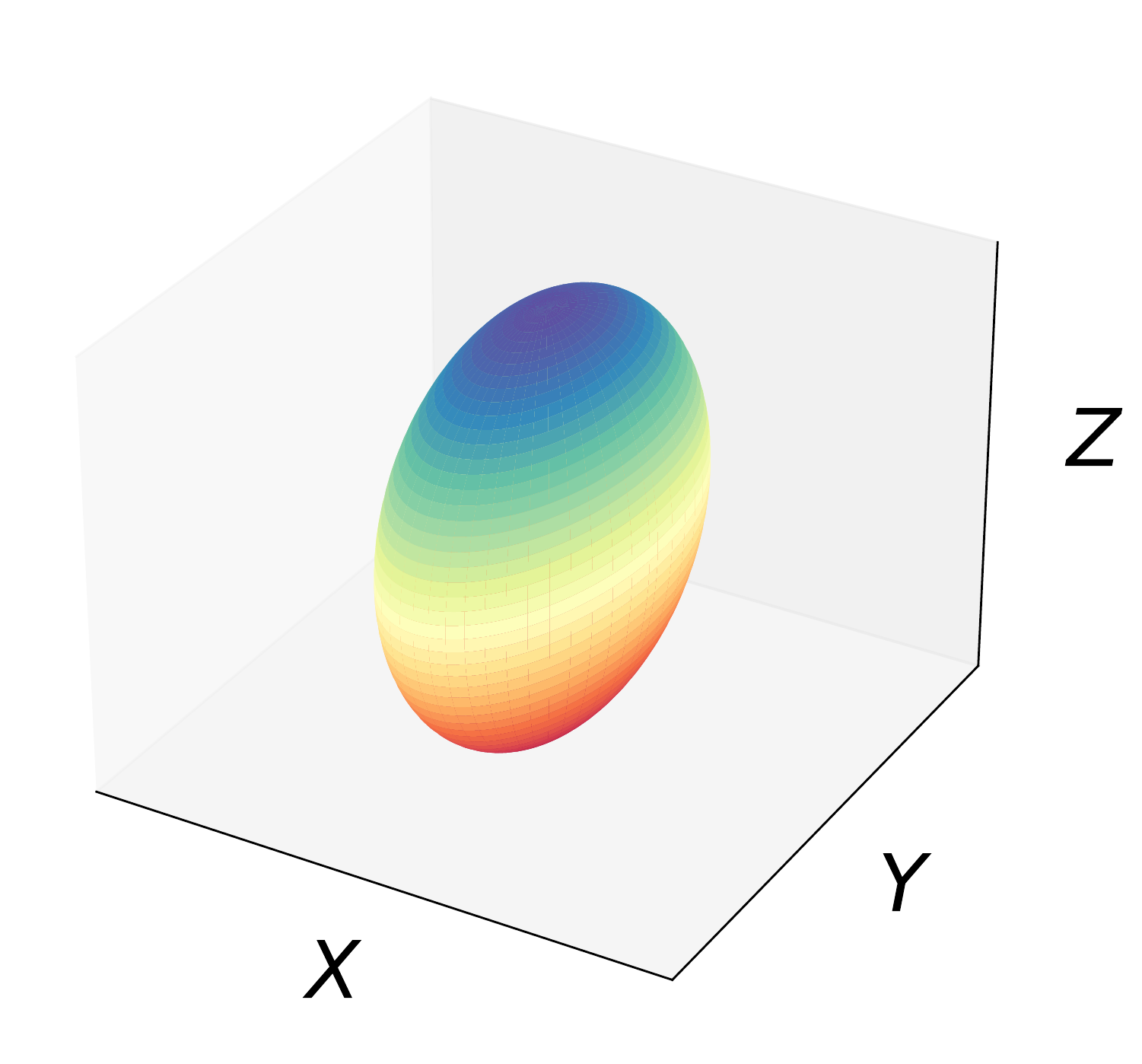}
\includegraphics[scale=.2]{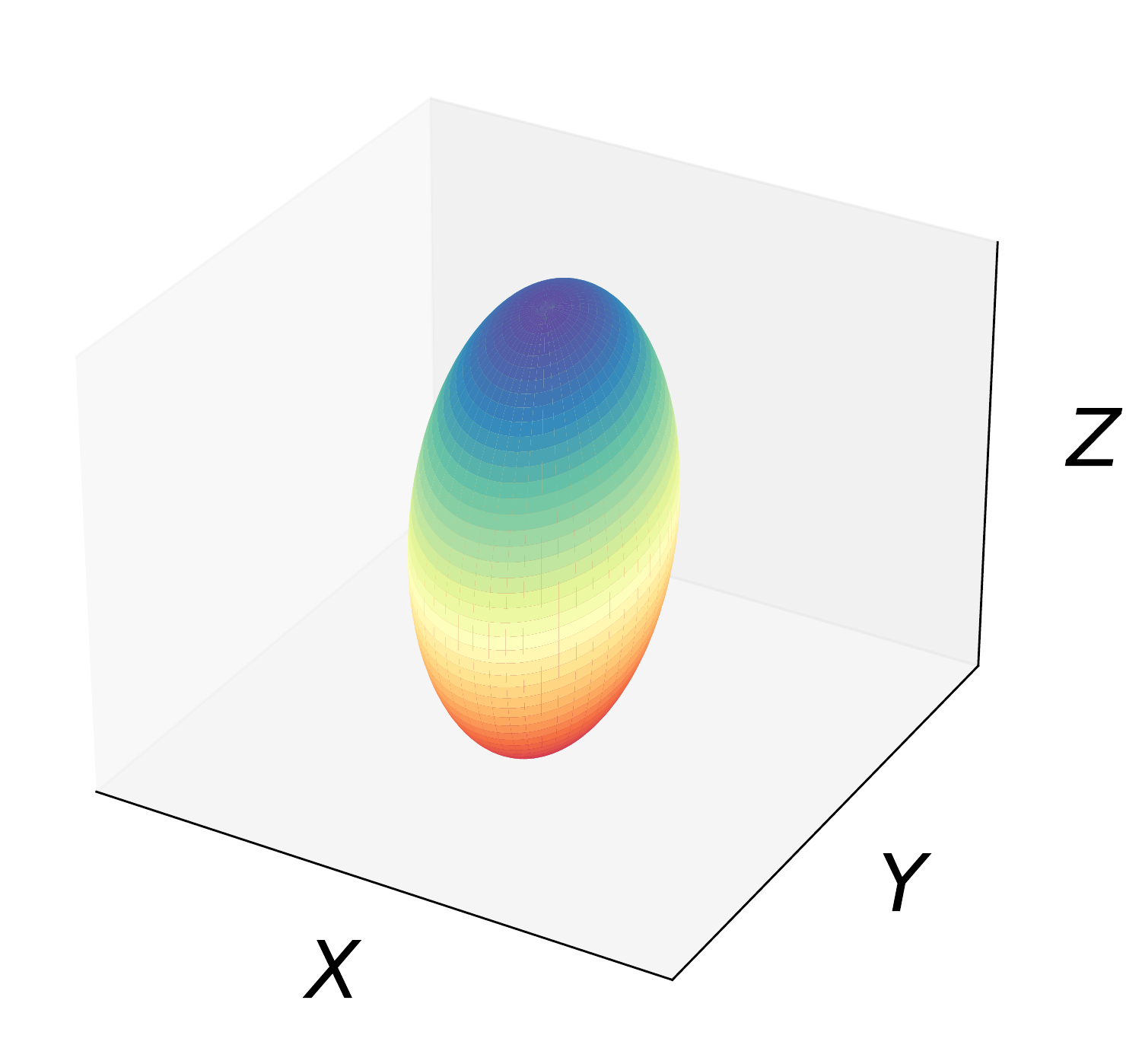}
\includegraphics[scale=.2]{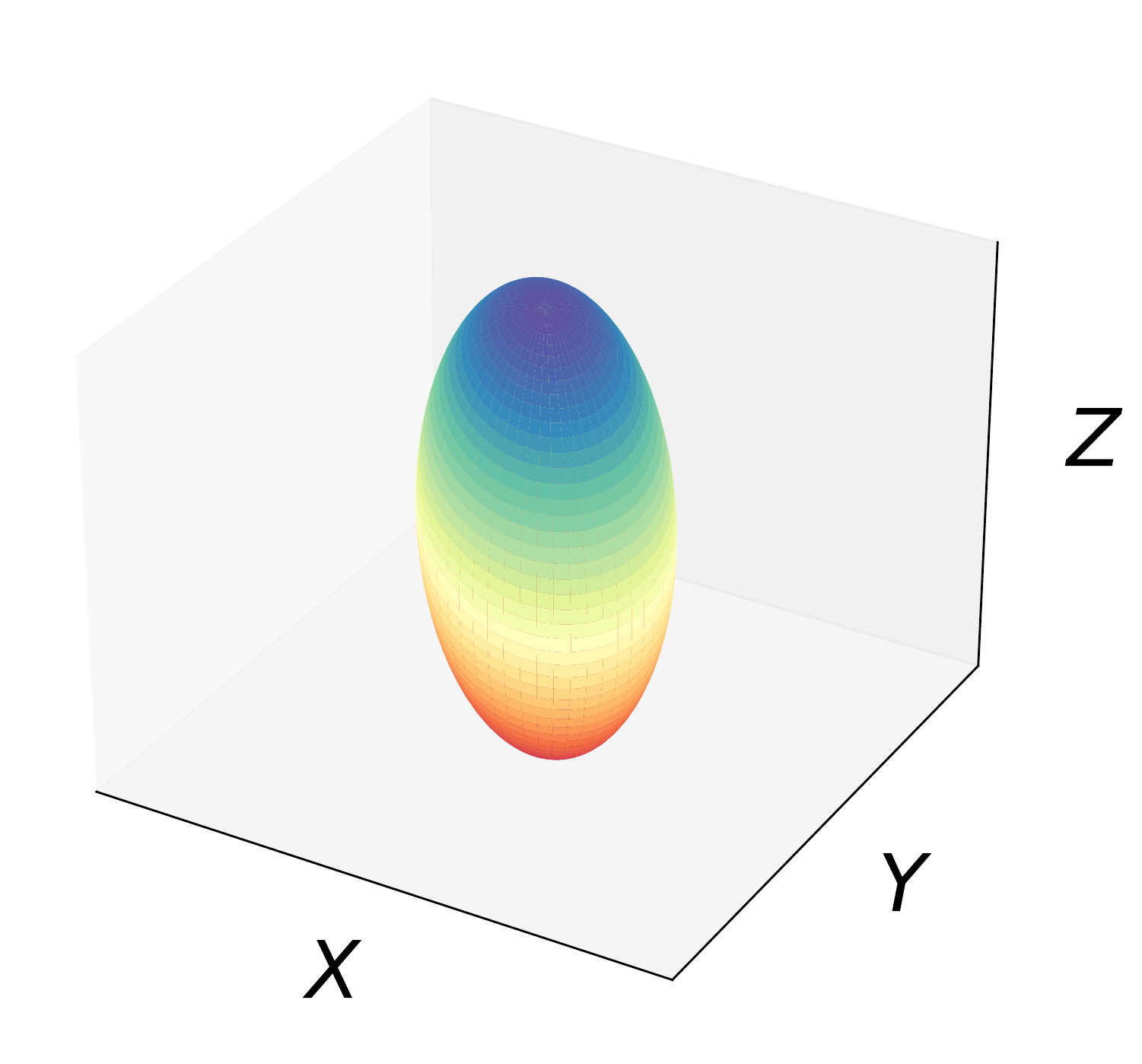}
\includegraphics[scale=.2]{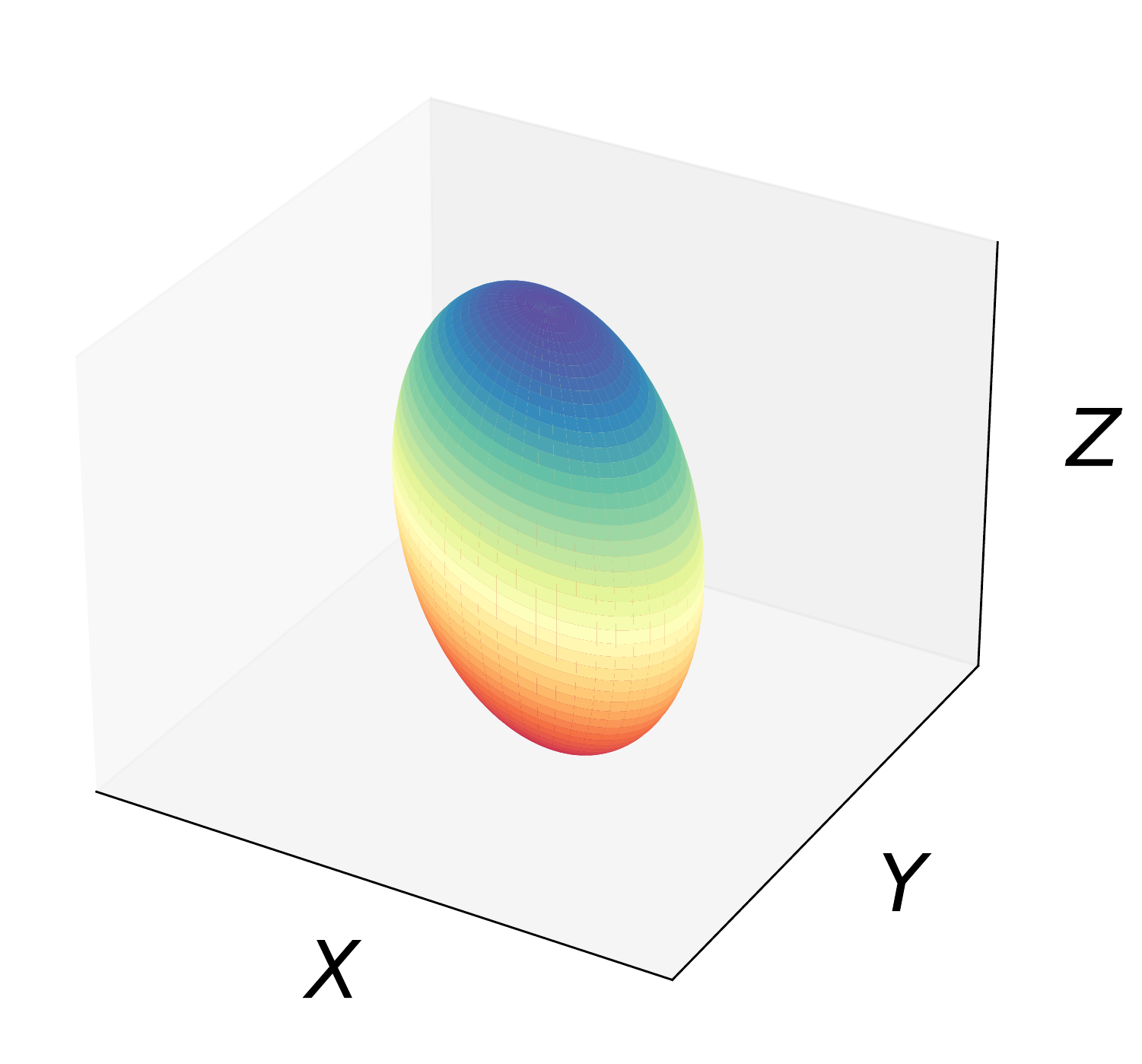}
\includegraphics[scale=.2]{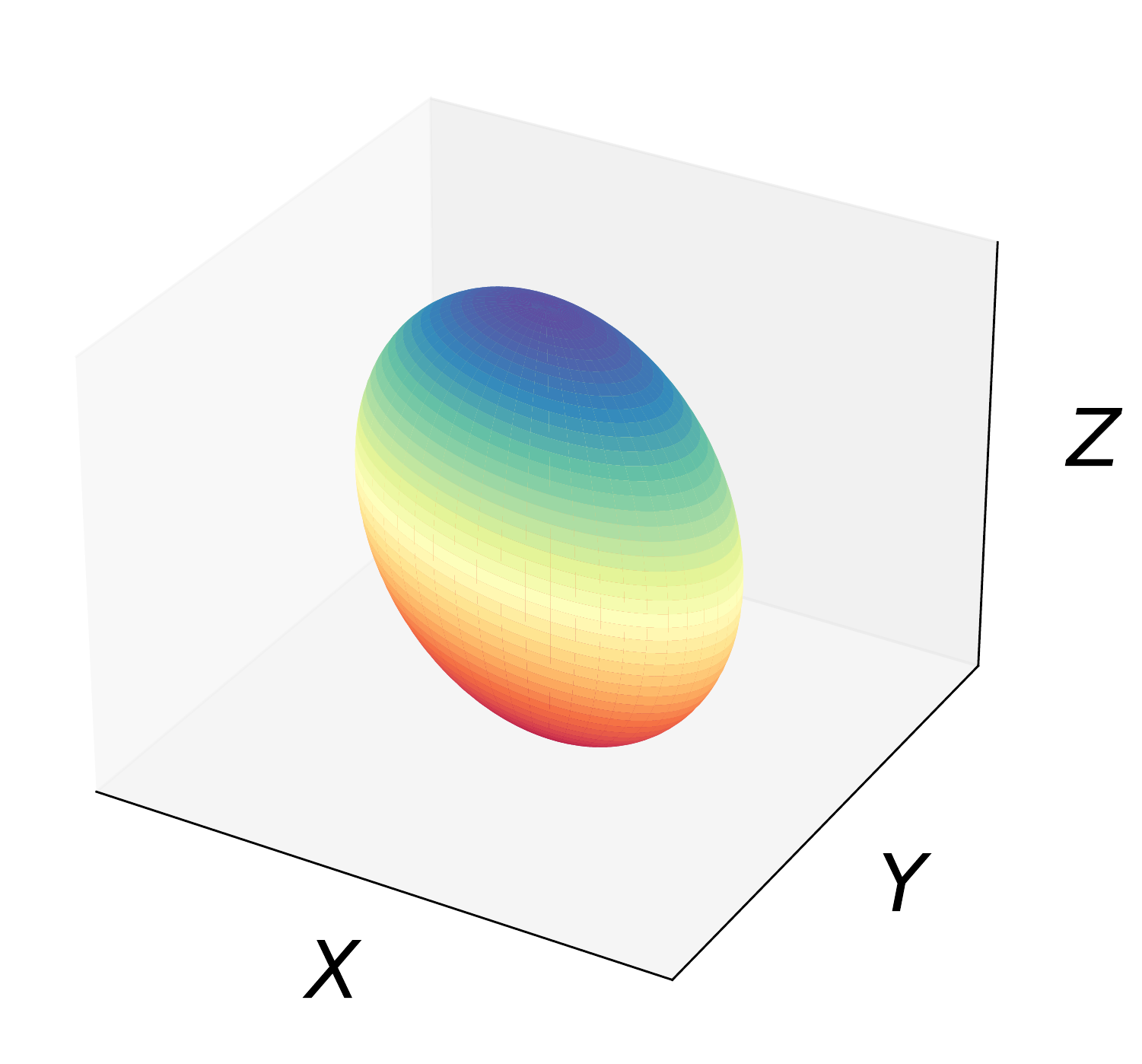}
\caption{Rotating Morse graph along z-axis (see top figures) will make the prediction of 
equivariant neural networks rotate accordingly (see bottom figures).
The second-order tensor predicted by the \enn is visualized as an ellipsoid}
\label{fig:rot-z}
\end{center}
\end{figure}

\textbf{Visualization}:
Given a positive semi-definite matrix $M$, we can visualize it as an ellipsoid located at the origin, where the distance away from boundary in direction $v$ (unit length) is $v^{T}Mv$. We visualize the prediction of \enn for Morse graphs of different orientations, shown in figure \ref{fig:rot-y} and \ref{fig:rot-z}. It can be seen that as we rotate the Morse graph, the output of \enn, visualized as an ellipsoid, also rotates accordingly. 

\begin{figure}[htbp]
\begin{center}
\includegraphics[scale=.4]{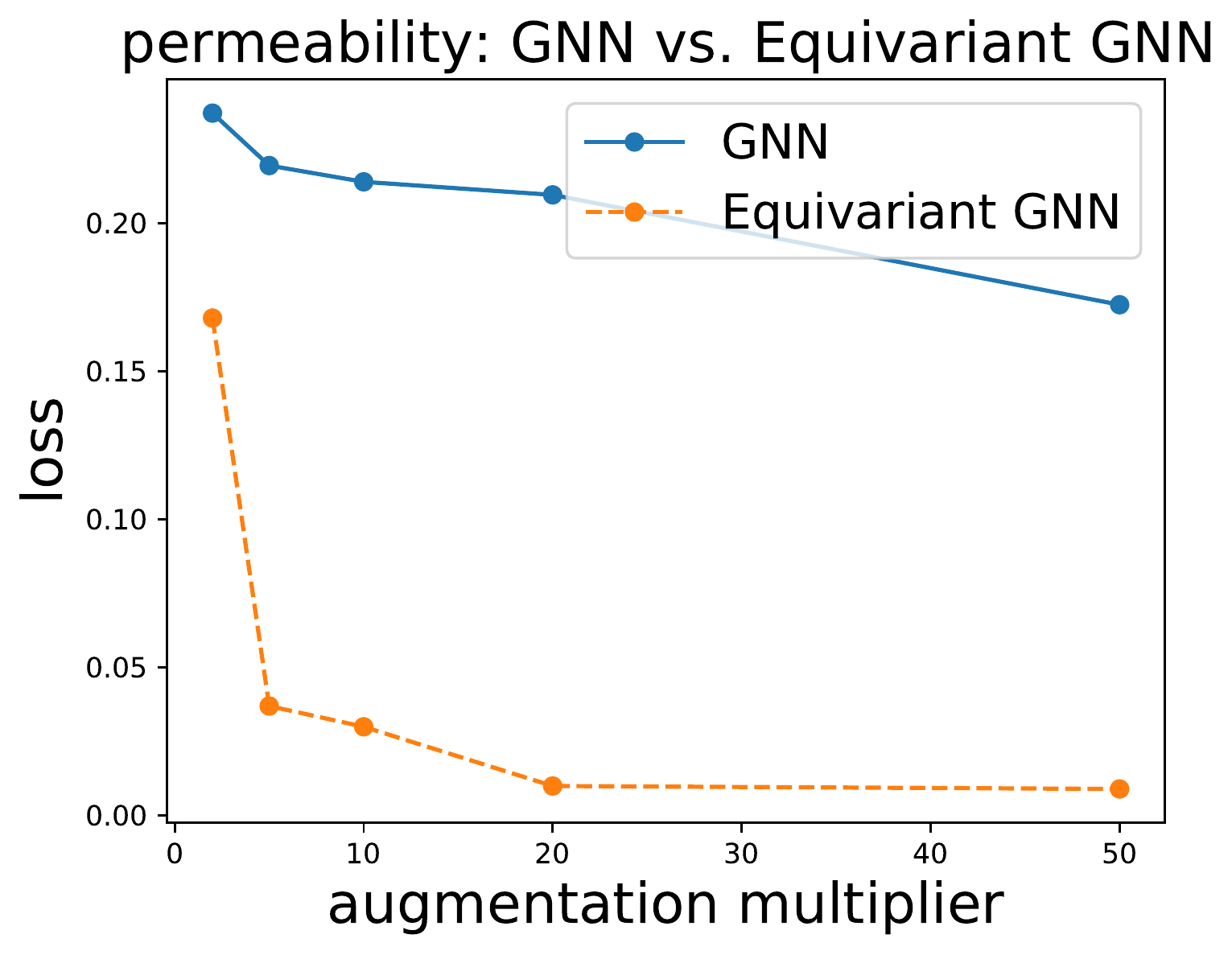}
\includegraphics[scale=.4]{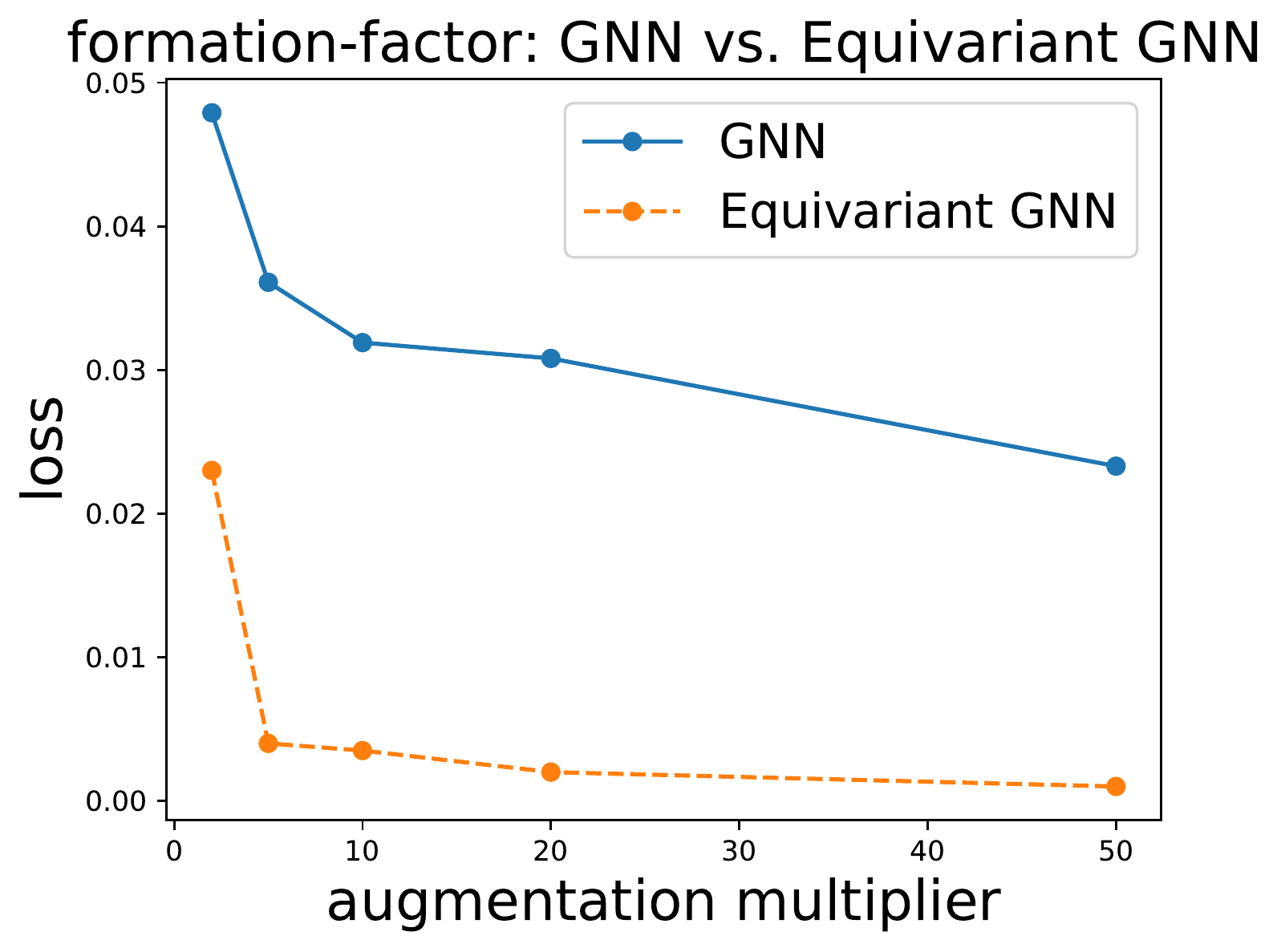}
\caption{Augment all dataset. The loss after augmenting all data by $k$ copies (shown in x-axis as augmentation multiplier). The blue/orange curve indicates the performance of GNN under different number of copies ($k= 2, 5, 10, 20, 50$). Equivariant neural network outperforms GNN by a large margin.}
\label{fig:aug-all}
\end{center}
\end{figure}

\begin{figure}[htbp]
\begin{center}
\includegraphics[scale=.4]{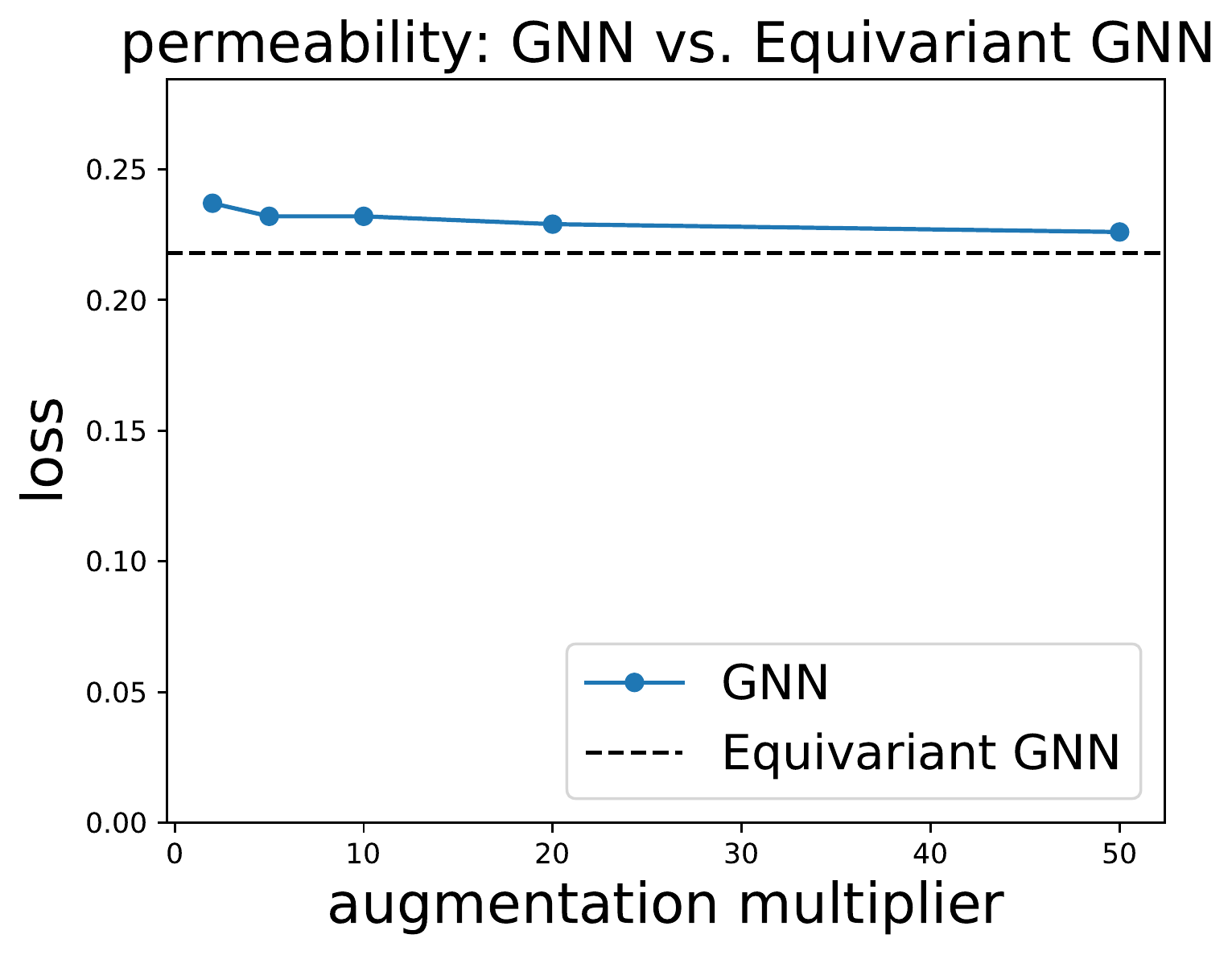}
\includegraphics[scale=.4]{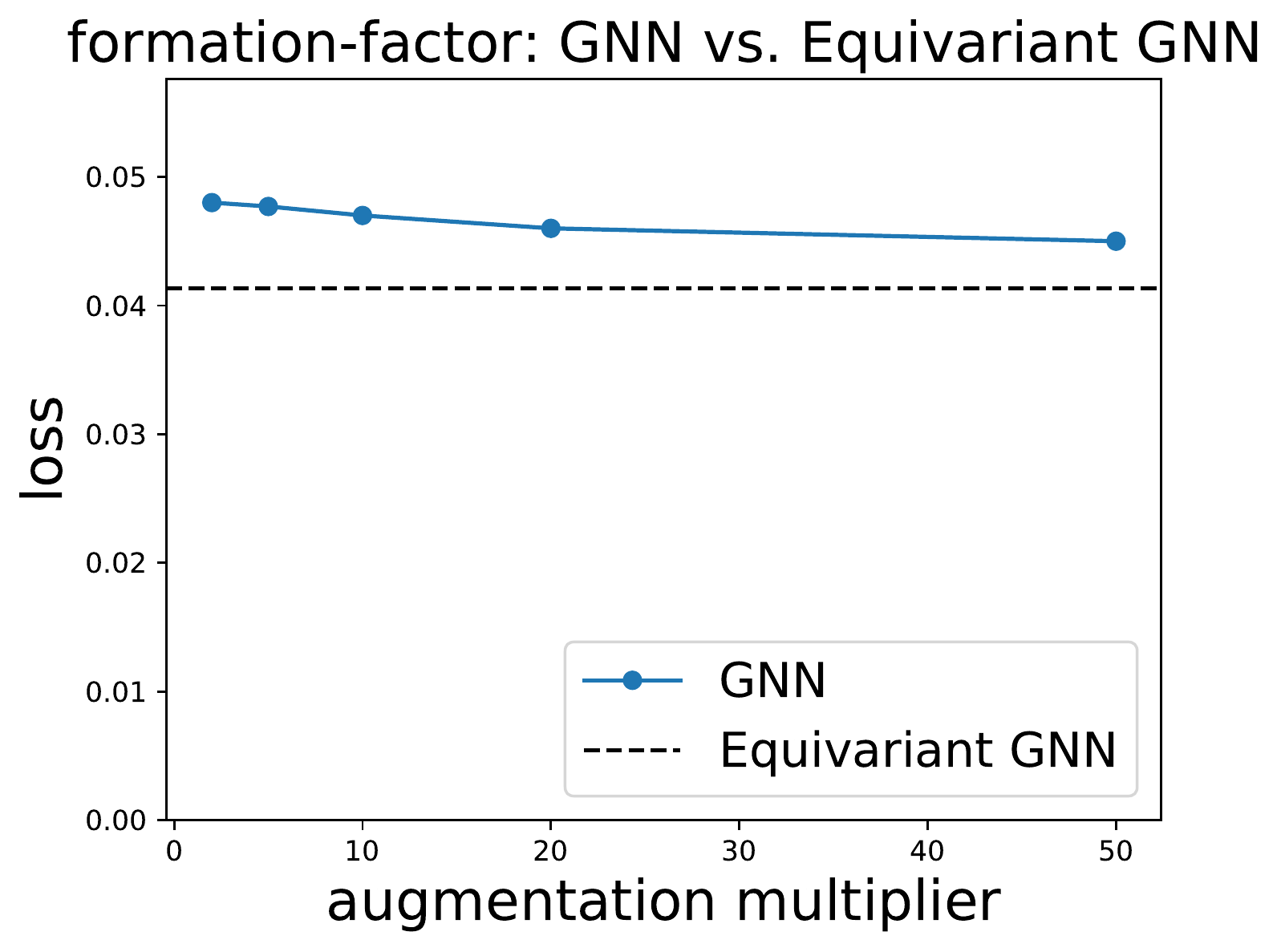}
\caption{Augment only \textit{training} data. The loss after augmenting all data by $k$ copies (shown in x-axis as augmentation multiplier). We can see as $k$ increases, the loss of GNN on test data slightly decreases, but still not better than the \enn without data augmentation, whose result is shown in dashed line. }
\label{fig:aug-train}
\end{center}
\end{figure}

\textbf{Results on the augmented dataset}:
As the dataset is small, and may not play an advantage to the equivariant neural network, we augmented each graph with property $P$ by randomly rotating the input by $\tensor{R}$, and take the $\tensor{R}P\tensor{R^{T}}$ as the ground truth property of the rotated input. Results are shown in Figure \ref{fig:aug-all}. We tested two ways of augmentation 1) augment all datasets and 2) only augment training dataset. The results are shown in Figures \ref{fig:aug-all} and \ref{fig:aug-train}.

We find for the equivariant neural network, the test error is significantly lower than the non-equivariant ones. As the argumentation multiplier increase from 2 to 5, the error reduces significantly to nearly perfect performance. For non-equivariant neural networks, although the data augmentation improves the performance measured by the metric, the result is still far away from (one order of magnitude worse) the results obtained from the \ennend 

In the case of augmenting only training data, we find that the performance slowly improves as the number of copies increases for non-equivariant GNN, but the \enn without data argumentation is still better than the GNN. This demonstrates the data efficiency of \ennend

\section{Prediction results compared with CNN} \label{sec:result}
The permeability and formation factor predictions made by the trained CNN, GNN, and Equivariant GNN are shown in Tables \ref{tab:comppermeability} and \ref{tab:compformationfactor} respectively. Generally speaking, one may observe that GNN outperforms CNN, and Equivariant GNN outperforms GNN. We believe the Morse graph representation, although only being a rough representation of 3D images, captures the essence of the geometrical and topological information that aids downstream tasks. This trend is more obvious in the formation factor calculation 
than that of the effective permeability.

Another advantage of GNN over CNN is the computational efficiency: performing convolution on 3D images takes about 100s per epoch while GNN takes only 4s. Even though using Morse graph comes with the overhead of computing Morse graphs (takes roughly 3 minutes per graph), the overall computational time of working with graph representation is still desirable.

\begin{table}[]
\centering
\caption{Permeability. Comparison between three models. The error for baseline CNN, baseline GNN, equivariant GNN are shown for five different splits of data. The improvement indicates the improvement of equivariant GNN over GNN. }
\label{tab:comppermeability}
\begin{tabular}{@{}lllll@{}}
\toprule
seed & CNN & GNN & Equivariant GNN & Improvement \\ \midrule
1 & 0.312 & 0.246 & 0.218 & 11.3\% \\
2 & 0.354 & 0.247 & 0.221 & 10.5\% \\
3 & 0.339 & 0.251 & 0.226 & 10.0\% \\
4 & 0.379 & 0.247 & 0.228 & 7.7\% \\
5 & 0.382 & 0.298 & 0.252 & 15.4\% \\ \midrule
mean & 0.353 & 0.258 & 0.229 & 11.2\% \\ \bottomrule
\end{tabular}
\end{table}

\begin{table}[]
\centering
\caption{Formation factor. Comparison between three models. The error for baseline CNN, baseline GNN, equivariant GNN are shown for five different splits of data. The improvement indicates the improvement of equivariant GNN over GNN. }
\label{tab:compformationfactor}
\begin{tabular}{@{}lllHlHl@{}}
\toprule
seed & CNN & GNN & equivariant CNN & Equivariant GNN & BL & Improvement \\ \midrule
1 & 0.081 & \vphantom{\sout{0.0895}} 0.048 & 0.0390 & 0.039   &  0.0335 & 17.6\%\\
2 & 0.091 & \vphantom{\sout{0.1090}} 0.049 & 0.0392 & 0.044   &  0.0345 & 9.9\%\\
3 & 0.129 & \vphantom{\sout{0.0706}} 0.050 & 0.0395 & 0.043   &  0.0344 & 14.6\%\\
4 & 0.127 & \vphantom{\sout{0.0704}} 0.051 & 0.0393 & 0.042   & 0.0335 & 18.3\%\\
5 & 0.151 & \vphantom{\sout{0.0684}} 0.047 & 0.0397 & 0.039   &  0.0333 & 17.4\% \\ \midrule
mean & 0.116 & \vphantom{\sout{-}} 0.049 & c & 0.041 & a & 15.6\%\\ \bottomrule
\end{tabular}
\end{table}

Previously published works that employ convolutional neural networks for effective permeability predictions are often trained 
with data set way smaller than those used
in other scientific disciplines. For instance, 
the convolutional neural network predictions on velocity field employ only 
1080 3D images with $80^3$ voxels in \citet{santos2020poreflow} whereas similar research work
on predicting effective permeability by \citet{srisutthiyakorn2016deep} and \citet{sudakov2019driving}
employs database that consists
of 1000 images with $100^3$ voxels and 9261 3D images with $100^3$ voxels respectively. In our cases, we employ 300 3D images with $150^3$ voxel data for the training and test where only 2/3 of the data are used for training the neural network and 1/3 of the data are used in the test cases that measure the performance reported in Tables \ref{tab:comppermeability} and \ref{tab:compformationfactor}. As a comparison,  the common datasets showcasing the advantage of \enn is QM9 are much larger (133,885 small molecule graphs) and more diverse than the dataset employed in this study. 
However, since both the pore-scale numerical simulations and experiments are expensive \citep{arns2004virtual, sun2011connecting, andra2013digitalb, sun2011multiscale, vlassis2020geometric, wang2021non, fuchs2021dnn2, heider2021offline}, one may argue that the ability of the neural network to function properly with a relatively smaller dataset is crucial for practical purposes. 
 Although the data set may not play to our advantage, \enn still outperforms GNN. The advantage of \enn over GNN is much more prominent in the case of the augmented dataset, shown in Figure \ref{fig:aug-all}. In addition, equivariance \wrt SE(3) group guarantees the material frame indifference. 

\begin{figure}[htbp]
\begin{center}
\includegraphics[scale=.45]{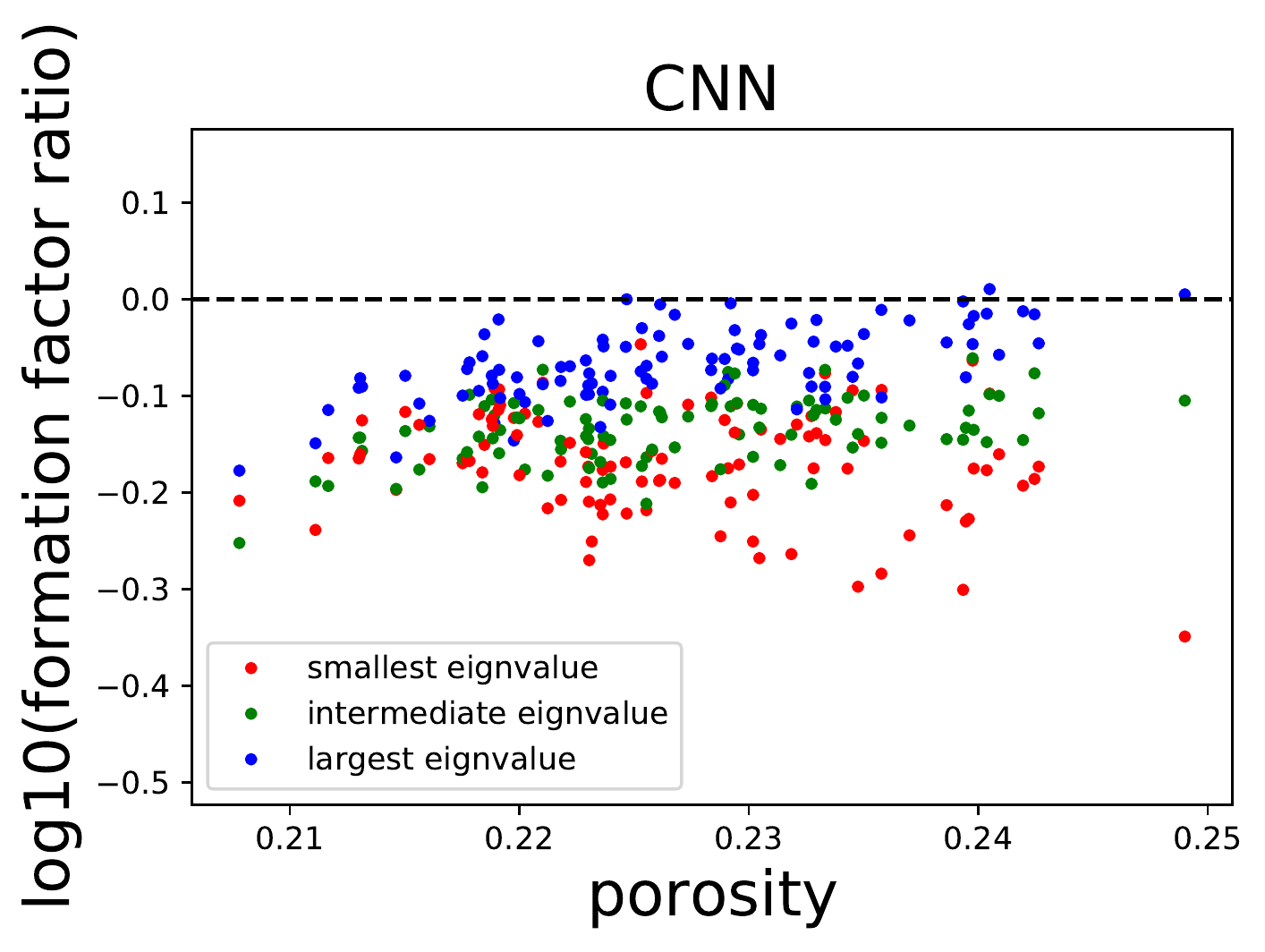}
\includegraphics[scale=.45]{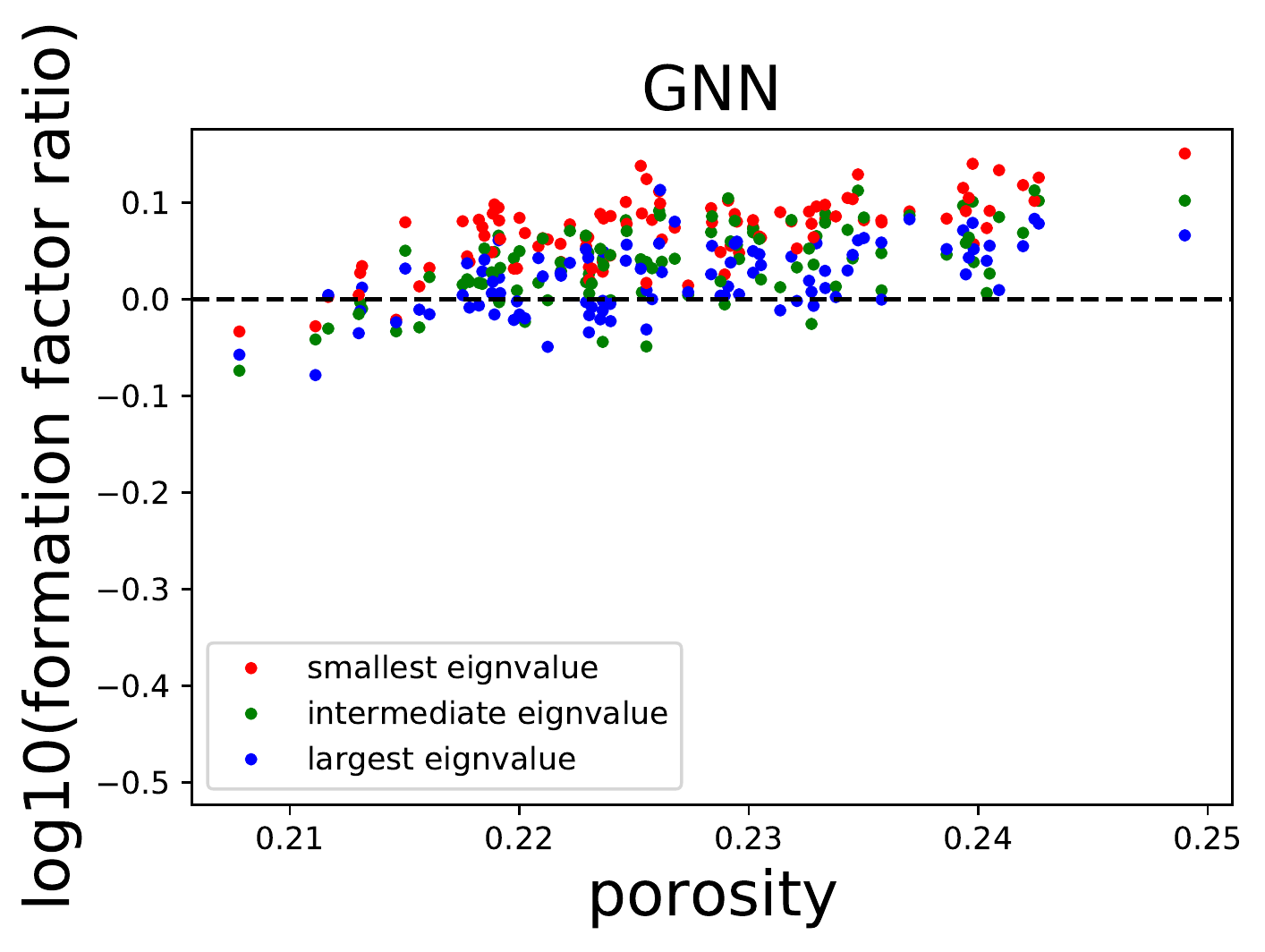}
\includegraphics[scale=.45]{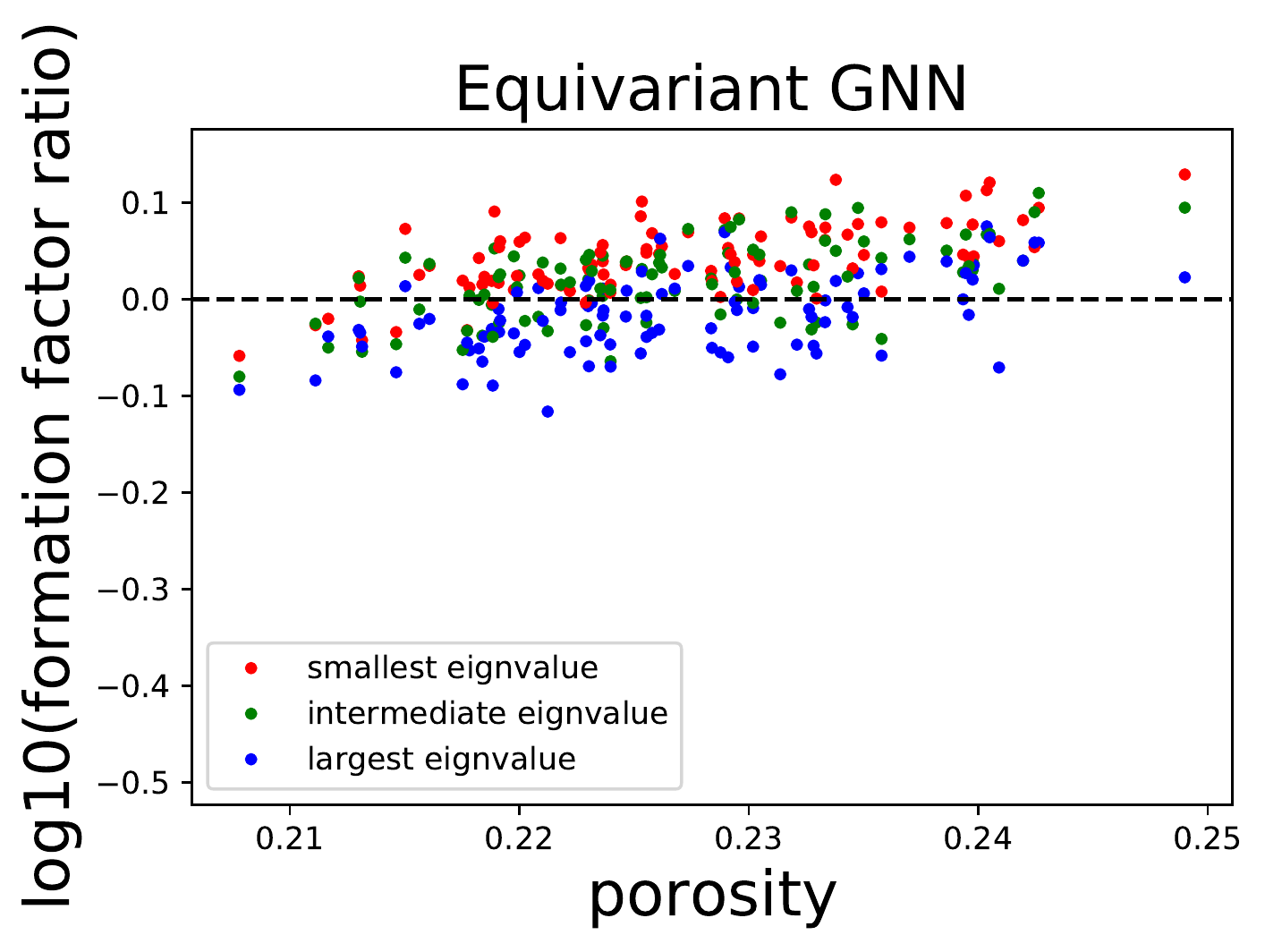}
\caption{For each test data, we plot the ratio of predicted eigenvalues over true eigenvalues of the formation factor tensor. From left to right: prediction of CNN, GNN, and equivariant GNN.}
\label{fig:eigenformationfactor}
\end{center}
\end{figure}

\begin{figure}[htbp]
\begin{center}
\includegraphics[scale=.45]{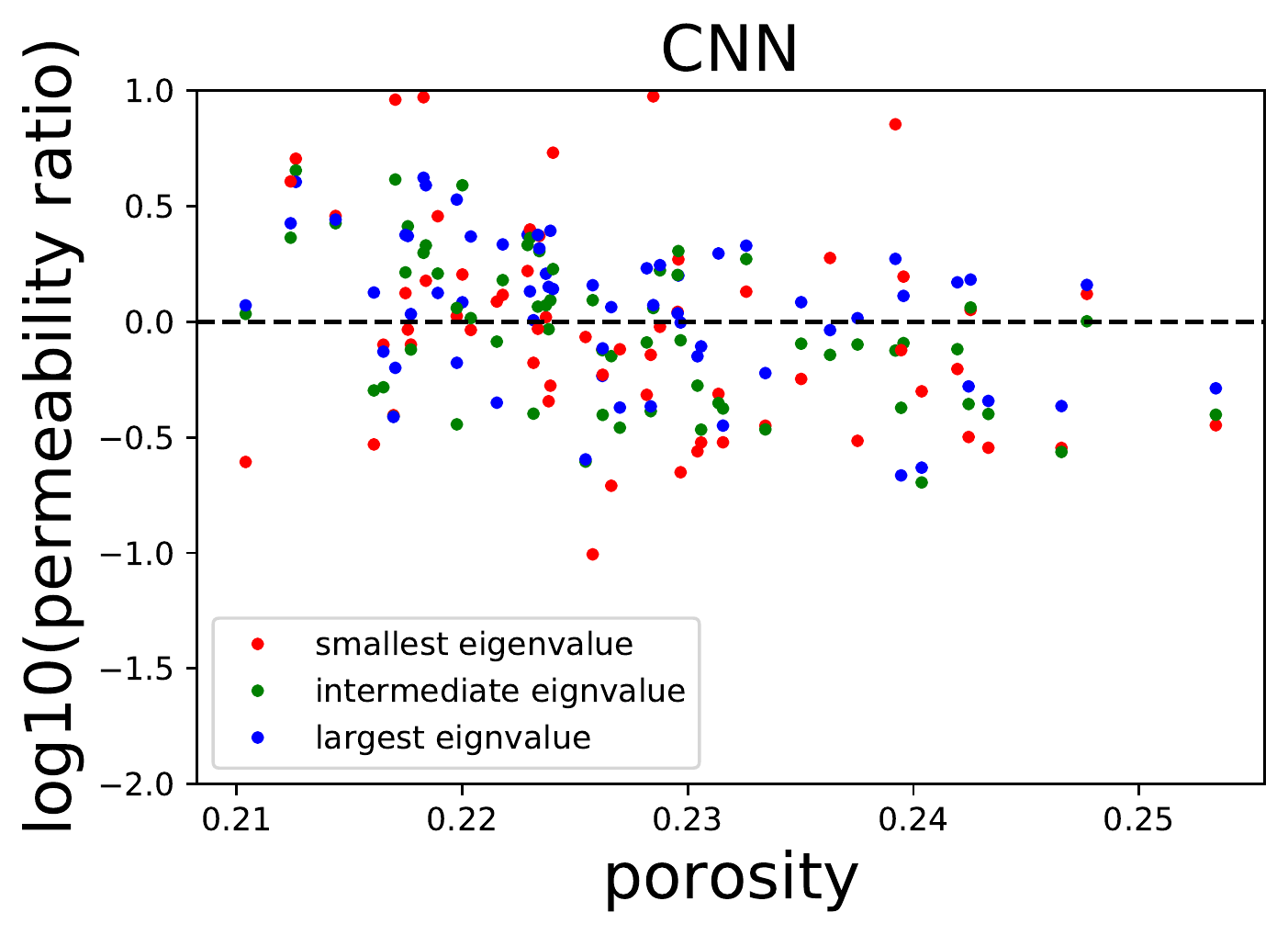}
\includegraphics[scale=.45]{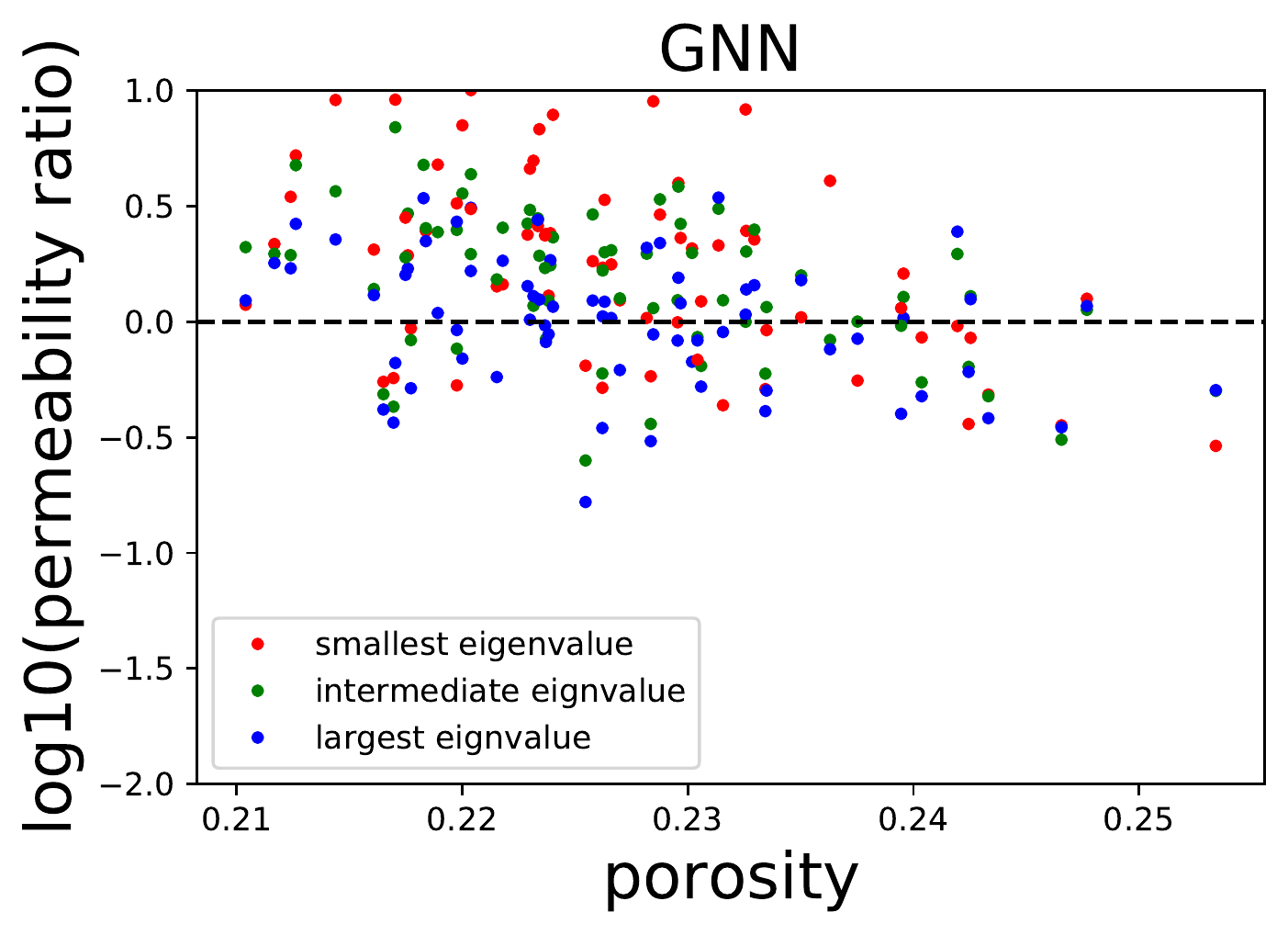}
\includegraphics[scale=.45]{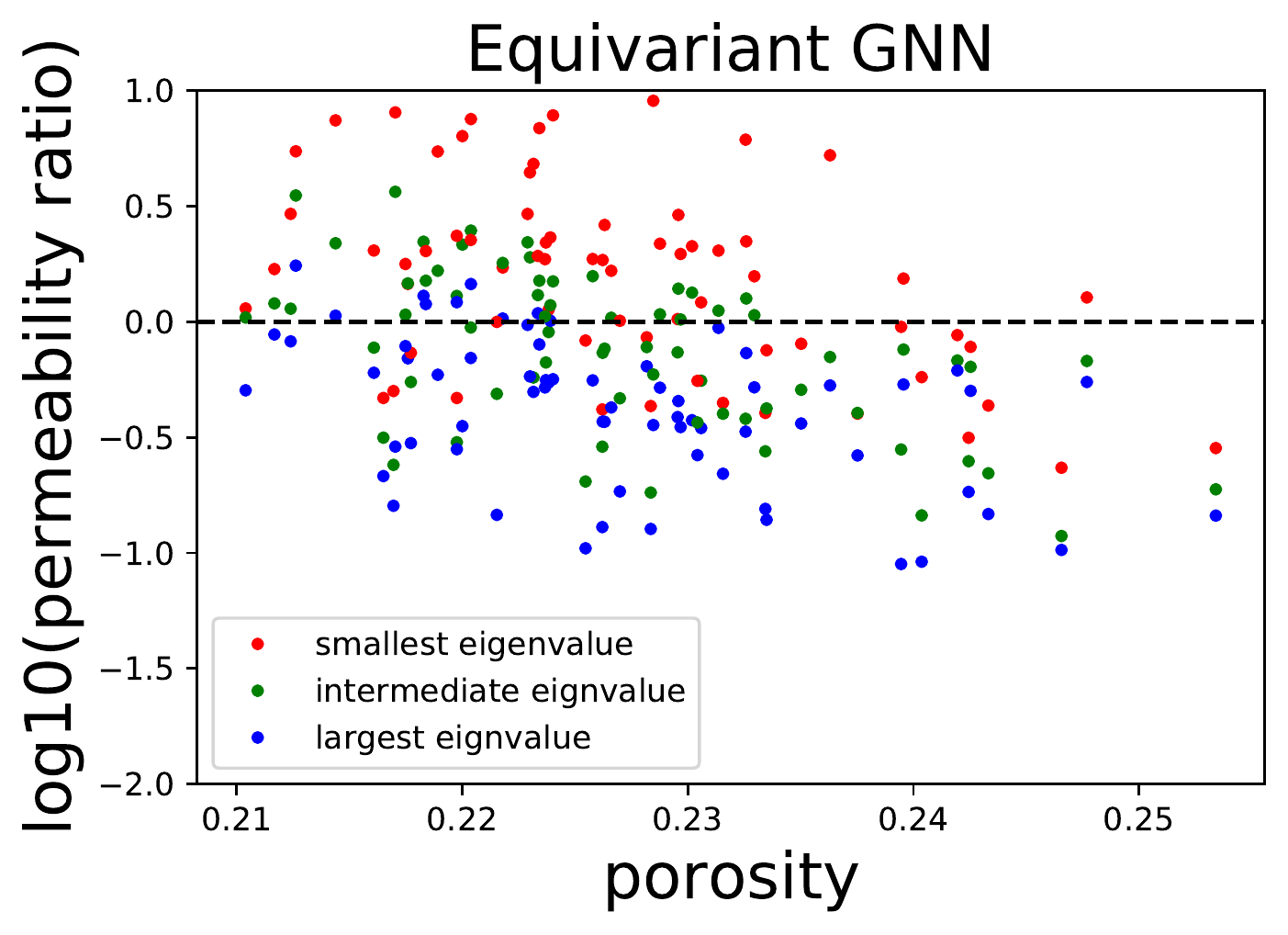}
\caption{For each test data, we plot the ratio of predicted eigenvalues over true eigenvalues of the effective permeability tensor. From left to right: prediction of CNN, GNN, and equivariant GNN. } 
\label{fig:eigenpermeability}
\end{center}
\end{figure}

Figures \ref{fig:eigenformationfactor} and \ref{fig:eigenpermeability} showcase results of 100 \textbf{blind predictions} on formation factor and effective permeability on binary micro-CT images that are excluded from the training database. 
In both cases, we compare the ratio of the major, immediate and minor principle values of the effective permeability predicted by CNN, GNN, and equivariant GNN over the benchmark calculation done by the FFT solver. If the results are perfect, the ratio is 1. In both formation factor and effective permeability predictions, there are few trends that worth noticing. First, 
the formation factor predictions are generally more accurate than that of the effective permeability. Second, the benefits of using the GNN and equivariant GNN are more obvious in the formation factor predictions than that of the effective permeability.
This might be attributed to the fact that the surface conductivity leads to the topology of the flux in the inverse problem correlates more strongly with the Morse graph. 

Another common trend we noticed is that the predictions on the two graphs neural network tend to underestimate the largest eigenvalues of the formation factor and more likely to overestimate the middle and smallest eigenvalues, whereas the CNN generally leads to underestimation of all eigenvalues with the smallest eigenvalues being the most underestimated. 
This trend, however, is not observed in the effective permeability predictions in Figure \ref{fig:eigenpermeability}. 

On the other hand, the CNN predictions on effective permeability do not exhibit a similar trend. While the errors for the 
CNN permeability predictions do not exhibit a clear pattern, the equivariant GNN and, to a lesser extent, the GNN predictions both show that the largest eigenvalues tend to be more likely to be underestimated whereas the smallest one is likely to be overestimated. These results suggest that the prediction seems to be less accurate when the anisotropic effect is strong. 

\begin{figure}[htbp]
\begin{center}
\includegraphics[scale=.5]{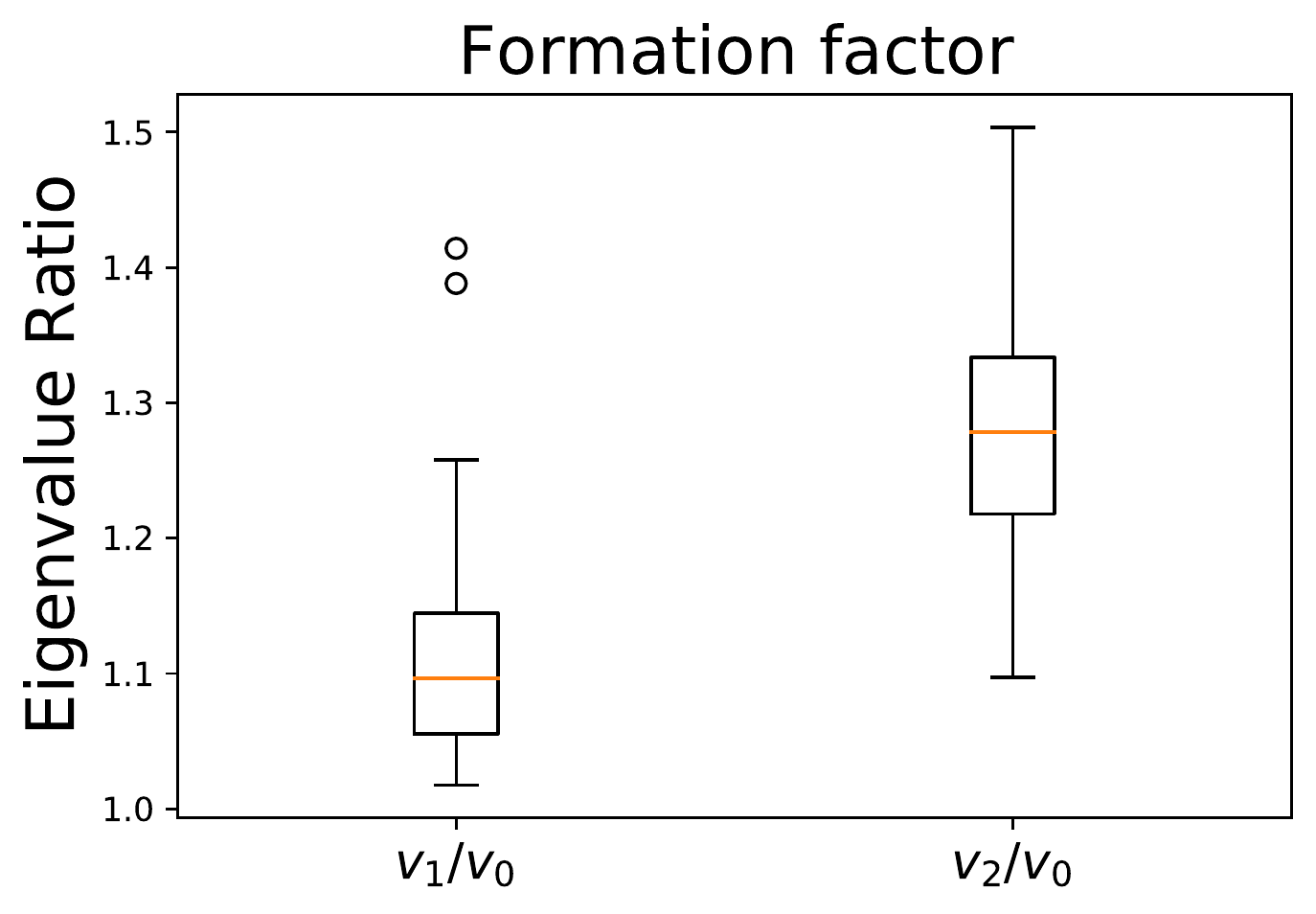}
\includegraphics[scale=.5]{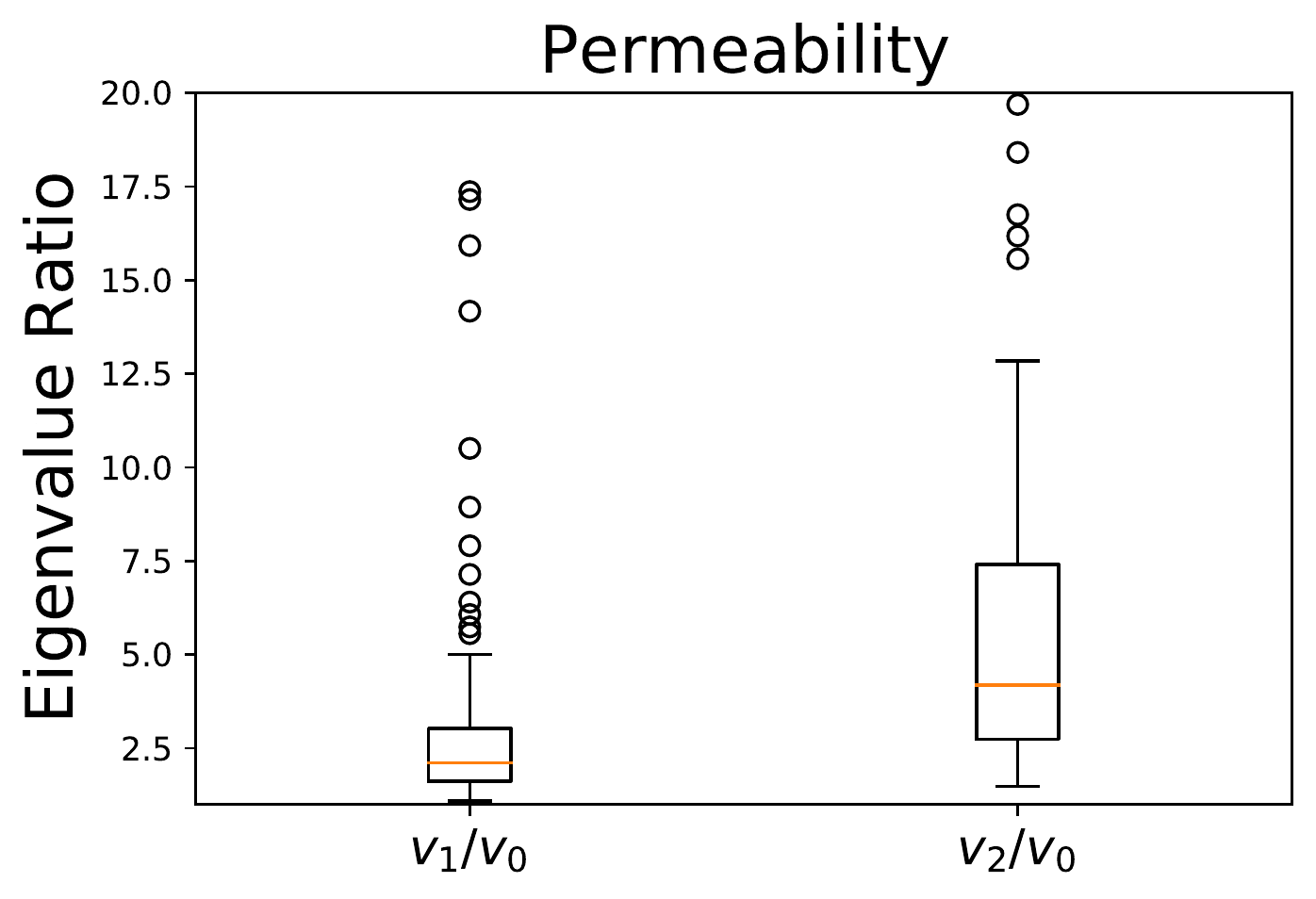}
\caption{Let $v_{0}, v_{1}, v_{2}$ denote the eigenvalues of true permeability/formation factor from small to large. We plot the distribution of eigenvalue ratios over all data.}
\label{fig:eig_ratio}
\end{center}
\end{figure}

To examine this issue further, we compare the spectrum of the eigenvalues of both formation factor and effective permeability in Figure \ref{fig:eig_ratio}. 
It can be seen that formation factor has a much smaller eigenvalue ratio compared to permeability. While the ratio of the eigenvalues of the effective permeability may vary by about an order of magnitude, that of formation factor is generally within a factor of 2 and hence suggesting the nearly isotropic nature of the formation factor. Since the formation factor predictions are about one order more accurate than the effective permeability predictions, these results suggest that the degree of anisotropy may play an important role in the accuracy of the GNN and equivariant predictions. This limitation might be circumvented by including more anisotropic data in the training dataset or different graph representations that better captures the anisotropy of the pore space. Researches on both topics are ongoing but are out of the scope of this study. 

\begin{figure}[htbp]
\begin{center}
\includegraphics[scale=.5]{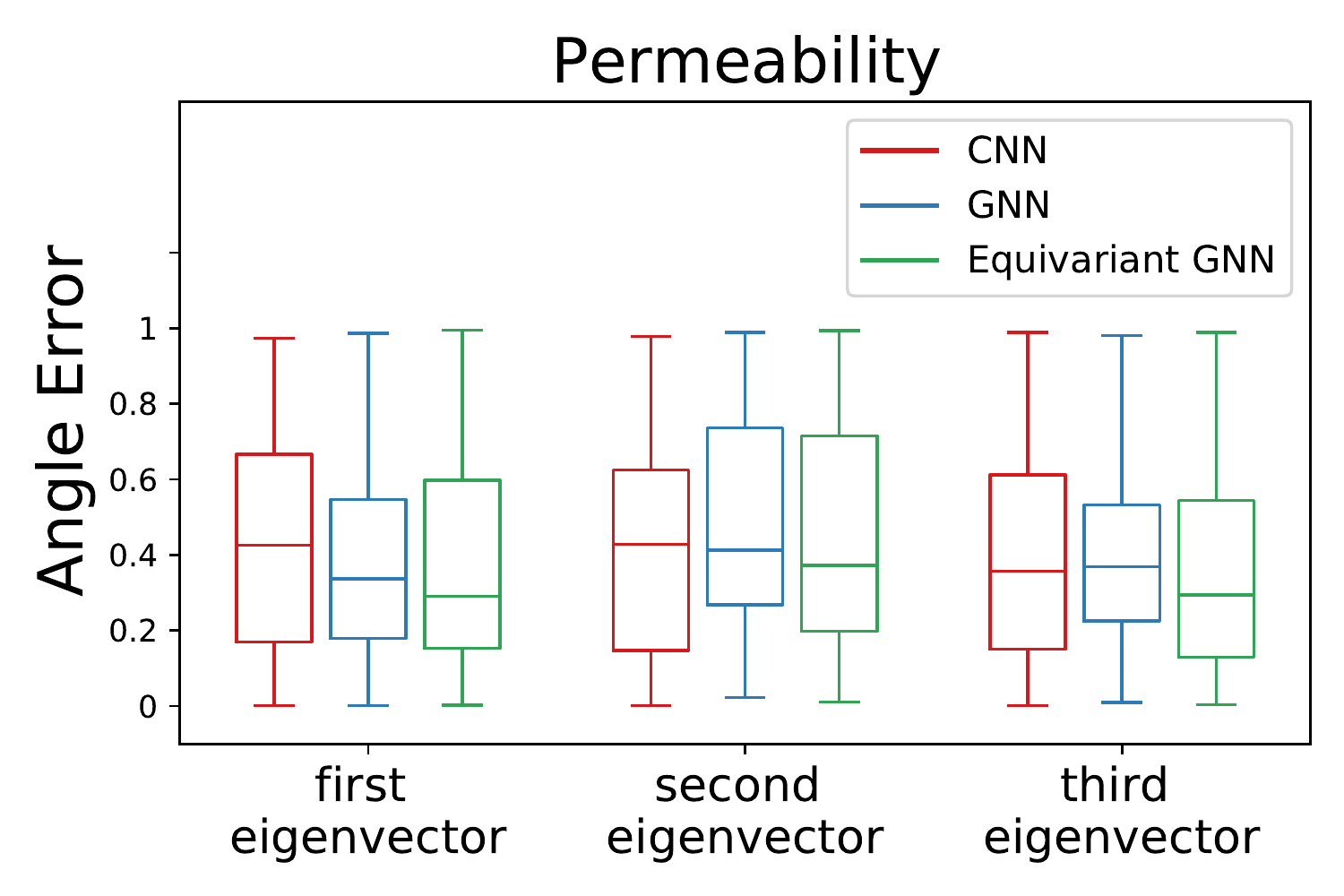}
\includegraphics[scale=.5]{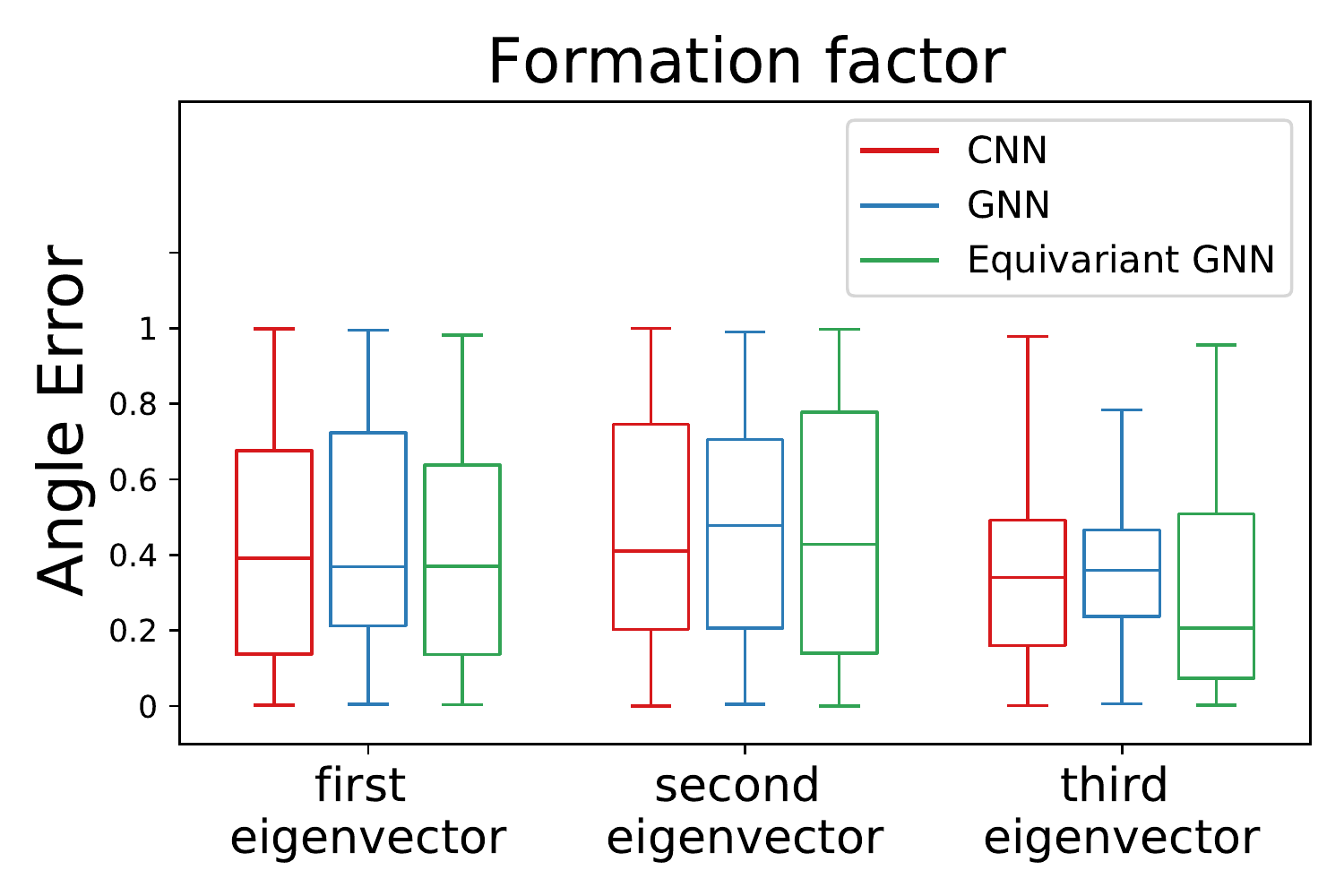}
\caption{The plot of angle error for different property and models. First/second/third eigenvector denotes the eigenvectors corresponding to the smallest, intermediate, and largest eigenvalues.} 
\label{fig:angle}
\end{center}
\end{figure}

In addition to the eigenvalues, we also compare the predictions of the principal directions of both formation factor and effective permeability. Notice that the permeability tensors obtained from the inverse problems are often non-symmetric (see \citet{white2006calculating} and \citet{sun2018prediction}) whereas the formation factors we obtained from the FFT are very close to symmetric despite machine precision error. 
Given the true and predicted permeability/formation factor tensor, we compute its eigenvectors (unit length) and show distribution (in boxplot) of angle error.
Angle error measures how close the predicted eigenvector $\hat{v}$ is to true eigenvector $v$, measured in terms of $1-| v \cdot \hat{v} | \in [0, 1]$. Equivariant GNN is better at predicting the eigenvector corresponding to the largest eigenvalue (but not necessarily best for other eigenvectors), which might be the reason that it performs better than the other two models.
Notice that the spectrum of the eigenvalues for the formation factor is very narrow and hence predicting the eigenvectors of the formation factor could be more difficult but of less importance. On the other hand, since the permeability exhibits a higher degree of anisotropy, the eigenvector predictions are more crucial for practical purposes. Closer examination of the eigenvector predictions performed by CNN, GNN, and the equivariant GNN indicates that 
the equivariant neural network may lead to modest improvements in the accuracy of the orientation of the principal directions (see Figure \ref{fig:angle}). 
Presumably, one may expect that this improvement can be more significant with more data. A systematic study on the relationship between the amount of training of data and performance may provide a fuller picture on this issue but is out of the scope of this research.

\section{Conclusions}
This paper introduces the equivariant GNN and Morse graph representation to enable a direct image-to-prediction workflow that provides formation factor and effective permeability images of a series of sandstone images with improved accuracy. There are several departures from the standard convolutional neural network that may attribute to the improved performance. First, the introduction of Morse graphs to represent the pore topology and geometrical information provide a more efficient way to represent and store the microstructures. Second, 
incorporating SE(3) equivariance constraint into the neural network is the key to enforce the material frame indifference and ensure that the machine learning predictions are not affected by the observer's frame. This work demonstrates how graph and group representations can be leveraged to enforce physics principles for mechanistic predictions relevant to image-based engineering analysis. 

\section{Acknowledgements}
Chen Cai would like to thank the helpful discussion on theory and implementation of \enn with Maurice Weiler, Pim de Haan, Fabian Fuchs, Tess Smidt and Mario Geiger.
This collaborative work is primarily supported by grant contracts OAC-1940203, OAC-1940125 and OAC-2039794. The additional efforts and labor hours of the Columbia Research team members are supported by the NSF CAREER grant from the Mechanics of Materials and Structures program at National Science Foundation under grant contract CMMI-1846875, the Dynamic Materials and Interactions Program from the Air Force Office of Scientific Research under grant contracts FA9550-19-1-0318 and Army Research Office under grant contract W911NF-18-2-0306. These supports are gratefully acknowledged.
\bibliography{permeabilityGNN}
\end{document}